\documentclass[fleqn,usenatbib]{mnras}

\usepackage{newtxtext,newtxmath}
\usepackage[T1]{fontenc}
\usepackage[hyphenbreaks]{breakurl}

\DeclareRobustCommand{\VAN}[3]{#2}
\let\VANthebibliography\thebibliography
\def\thebibliography{\DeclareRobustCommand{\VAN}[3]{##3}\VANthebibliography}

\usepackage{graphicx}	% Including figure files
\usepackage{amsmath}	% Advanced maths commands
\usepackage{adjustbox}
\usepackage{rotating}
\usepackage{caption}
\usepackage{lscape}
\title[Spectro-temporal investigation of GX 349+2]{Broadband spectro-temporal investigation of neutron star low-mass X-ray binary GX 349+2}

\author[U. Kashyap et al.]{
Unnati Kashyap$^{1}$\thanks{E-mail: phd1801121005@iiti.ac.in},
Manoneeta Chakraborty$^{1}$,
Sudip Bhattacharyya$^{2}$,
and Biki Ram$^{1}$
\\
$^{1}$ DAASE, Indian Institute of Technology Indore, Khandwa Road, Simrol, Indore - 452020, India\\
$^{2}$ Department of Astronomy and Astrophysics, Tata Institute of Fundamental Research, 1 Homi Bhabha Road, Colaba, Mumbai - 400005, India\\
}
\date{Accepted XXX. Received YYY; in original form ZZZ}

\pubyear{2015}

\begin{document}
\label{firstpage}
\pagerange{\pageref{firstpage}--\pageref{lastpage}}
\maketitle

\begin{abstract} 
We report a broadband investigation of the Z-type neutron star (NS) low mass X-ray binary (LMXB) GX 349+2 using {\em AstroSat} and {\em NICER}. {\em AstroSat} observed the source exhibiting large scale variability in its normal branch (NB) /flaring branch (FB) vertex and flaring branch (FB) and a moderate evolution during {\em NICER} observations. The power spectra exhibit very low-frequency noise (VLFN) and low-frequency noise (LFN)/flaring branch noise (FBN), described by a power law and an evolving Lorentzian. We investigate the energy dependence of variability components and their correlation with the spectral state to probe their origin. The joint spectra of GX 349+2 are modeled by two thermal and one non-thermal component. The source moves along the Z track, with the increasing accretion rate, further heating of the NS boundary layer, and increasing temperature/radius of the brightened hotspot at the disc-boundary layer interface/NS surface. A power law well represents the hard non-thermal coronal emission. As predicted by the gravitational redshift, we find a correlation between the line energy detected in {\em NICER} spectra and the inner disc radius with the Spearman rank correlation coefficient of 1. Using this correlation, we demonstrate the potential of a method to constrain the accreting compact object properties, including evolving continuum and line spectroscopy. We report the first detection of hard lag providing evidence of the VLFN originating from the accretion disc in NS LMXBs, representing fluctuation of propagation through the disc.

\end{abstract}


%%%%%%%%%%%%%%%%%%%%%%%%%%%%%%%%%%%%%%%%%%%%%%%%%%

%%%%%%%%%%%%%%%%% BODY OF PAPER %%%%%%%%%%%%%%%%%%

\section{Introduction}
\label{section1}

Based on the spectro-temporal properties, Neutron star Low Mass X-ray Binaries (NS-LMXBs) can be classified as Z-sources and atoll sources \citep{1989A&A...225...79H}. The Z-sources trace out Z-shaped tracks in the Hardness Intensity Diagrams (HIDs) consisting of three branches, called horizontal branch (HB), normal branch (NB), and flaring branch (FB). On the other hand, the atoll sources trace a  well-defined banana and island state in the  HIDs. The banana branch in atoll HID is further split into the upper banana with high luminosity and the lower banana with lower luminosity \citep{2004astro.ph.10551V}. So far, there are six known persistent Z sources, and they are further classified into two sub-classes: Cyg-like and Sco-like Z sources. Cyg X-2, GX 5-1, and GX 340+0 are identified as Cyg-like Z sources and Sco X-1, GX 17+2, and GX 349+2, are referred to as Sco-like Z sources \citep{1994A&A...289..795K,1997MNRAS.287..495K,2012A&A...546A..35C,2012MmSAI..83..170C}. However, the detections of NS LMXBs, IGR J17480-2446, XTE J1701-462, XTE J1806-246, Cir X-1, ,  which can apparently switch between atoll and Z-like behavior have also been reported \citep{2011ApJ...730L..23C,2011MNRAS.418..490C,2007ApJ...656..420H,1999ApJ...522..965W}. The mass accretion rate in the case of Z sources is thought to increase from the HB, through the NB, to the FB, and the Z track differs from source to source \citep{1989A&A...225...79H,2002ApJ...568L..35M,2007ApJ...656..420H}. The fast variability properties in the NS Z sources are highly dependent on their position within the Z in the HID. Kilohertz quasi-periodic oscillations (kHz QPOs) and a 15-60 Hz QPO also known as Horizontal Branch Oscillation (HBO), are proposed to occur on the HB and upper NB \citep{2004astro.ph.10551V}. Normal branch oscillation (NBO) at $\sim$ 6 Hz and power law noise $<$ 1 Hz are observed to occur on the lower NB and the FB \citep{2004astro.ph.10551V}. Moreover, the detection of low-frequency QPOs and kHz QPOs (200-1200 Hz) from Z-type sources have also been reported \citep{2000ARA&A..38..717V,1998ApJ...493L..87W,1998ApJ...499L.191J,1998ApJ...504L..35W,2020JHEAp..25....1J,1998ApJ...499L.191J,1998AIPC..431..381W} previously.  

The Sco-like Z source GX 349+2 is known to be peculiar among the Z sources \citep{1998A&A...332..845K}. All the 6 known Z sources are observed to display all three branches except for GX 349+2, which is observed either in the NB or in the FB \citep{1989A&A...225...79H,1995NYASA.759..344K}. A broad peaked noise with centroid frequency at 6 Hz on the FB of its Z track was observed from the NS-LMXB source \citep{1988MNRAS.231..999P}. During the RXTE study along the Z-track of GX 349+2, a broad peaked noise at 3.3–5.8 Hz on the FB and a new peaked noise feature at 11–54 Hz on the NB and FB were detected \citep{2002MNRAS.336..217O}. Although \cite{2003A&A...398..223A} mention the possibility that the peaked noise and  normal branch oscillation (NBO)/flaring branch oscillation (FBO) are of the same origin, \cite{2002MNRAS.336..217O} argues with the fact that the NBO/FBO peaks are much narrower than the broad peaks detected and FBOs are generally present in the 10-20 \% of the FB whereas the peaked noise is detected all along the Z track of GX 349+2. \cite{2002MNRAS.336..217O} further stated that the peak detected in NB and FB are low-frequency noise (LFN) and /flaring branch noise (FBN). Additionally, the LFN and FBN are thought to have different origins based on the properties of the LFN/FBN peak as well as the difference in NB/FB mass accretion rate along the Z track. But, there is a possibility of the evolution of LFN into FBN. The detections of twin kilohertz (kHz) QPOs with lower and upper frequencies of 712 and 978 Hz were also reported from this Sco-like Z source \citep{1998ApJ...500L.167Z}. A QPO with centroid frequency $\sim$ 3.8 Hz was reported during the correlated spectro-timing study of GX 349+2 \citep{2003A&A...398..223A}. \\

 In the detailed spectral study of  GX 349+2, using RXTE data, the spectra in 2.5–25 keV were observed to be well described by a two-component model consisting of a disc blackbody and a Comptonized component representing Comptonization in the hot central corona or the boundary layer \citep{2003MNRAS.346..933A}. A hard tail in the X-ray broadband (0.1-200 keV) spectrum was detected in the flaring branch of the source using BeppoSAX \citep{2001ApJ...554...49D}. In the cross-correlation study between the soft and hard state light curves of GX 349+2, positively correlated/anti-correlated time lags were detected throughout/outside the most extensive Z track traced by the source in its HID indicating different mechanisms behind the two types of time lags \citep{2016MNRAS.455.2959D}. The XMM spectra of GX 349+2 in the 0.7–10 keV energy band were observed to contain features in the Fe-$K_{\alpha}$ region and were proposed to form due to the reflection in the inner disc region \citep{2009A&A...505.1143I}. The relativistic Fe-K emission line detection from GX 349+2 has been reported in other studies as well \citep{2008ApJ...674..415C,2018ApJ...867...64C}.

In this work, we report a broadband investigation of the Sco-like peculiar Z-source GX 349+2 using {\em AstroSat} and NICER. The X-ray instruments onboard {\em AstroSat} observed the source in its NB/FB vertex (the soft apex known as the boundary between stable and
unstable nuclear burning \citep{2012MmSAI..83..170C}) and FB branch of its Z-track. The source exhibits large-scale variability, and the broadband capability of AstroSat helps us investigate the evolution of the source throughout different intensity states along the Z-track (see Section \ref{section3.1}). We also investigate different variabilities and their frequency/energy dependence to probe their physical origin (see Section \ref{section3.2}). The joint simultaneous broadband spectra of GX 349+2 are well modelled by two thermal and one non-thermal component, and we study and inspect the possible physical scenario corresponding to each of these model components (see Section \ref{section3.3.1}). {\em NICER}'s decent energy resolution and good sensitivity enable the detection of the iron (emission) line and time lag from GX 349+2 (see Section \ref{section3.3.2}). This kind of broadband and sensitive spectro-temporal correlation study helps us to get an insight into the source evolution along the different tracks traced by the NS Z sources.

\section{Observations and Data reduction}
\label{section2}

\begin{figure}
     \includegraphics[width=0.50\textwidth]{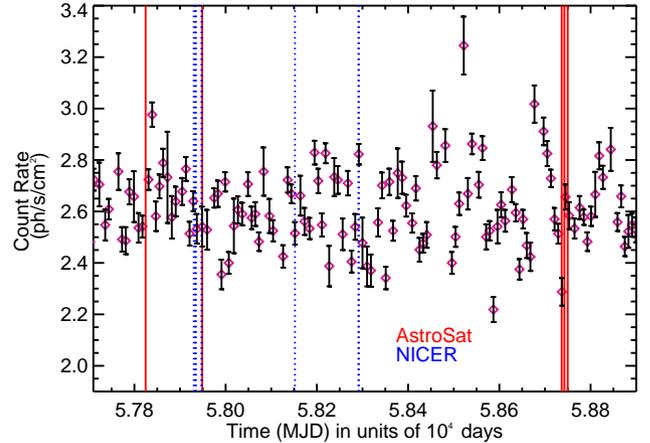}
     \caption{The two-days binned MAXI light curve of GX 349+2. The red and blue vertical lines represent {\em AstroSat} and {\em NICER} observations considered in this work (see Section \ref{section2}).}
     \label{figure1}
\end{figure}

\begin{table*}
\centering
%\begin{tabular}{|c|c|c|c|c|c|p{1cm}p{1cm}p{1cm}p{1cm}p{1cm}p{1cm}p{1cm}p{1cm}p{1cm}|}
\caption{ {\em AstroSat} Observations of GX 349+2 (see Section \ref{section2.1}). }

\begin{tabular}{|c|p{3cm}|c|c|c|}
\hline
Observation (Instrument) & Observation ID & Date (dd-mm-yyyy) & Start time (hh:mm:ss.ss) &Exposure time (s)  \\ \hline

A1 (LAXPC) &9000001078 & 12-03-2017  & 14:06:43  &162500\\
A1 (SXT) &9000001078&12-03-2017&15:15:10& 147100\\
A2 (LAXPC) &9000001384 &14-07-2017 &16:27:41&111700\\
A2 (SXT) &9000001384&14-07-2017&16:21:24&106000\\
A3 (LAXPC) &9000003154& 10-09-2019 &15:09:54&33100\\
A3 (SXT) &9000003154&10-09-2019&15:08:35&34100\\
A4 (LAXPC) & 9000003164 &16-09-2019 &09:58:38&7100\\
A4 (SXT) &9000003164&16-09-2019& 10:26:13&1400\\
A5 (LAXPC) & 9000003196 &23-09-2019 &19:26:15& 23100\\
A5 (SXT) &9000003196&23-09-2019&19:26:14&23400\\
\hline
\label{table1}
\end{tabular}
%\caption{Observation Details (SXT)}
%\label{table2}
\end{table*}

\begin{table*}
\centering
\caption{ {\em NICER} Observations of GX 349+2 (see Section \ref{section2.2}). }

\begin{tabular}{|c|p{3cm}|c|c|c}
\hline
Observation & Observation ID & Date (dd-mm-yyyy) & Start time (hh:mm:ss.ss)&Exposure time (s)   \\ \hline

N1&0034090106 & 26-06-2017 &03:39:53& 46000\\ 
N2&0034090108&29-06-2017&04:34:42&22600\\ 
N3&0034090110&01-07-2017&02:19:40&46800\\
N4&0034090117&13-07-2017&00:24:51&76500\\
N5&0034090118&14-07-2017& 03:47:30&45300\\
N$6^{a}$&1034090101&02-02-2018&23:18:07&24800\\
N7&1034090111&22-06-2018&00:15:26&84700\\ 
N8&1034090112&23-06-2018&00:38:08&2500\\ 
\hline
\end{tabular}
\label{table2}
\begin{flushleft}
\footnotesize{$^a$ Corrected from the geocenter to the solar system barycenter  by setting orbitfiles=GEOCENTER, as there is a mismatch between the orbit file and event file timestamps}\\
\end{flushleft}
\end{table*}

\subsection{{\em AstroSat}}
\label{section2.1}

 We analyse the publicly available {\em AstroSat} archival data of GX349+2 starting from 12 March 2017 to 23 September 2018  over five observational epochs. The observations used in this work are marked on the MAXI light curve, shown in Figure~\ref{figure1}. The details of the observations are mentioned in Table \ref{table1}. 

\subsubsection{Large Area X-ray Proportional Counter (LAXPC)}
\label{section2.1.1}

 One of the primary payloads on {\em AstroSat} is the  Large Area X-ray Proportional Counter (LAXPC)\footnote{\url{https://www.tifr.res.in/~astrosat_laxpc/astrosat_laxpc.html}} instrument. It consists of three identical proportional counter detector units named LAXPC10, LAXPC20, and LAXPC30. It has the largest effective area in the mid-X-ray range, 3-80 keV, and a large area collection of 6000 $cm^{2}$ at 15 keV. The collimator of the LAXPC detectors have a field of view of about $1^\circ \times 1^\circ $. The time resolution of LAXPC is 10 $\mu$s and the good time resolution makes LAXPC ideal for fast timing analyses \citep{2016ApJ...833...27Y,2017CSci..113..591Y, 2017JApA...38...30A}. The dead time of the detector is $\sim$42 $\mu$s.
 
 For this analysis, we processed LAXPC event analysis (EA) mode data using {\tt LaxpcSoft}\footnote{\url{https://www.tifr.res.in/~astrosat_laxpc/software.html}} software \citep{2017ApJS..231...10A}. In the event mode, the arrival times and the energies of each detected photon are recorded. For the LAXPC analysis, we use LAXPC20 data as it provides the maximum gain and sensitivity. To extract lightcurves, spectra, background, as well as response files {\tt LaxpcSoft} software is used. We apply Barycenter correction to the LAXPC level 2 data using the {\tt as1bary} tool at RA of $256.4354^\circ$ DEC of $-36.4230^\circ$.

\subsubsection{Soft X-ray Telescope (SXT) }
\label{section2.1.2}

The Soft X-ray Telescope (SXT) \footnote{\url{https://www.tifr.res.in/~astrosat_sxt/index.html}} \citep[SXT; ][]{2017JApA...38...29S,2021JApA...42...17B} onboard {\em AstroSat} provides spectra in the soft X-ray energy range (0.3-7 keV). The effective area of SXT is 90  $cm^{2}$  at 1.5 keV. Its field of view is  $40\:'$, with a focal length of 2 meters. The time resolution of SXT is 2.37 s, and the point spread function (PSF) size is $3-4\:'$. The good spectral resolution ($\sim$ 150 eV at 6 keV) and sensitivity of SXT enable spectral analysis and variability observations in the soft X-ray regime \citep{2017JApA...38...29S, 2016SPIE.9905E..1ES}.  

 In order to extract the image, light curves, as well as spectrum, cleaned level 2 SXT event files are collected. Then we use {\tt SXTEVTMERGERTOOL} (Julia-based module) to merge different orbits and produce a single level-2 Event file for each observation epoch. For extracting images, light curves, and spectra, {\tt XSELECT} available as a part of the Heasoft 6.29 package is used. The appropriate ancillary response files are produced using {\tt sxtARFmodule}. During spectral analysis, the response file (sxt\_pc\_mat\_g0to12.rmf) and blank sky background spectrum file (SkyBkg\_comb\_EL3p5\_Cl\_Rd16p0\_v01.pha) provided by the SXT team are used. The data are in PC (Photon Counting) mode and FW (Fast Window) mode. In PC mode, data from the entire CCD ($\approx$ 36000 pixels) above specified threshold energy are collected. The time resolution of SXT in this mode is 2.37 s. The FW (Fast Window) mode provides a better timing resolution (278 ms). A circular region of $5'$ radius is used as a source region around the source location using {\tt ds9}.

\subsection{Neutron star Interior Composition Explorer ({\em NICER})}
\label{section2.2}

In this work, we analyse the publicly available {\em NICER} archival data of GX349+2 starting from 26 June 2017 to 23 June 2018 over eight observational epochs. The details of the {\em NICER} observations are mentioned in Table \ref{table2}. 

{\em NICER} is an International Space Station (ISS) payload \citep{10.1117/12.926396}. {\em NICER}'s X-ray Timing Instrument (XTI) comprises a collection of 56 X-ray concentrator optics (XRC) and silicon drift detector (SDD) pairs. {\em NICER} provides good spectral information in the 0.2-12 keV energy range \citep{10.1117/12.2231304}. 

We have analysed the data using Heasoft v6.29c, {\em NICER} software version 2021-08-31\_V008c, and {\em NICER} CALDB version of 20210707. Cleaned event files are generated using {\texttt nicerl2} pipeline and by applying standard filtering criteria. We further exclude the detector IDs 14 and 34 from our analysis and generate response and ancillary response files using the {\tt nicerrmf} and {\tt nicerarf} tools for each observation individually. We create background files for each observation using the {\tt nibackgen3C501} tool \footnote{\url{https://heasarc.gsfc.nasa.gov/docs/nicer/tools/nicer_bkg_est_tools.html}}. To apply barycenter correction for each {\em NICER} observation, we use the ftool {\tt barycorr}.

%NICER version (2021-08-31\_V008c)\\
%CALDB version caldb.indx20210707

\begin{figure*}
     \includegraphics[width=0.33\textwidth]{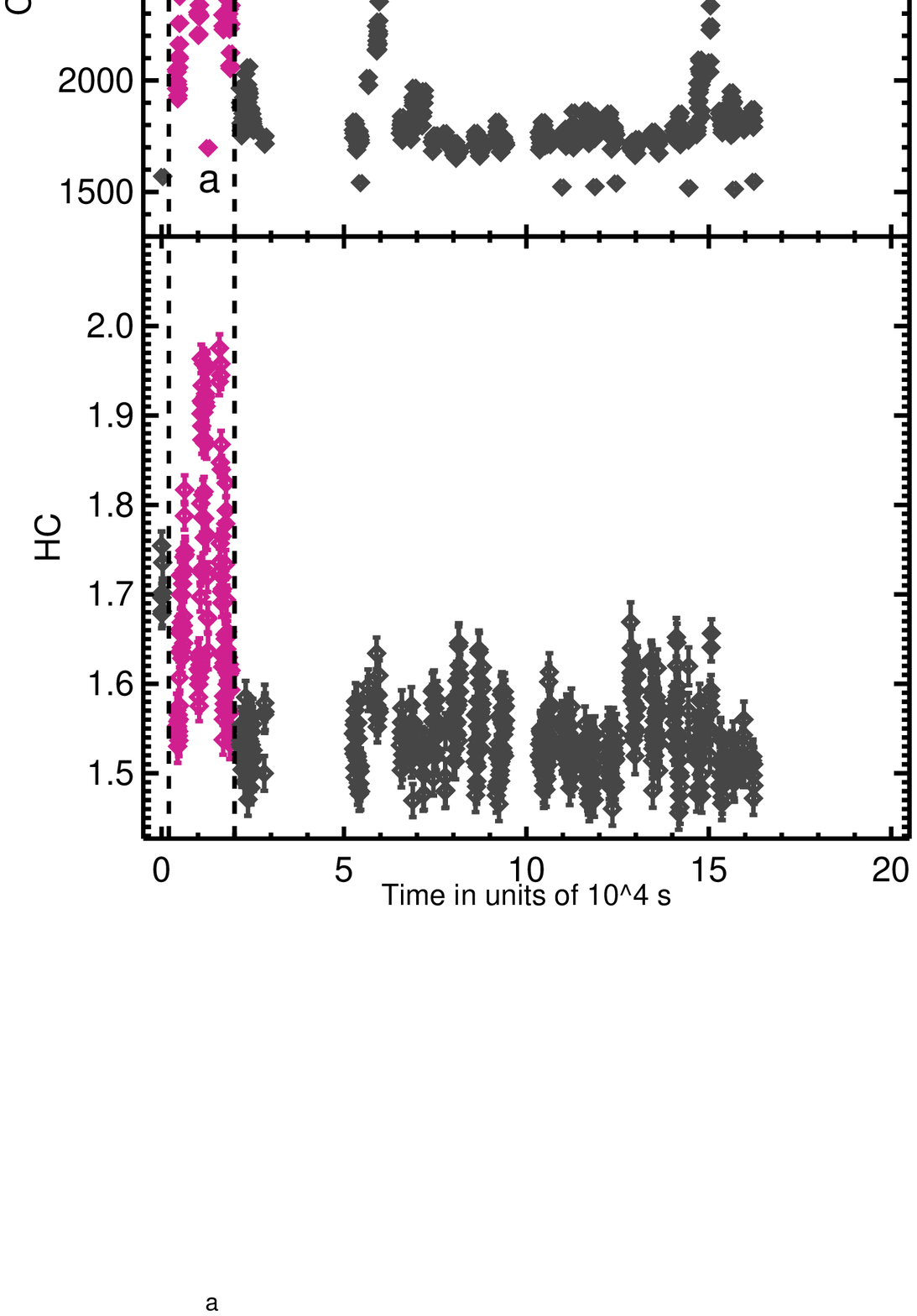}
     \includegraphics[width=0.33\textwidth]{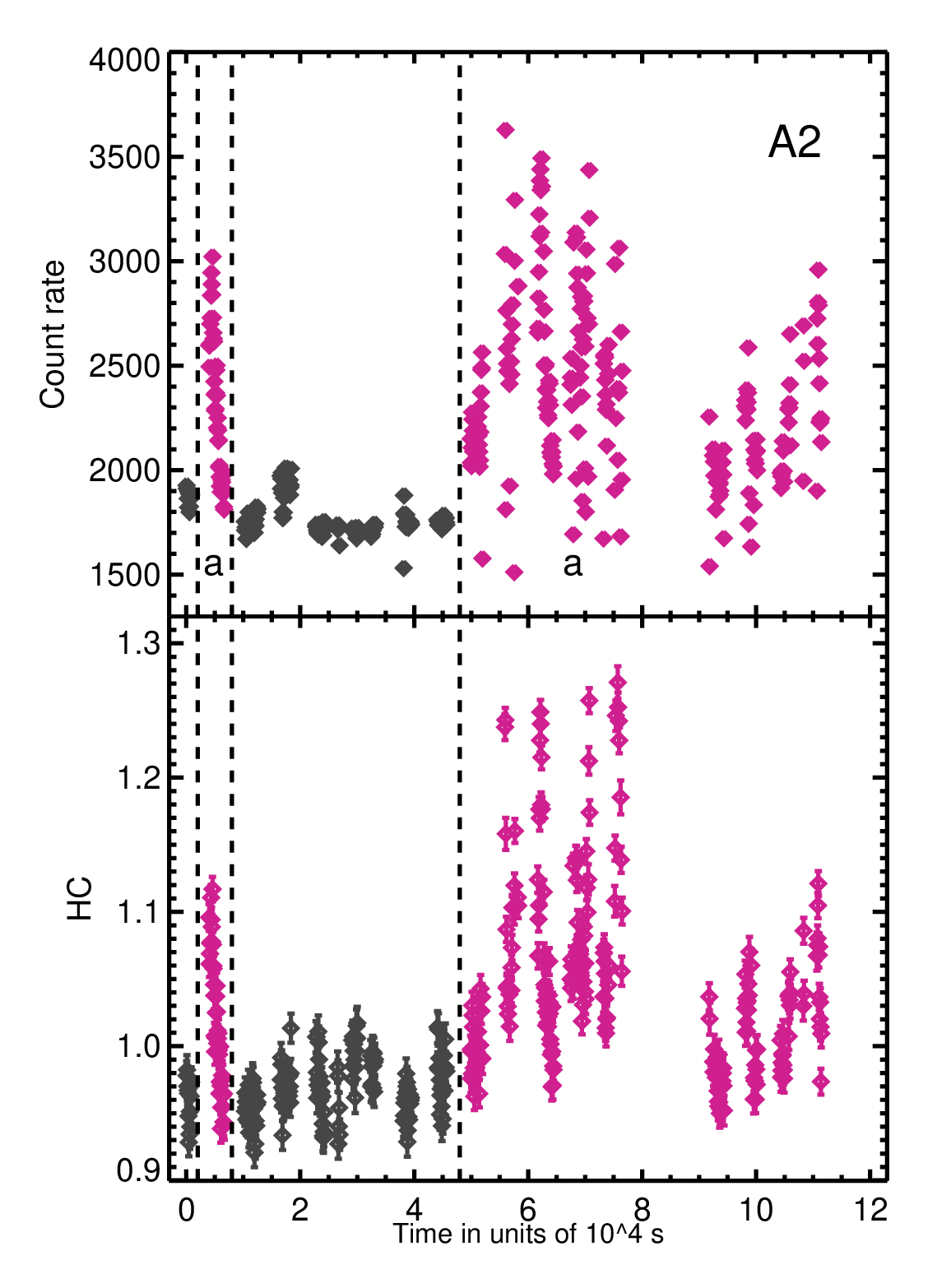} \includegraphics[width=0.33\textwidth]{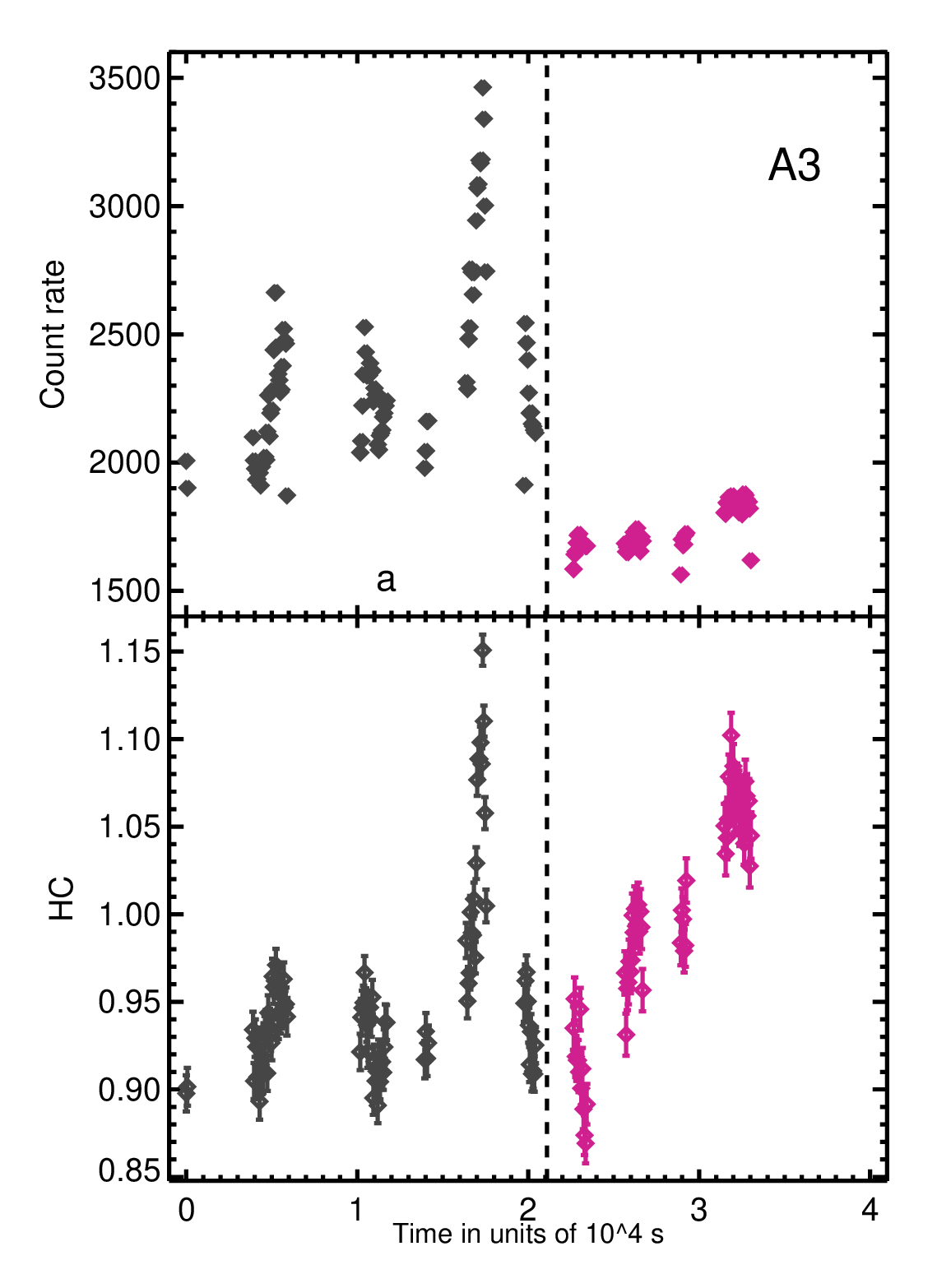}
     \includegraphics[width=0.33\textwidth]{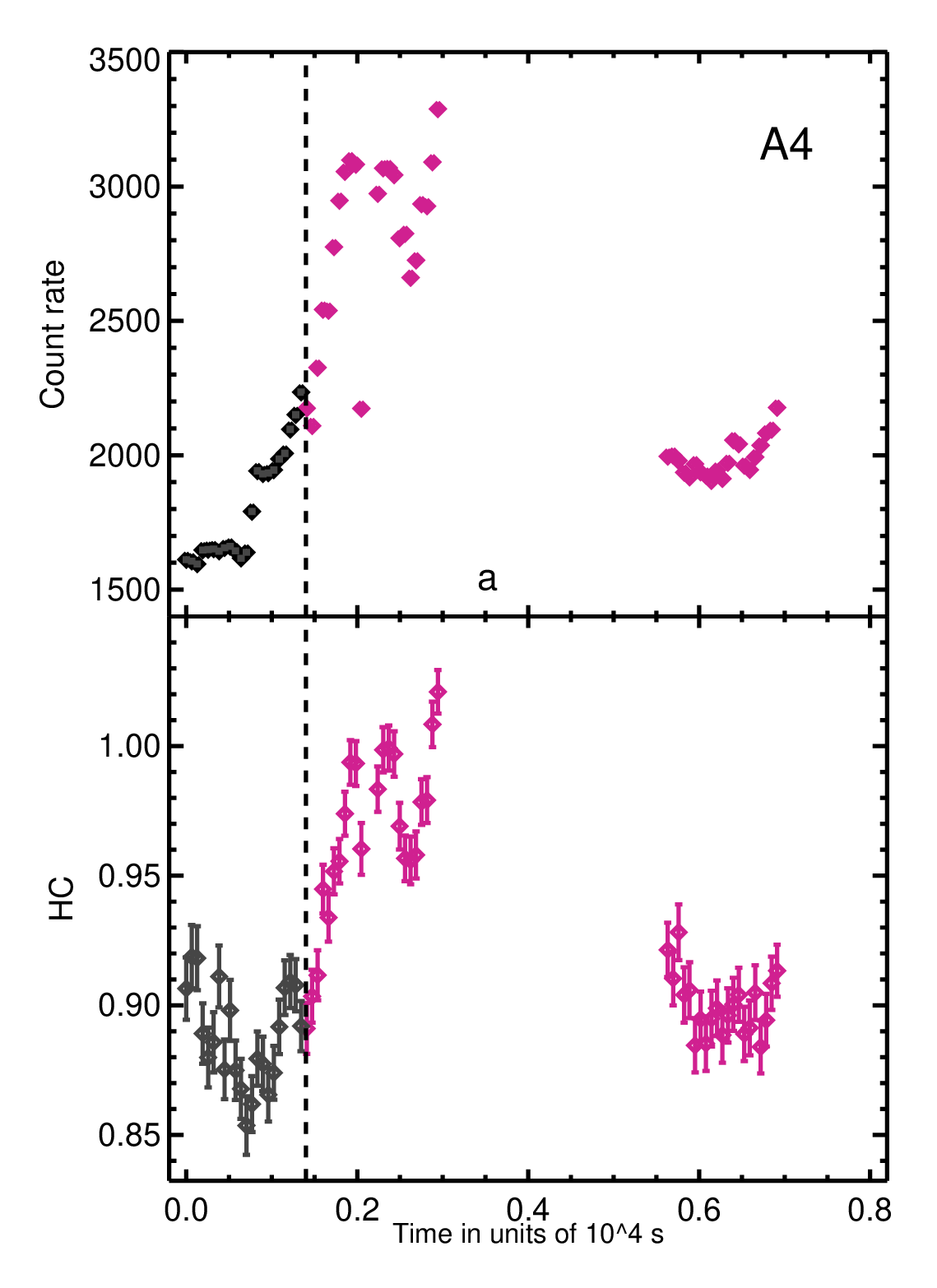}
     \includegraphics[width=0.33\textwidth]{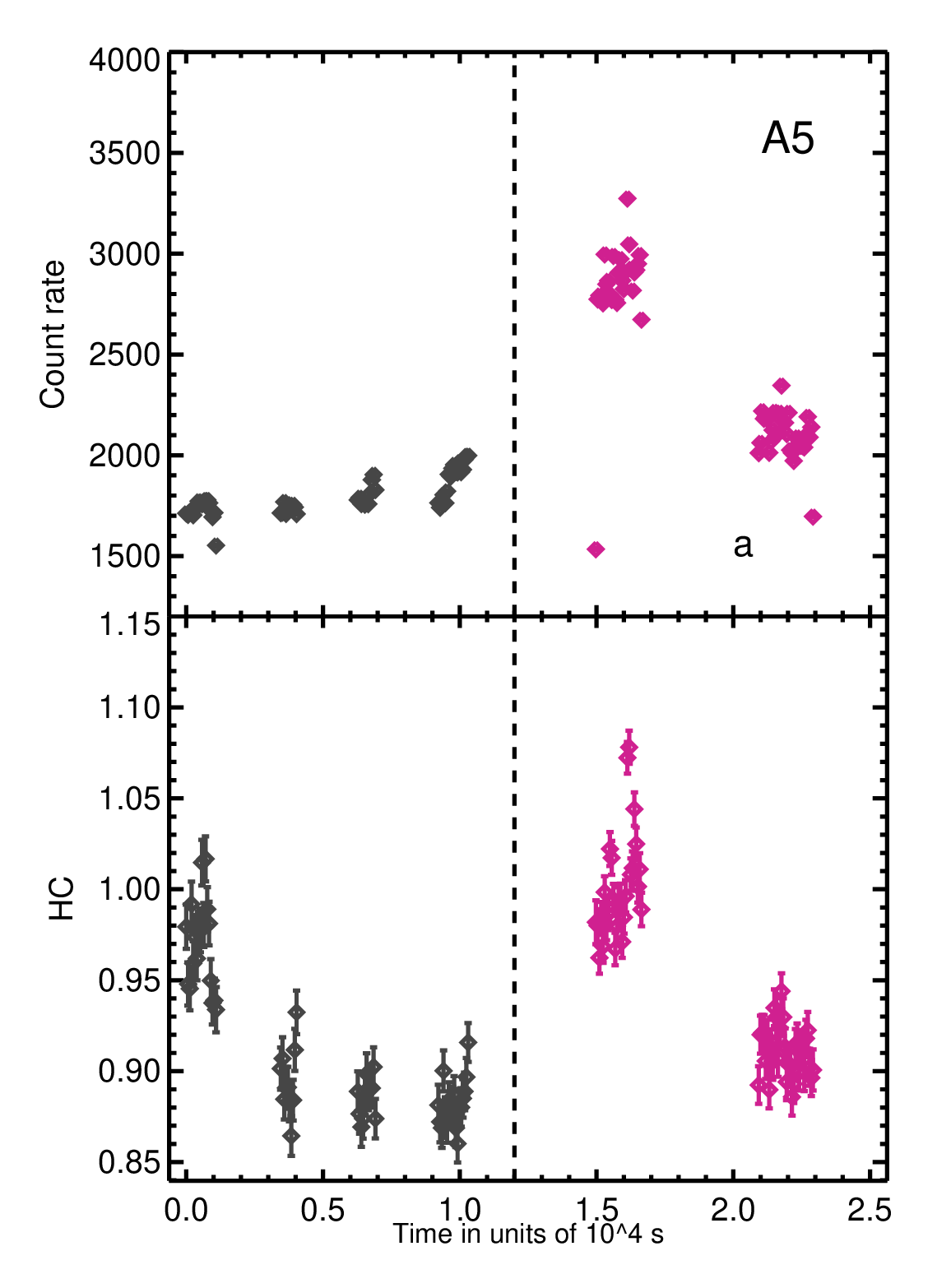}
     \caption{The  Light curves in the whole energy band (LAXPC) of GX 349+2 during observation 1 (A1), observation 2 (A2), observation 3 (A3), observation 4 (A4) and observation 5 (A5) and the background subtracted hard color. The hard color is defined as the ratio of the count rates in the 6.2-8.7 and 8.7-19.7 keV energy bands. The dashed vertical line represents the two intensity levels. 'a' represents the high intensity state of GX 349+2 during the observations. At each point, 64 s is averaged (see Section \ref{section3.1}).}
    \label{figure2}
\end{figure*}

\begin{figure*}
     \includegraphics[width=0.35\textwidth]{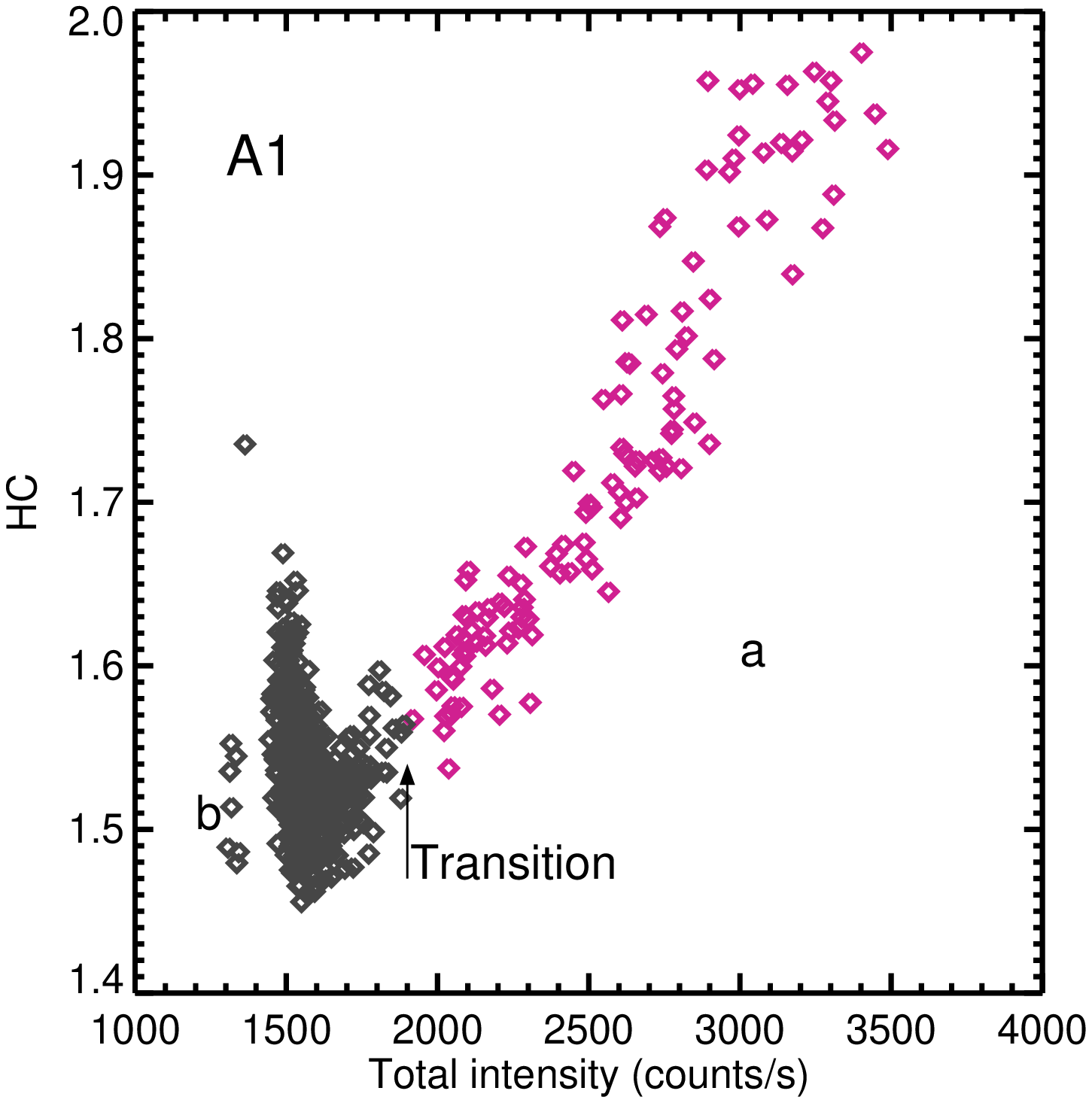}
     \includegraphics[width=0.35\textwidth]{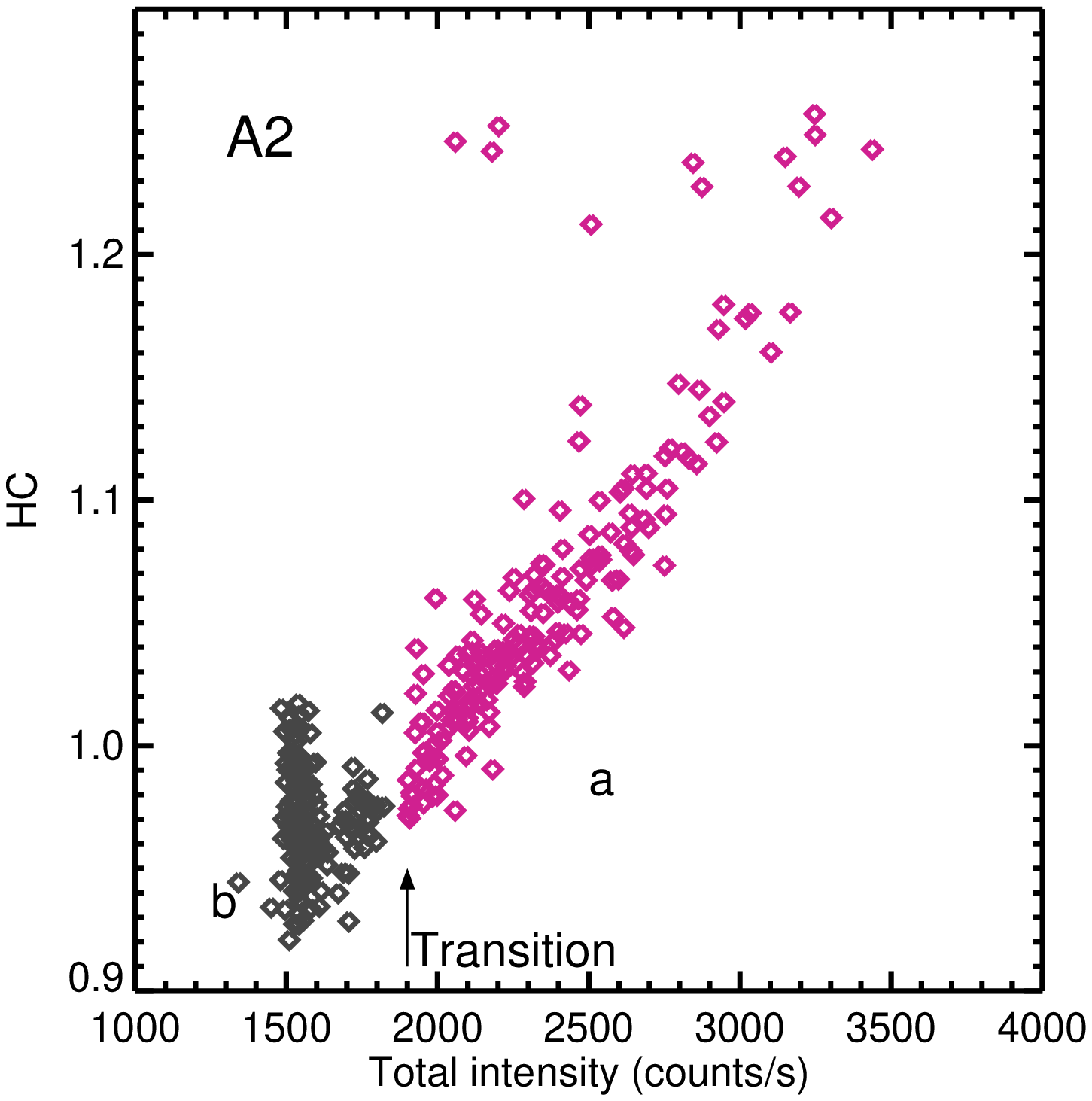}
     \includegraphics[width=0.35\textwidth]{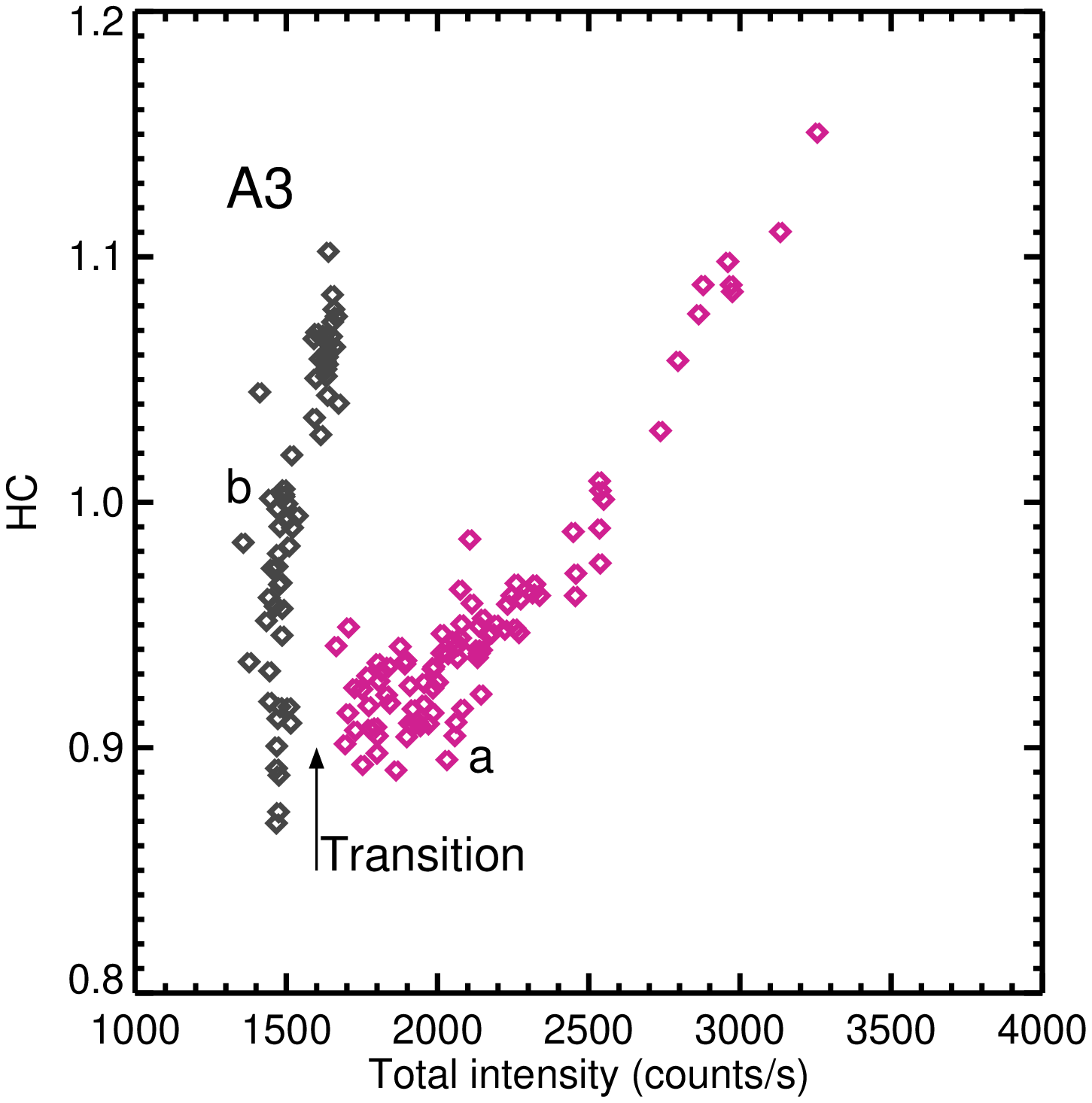}
     \includegraphics[width=0.35\textwidth]{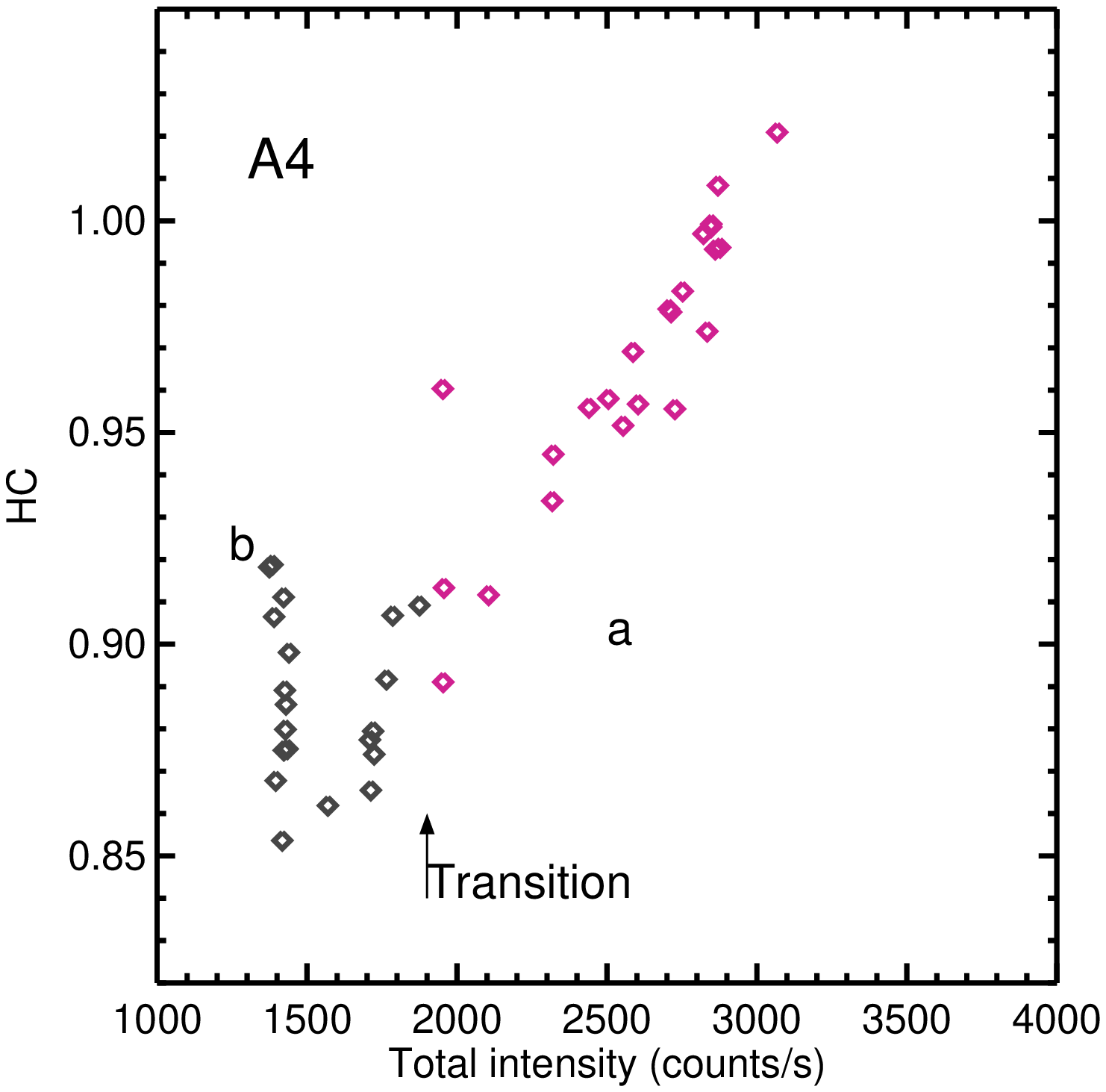}
     \includegraphics[width=0.35\textwidth]{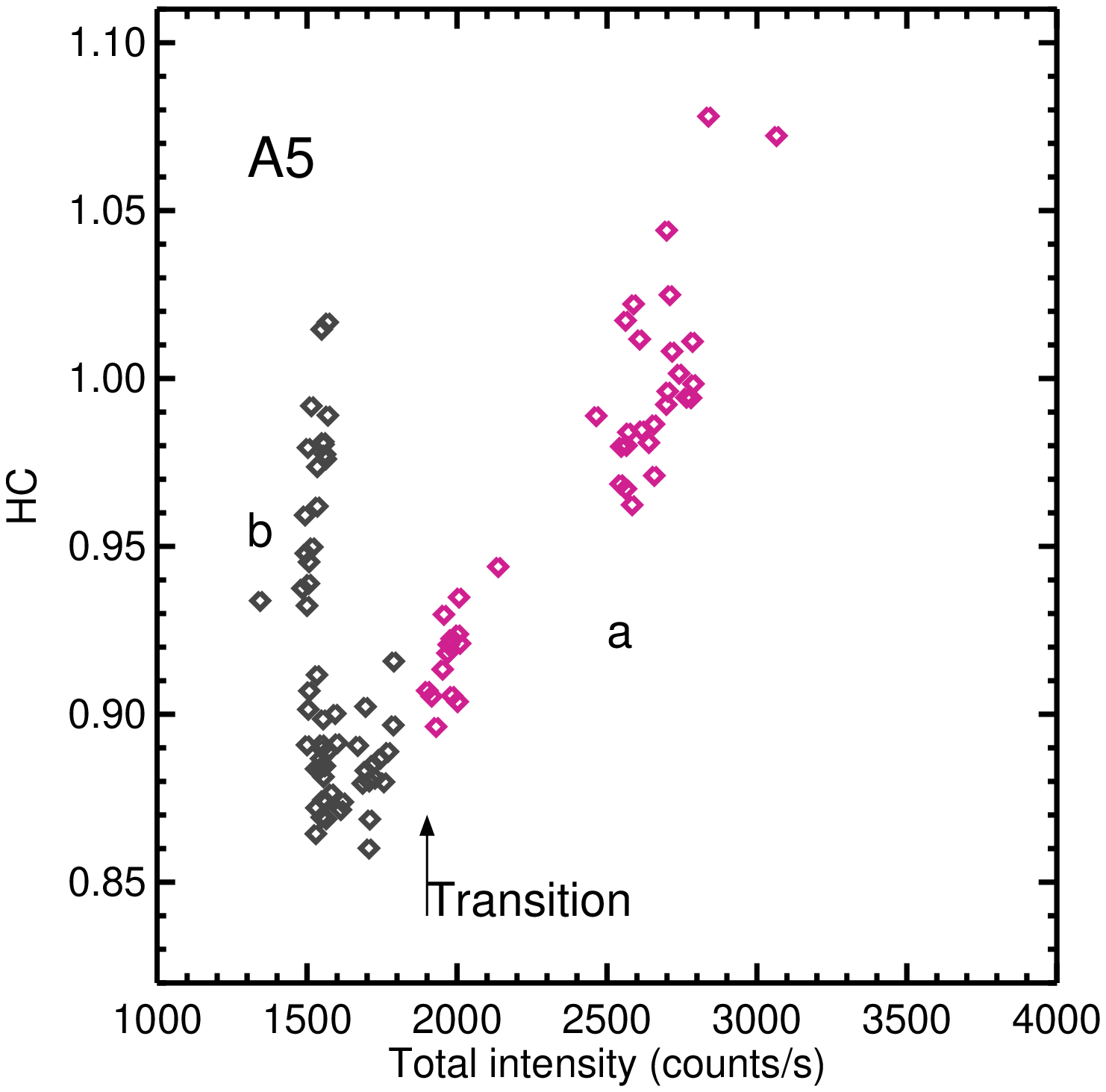}
     \caption{The hardness-intensity diagram of the LAXPC observations of GX 349+2 during observation 1 (A1), observation 2 (A2), observation 3 (A3), observation 4 (A4) and observation 5 (A5). Hard color is defined as the 19.7-8.7 keV/8.7-6.2 keV background subtracted count rate ratio and intensity as the whole energy band background subtracted count rate. 'a' and 'b' represents the high and low intensity state of GX 349+2 during the observations. Each point corresponds to a 64 s bin size (see Section \ref{section3.1}).}
    \label{figure3}
\end{figure*}

\newpage

\section{Results}
\label{section3}

\subsection{Light curves and Hardness Intensity Diagram}
\label{section3.1}

The top panels of Figure~\ref{figure2} show that the {\em AstroSat} LAXPC20 light curves of GX 349+2 exhibit a large-scale variability. The corresponding background subtracted hard colors (HC) are shown in the bottom panels of  Figure~\ref{figure2}. The hard color (HC) is defined as the ratio of the background subtracted count rates in the 6.2-8.7 and 8.7-19.7 keV energy bands  \citep{2016MNRAS.455.2959D}. Figure~\ref{figure3} shows the Hardness Intensity Diagram (HID) of GX 349+2 during all five {\em AstroSat} observations. The source hardness shows a descending trend from observation A1 to A2 and modest variations for observations A3, A4, and A5. Furthermore, the light curves in Figure \ref{figure2} clearly show the source transitions between two intensity levels, as shown by the vertical dashed lines. The arrow in the HID (Figure~\ref{figure3}) of GX 349+2 for each of the five observations represents the transition between two intensity levels.  We define the high intensity ($\sim$ >1900 c/s for A1, A2, A3, and A5 and $\sim$ > 1600 c/s for A3) flaring branch (FB) as state a  and the vertex between the normal and flaring branches (NB/FB vertex) with relatively lower intensity ($\sim$ < 1900 c/s for A1, A2, A3, and A5 and $\sim$ < 1600 c/s for A3) as state b (also referred as soft apex) \citep{2012MmSAI..83..170C,2018ApJ...867...64C}. Thus, we investigate the broadband spectro-temporal variability of GX 349+2 for these two states separately.

\begin{figure*}
     \includegraphics[width=0.45\textwidth]{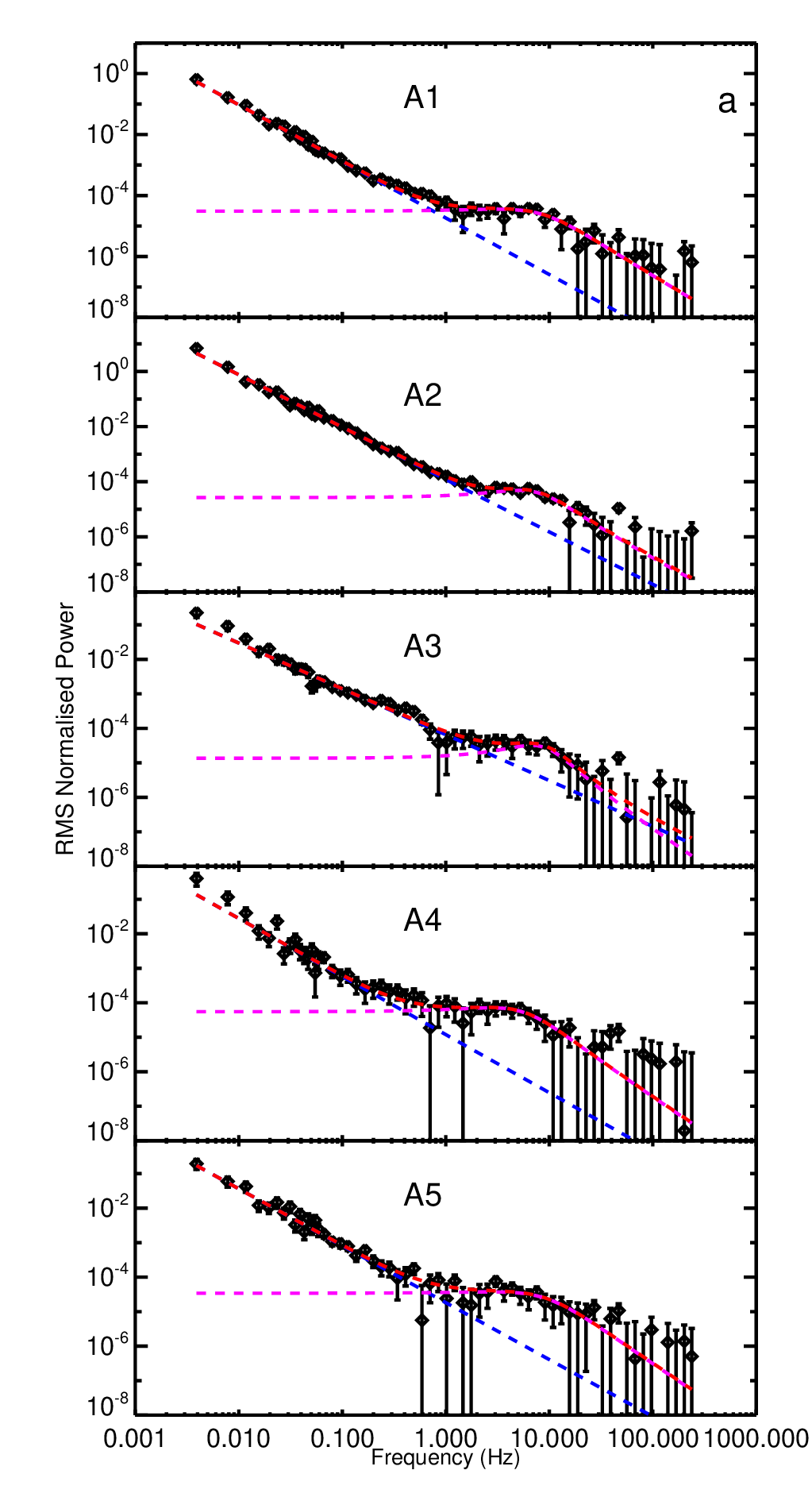}
     \includegraphics[width=0.45\textwidth]{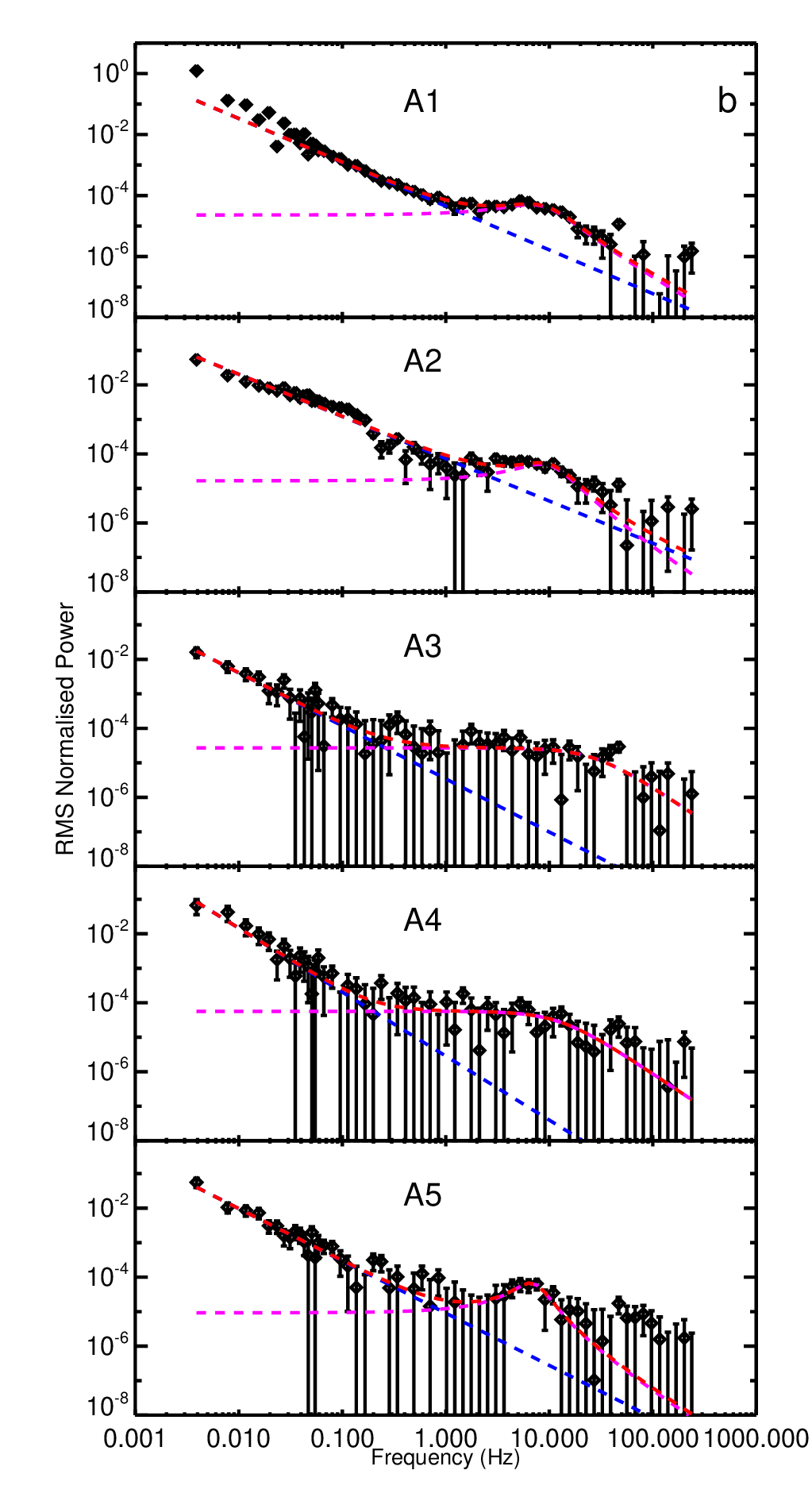}
     \caption{The Logarithmically rebinned power spectra of GX 349+2 from {\em AstroSat}/LAXPC observations in the RMS normalised power representations. All the power spectra are fitted with power law+Lorentzian model. The individual model components power law (blue) and Lorentzian (magenta) are shown by the dotted lines. The Poisson level is subtracted (see Section \ref{section3.2.1}).}
     \label{figure4}
\end{figure*}

\begin{table*}
  \centering
  \caption{Best-fitting parameters: Leahy normalized power spectra fit to models power law ($\alpha \times (f/0.01)^{-\nu}$) + Lorentzian ($\frac{r^{2} \delta}{\pi}$ $\frac{1}{\delta^{2}+(f-f_{0})^2}$), degrees of freedom and reduced $\chi^{2}$ value for the A1, A2, A3, A4, and A5 ({\em AstroSat}) observations of GX 349+2. The uncertainties mentioned are 1 $\sigma$ error, and  $<\pm$ represents the upper (+) and lower (-) limit of the error bars. For certain cases, only limits of the errors (up to two significant digits) are mentioned, where the computed statistical uncertainty becomes smaller than the threshold on the uncertainty set by the instrumental systematics. (see Section \ref{section3.2.1}).}
  \renewcommand{\arraystretch}{1.2}
  \begin{tabular}{|p{0.5cm}|p{1cm}|l|l|l|l|l|p{1.5cm}|}
    \hline
    {\textbf{Obs}}& \textbf{State} & \multicolumn{5}{|c|}{\textbf{Parameters}} &\textbf{ $\chi^{2}$ (DOF)} \\
    \cline{3-7}
    &&\hspace{1cm}\textbf{$\alpha$}&\hspace{1cm}\textbf{$\nu$}& \hspace{1cm}\textbf{$r$}& \hspace{1cm}\textbf{$\delta$}& \hspace{1cm}\textbf{$f_{0}$} &  \\
    %\hhline{~--}
    \hline
   {A1}&a& $0.54\pm 0.07$ & $ 1.85\pm 0.05$ &$0.03< \pm0.01$  &$7.96\pm2.12$ & $3.11\pm2.06$& 0.75 (56)\\
      &b&$0.13 \pm0.01$&$1.44\pm0.03$&$0.03<\pm0.01$&$6.13\pm 0.79$&$6.63\pm0.52$&7.77 (56)\\
   \hline
   {A2}&a&$ 4.56\pm 0.42$&$1.90\pm0.02$&$0.03< \pm0.01$&$5.67\pm1.10$&$5.16\pm0.78$& 1.09 (56)\\
   &b&$0.06<\pm0.01$&$1.22\pm0.03$&$ 0.03<\pm0.01$&$5.87\pm1.37$&$8.38\pm0.87$&2.84 (56)\\
   
   \hline
   {A3}&a&$0.10\pm0.02$&$1.33\pm0.05$&$0.02<\pm0.01$&$ 5.79\pm2.31$&$6.80\pm 1.52$&1.22 (56)\\
   &b&$0.02< \pm0.01$&$ 1.54\pm0.16$&$0.05 < \pm 0.01$ & $27.23\pm 11.47$&0&0.68(56)\\
   \hline
   {A4}&a&$ 0.14\pm0.05$&$1.68\pm0.14$&$0.03<\pm0.01$&$5.00\pm1.89$&$2.72\pm1.60$&0.67 (56)\\
   &b&$0.08<\pm0.03$&$ 1.85\pm0.18$&$ 0.05 <\pm 0.01$&$12.49\pm 5.92$&$0$&0.51 (56)\\
   \hline
   {A5}&a&$0.17\pm0.03$&$1.65\pm0.07$&$0.03<\pm0.01$&$8.94\pm3.82$&$2.69\pm4.16$&0.69 (56)\\
   &b&$0.04<\pm0.01$&$1.52\pm 0.11$&$0.02<\pm0.01$&$2.62\pm1.03$&$ 6.58\pm 0.70$ &0.62 (56) \\
   \hline
   \label{table3}
  \end{tabular}
\end{table*}

\newpage

\begin{figure*}
    \hspace*{-1.4cm}
    \begin{tabular}{lr}    
    \includegraphics[width=0.43\textwidth]{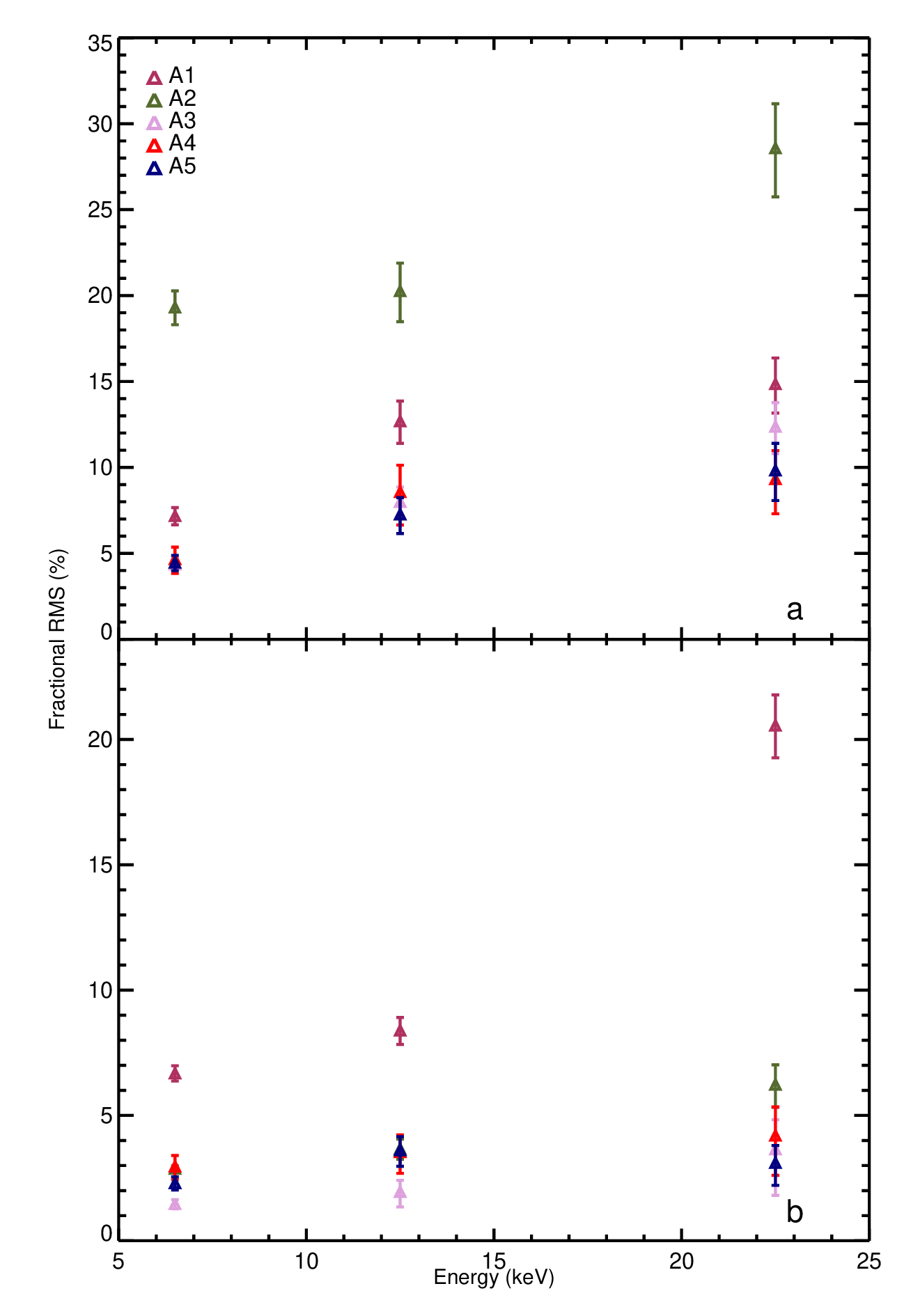} &
    \includegraphics[width=0.43\textwidth]{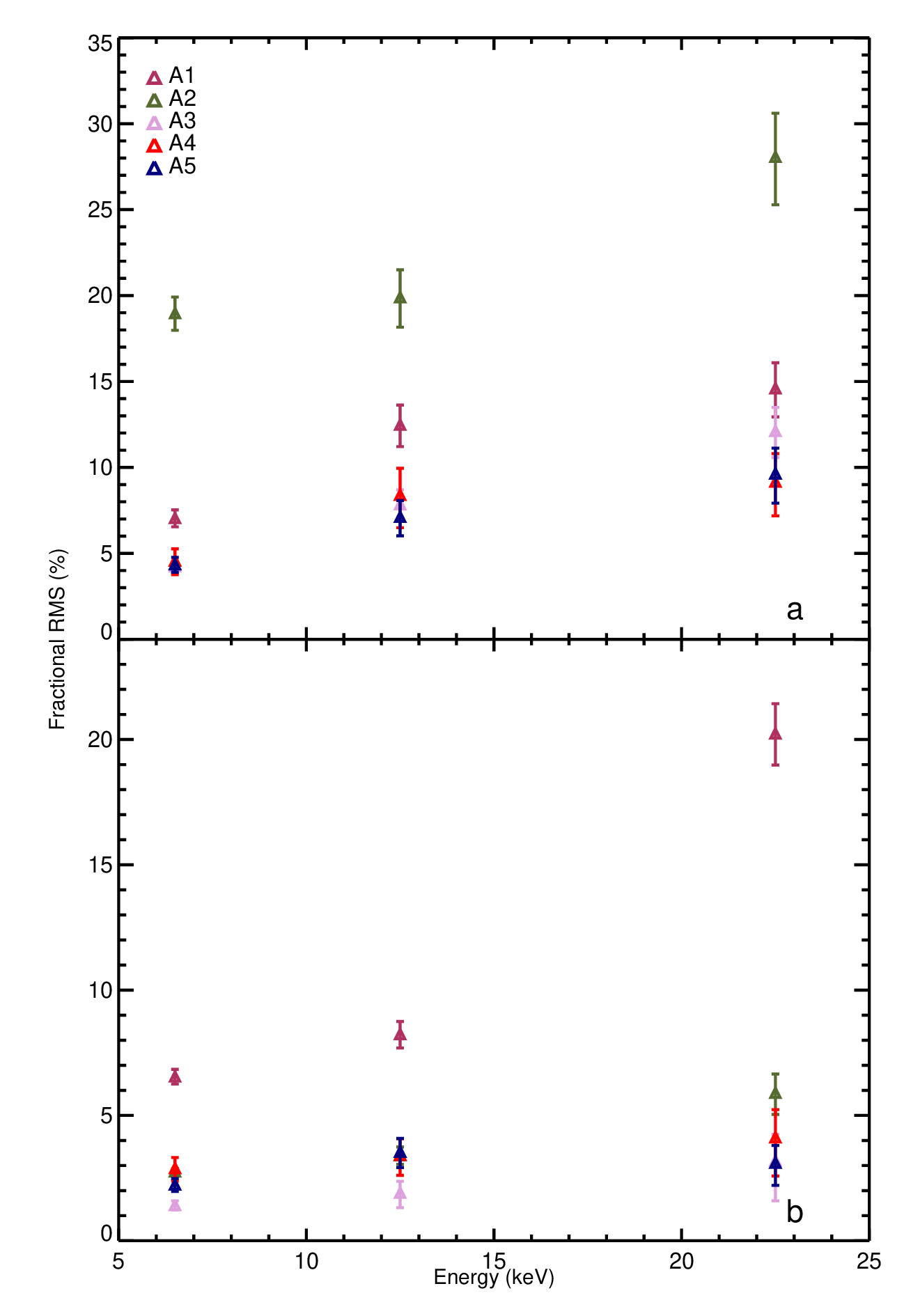} \\
    \end{tabular}
    \caption{ Fractional rms as a function of photon energy in the frequency range 0.01-1.2 Hz (left) and 1-12 Hz (right) for state a and state b for all five {\em AstroSat} observations. `a' and `b' represents the high and low intensity state of GX 349+2 during the observations (see Section \ref{section3.2.1}).} 
    \label{figure5}
\end{figure*}

\begin{table*}
  \centering
  \caption{Best-fitting parameters: Leahy normalized power spectra fit to models power law  ($\alpha \times (f/0.01)^{-\nu}$) + Lorentzian ($\frac{r^{2} \delta}{\pi}$ $\frac{1}{\delta^{2}+(f-f_{0})^2}$), degrees of freedom and reduced $\chi^{2}$ value for the N1, N2, N3, N4, N5, N6, N7, and N8 ({\em NICER}) observations of GX 349+2. The uncertainties mentioned are 1 $\sigma$ error, and  $<\pm$ represents the upper (+) and lower (-) limit of the error bars. For certain cases, only limits of the errors (up to two significant digits) are mentioned, where the computed statistical uncertainty becomes smaller than the threshold on the uncertainty set by the instrumental systematics. (see Section \ref{section3.2.2})}
  \renewcommand{\arraystretch}{1.2}
  \begin{tabular}{|p{0.5cm}|p{5cm}|l|l|l|l|l|p{1.5cm}|}
    \hline
    {\textbf{Obs}} & \textbf{Model} & \multicolumn{5}{|c|}{\textbf{Parameters}} &\textbf{ $\chi^{2}$ (DOF)} \\%& {\textbf RMS (\%)} \\
    \cline{3-7}
    &&\hspace{1cm}\textbf{$\alpha$}&\hspace{1cm}\textbf{$\nu$}& \hspace{1cm}\textbf{$r$}& \hspace{1cm}\textbf{$\delta$}& \hspace{1cm}\textbf{$f_{0}$} &  \\
    %\hhline{~--}
    \hline
   {N1}&$\alpha \times (f/0.01)^{-\nu}$+$\frac{r^{2} \delta}{\pi}$ $\frac{1}{\delta^{2}+(f-f_{0})^2}$& $0.02<\pm 0.01$ & $ 1.83\pm 0.36$ &$0.03< \pm0.01$  &$9.19\pm6.58$ & $12.65\pm4.22$& 0.50 (38)\\
   {N2}&$\alpha \times (f/0.01)^{-\nu}$&$0.13\pm 0.04$&$1.84\pm 0.13$ & &&&0.50 (41)\\
  {N3}&$\alpha \times (f/0.01)^{-\nu}$&$0.10\pm0.02$&$1.86\pm0.10$&&&&0.72 (41)\\
  {N4}&$\alpha \times (f/0.01)^{-\nu}$& $0.06\pm0.02$ &$1.83\pm0.22$&&&&0.52 (41)\\
  {N5}&$\alpha \times (f/0.01)^{-\nu}$&$0.01<\pm0.01$&$1.50\pm0.26$&&&&0.54 (41)\\ 
  {\textbf{N6}}&$\alpha \times (f/0.01)^{-\nu}$&$0.04<\pm0.01$&$1.49\pm0.06$&&&&0.75 (41)\\
  {N7}&$\alpha \times (f/0.01)^{-\nu}$+$\frac{r^{2} \delta}{\pi}$ $\frac{1}{\delta^{2}+(f-f_{0})^2}$&$0.05<\pm0.01$&$1.66\pm0.05$&$0.02<\pm0.01$&$7.91\pm3.03$&$2.59\pm3.07$&0.72 (38)\\
  {\textbf{N8}}&$\alpha \times (f/0.01)^{-\nu}$&$0.01<\pm0.01$&$0.94\pm0.16$&&&&1.95 (41)\\
  \hline
  \label{table4}
  \end{tabular}
\end{table*}

\begin{figure}
    \hspace*{-1.4cm}
    \begin{tabular}{lr}   
    \includegraphics[width=0.55\textwidth]{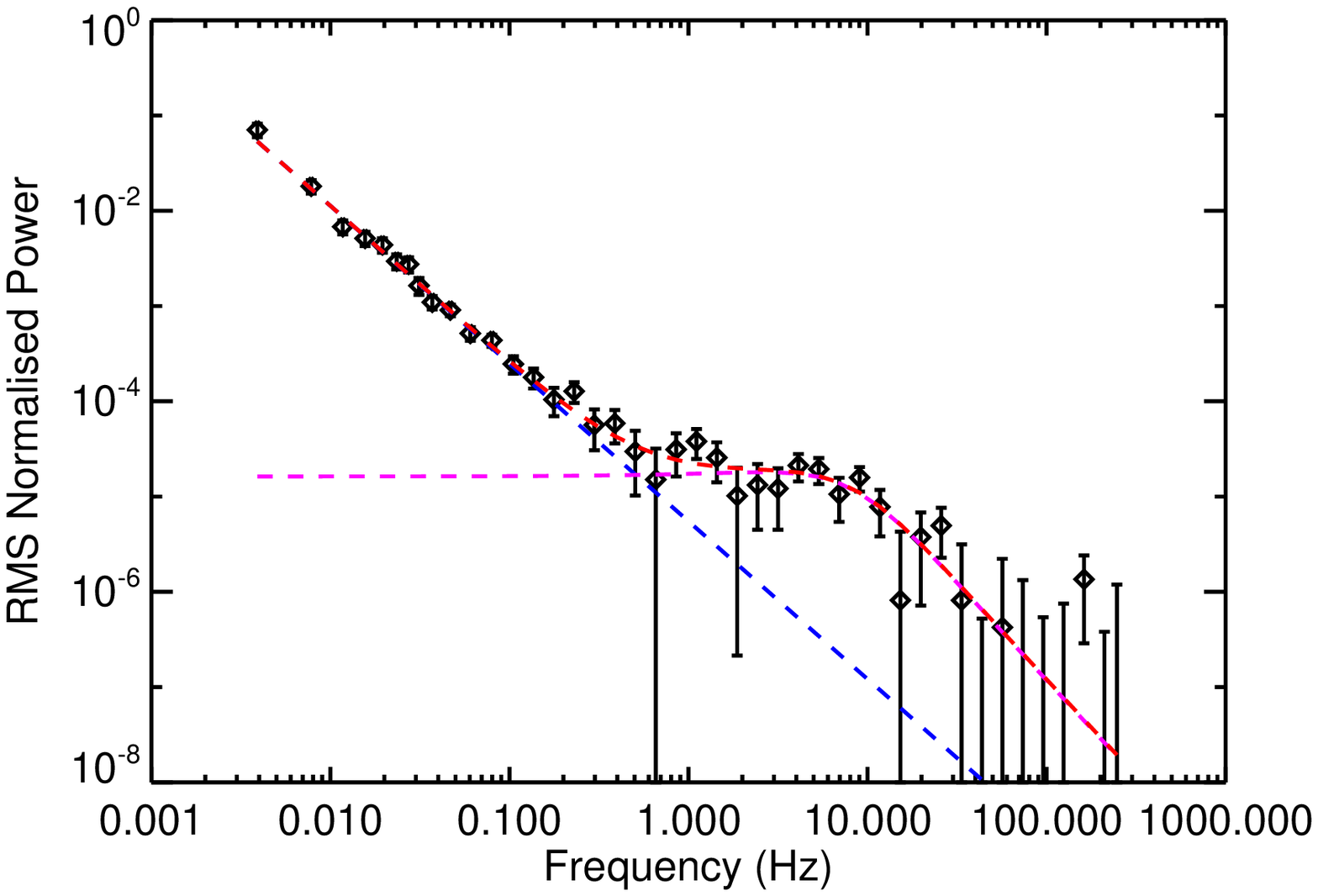} 
    \end{tabular}
    \caption{The Logarithmically rebinned power spectrum of GX 349+2 from {\em NICER} observation (N7) in the RMS normalised power representation. The power spectrum is fitted with power law+Lorentzian model. The individual model components power law (blue) and Lorentzian (magenta) are shown by the dotted lines. The Poisson level is subtracted (see Section \ref{section3.2.2}).} 
    \label{figure6}
\end{figure}

\newpage

\subsection{Timing Analysis}
\label{section3.2}

\subsubsection{{\em AstroSat}}
\label{section3.2.1}

To investigate the variability of the NS LMXB GX 349+2, we have obtained the Leahy Normalized power spectra from the light curves of each 256 s segment considering a Nyquist frequency of 256 Hz \citep{1983ApJ...266..160L}. Typically, these power spectra appear to be noisy as the noise power follows $\chi^{2}$ distribution with 100\% power uncertainties \citep{2004astro.ph.10551V}. To reduce the noise in the Leahy normalized powers, the power spectra for each 256 s segment are averaged over to obtain resultant power spectra for each state of all five observations. For investigating the variabilities and noise components present in the power spectra at different frequency ranges, we use a logarithmic frequency rebinning scheme, resulting in power spectra with high signal-to-noise ratio (SNR). To further study and quantify the variabilities, we obtain the RMS (root-mean-squared) normalized power spectra by scaling the time-averaged power spectra by corresponding intensity. 

Figure~\ref{figure4} shows the Poisson noise subtracted power spectra extracted for each spectral state from all five {\em AstroSat} observations of GX 349+2. The power spectra extracted for high-intensity FB (state a) and low-intensity NB/FB vertex (state b) from all five observations are shown in the left and right columns in Figure~\ref{figure4}. To further investigate the nature of different variability components present at different frequencies, we fit the low-frequency variability present in the power spectra with a power law (PL) model ($\alpha \times (f/0.01)^{-\nu}$; where $\alpha$ and f represents the normalisation and frequency). We add a Lorentzian ($\frac{r^{2} \delta}{\pi}$ $\frac{1}{\delta^{2}+(f-f_{0})^2}$; where r, $\delta$ and $f_{0}$ are the integrated fractional rms, HWHM, and centroid frequency of each Lorentzian) to represent the broad/peaked noise present at $\sim$ 3-8 Hz. Moreover, to check the statistical significance of the best-fitting models, we performed an F-test taking the corresponding fit statistic ( $\chi$ ) values of PL and Lorentzian models for each power spectra. We put a threshold of 3$\sigma$ significance on the chance probability obtained through the F-test. The second component is only considered in the cases where the chance probability of the improvement of the fit is less than 0.27 \%. The F-test condition is satisfied for the power spectra for each state for all five AstroSat observations requiring a Lorentzian component in addition to a power-law model. The best-fitting parameters for the state a and state b for all five {\em AstroSat} LAXPC observations are shown in Table \ref{table3}.

To examine the energy dependence of the variability, we extract the power spectra in the 3-10, 10-15, and 15-30 keV energy bands for states a and b for all five observations. But the low sensitivity of the data limits further investigations of the shape of the power spectra at different energy bands. However, the variability below 1 Hz is well represented by a power law noise, almost consistent in all the energy bands. We further calculate the energy-dependent fractional RMS in the 0.01-1.2 Hz and 1-12 Hz frequency ranges to understand the origin and evolution of the variability component present at the low frequency and the broad/peaked noise at $\sim$ 3--8 Hz. The left and right panel of Figure~\ref{figure5} shows the 0.01--1.2 Hz and 1--12 Hz fractional RMS amplitude evolution for states a and b during all five {\em AstroSat} observations of GX 349+2. We follow the method mentioned in \citep{1990A&A...230..103B} to obtain the fractional RMS. The total integrated fractional RMS and flux dependence of GX 349+2 are reported in Table~\ref{table5}. Additionally, we investigate the presence of frequency and energy-dependent time lags in the LAXPC data between different pairs of time series for multiple energy bands, considering the 3-10 keV energy band as a reference \citep{1998astro.ph..7278N}. We consider an integration time of 256 s and a Nyquist frequency of 256 Hz to compute the cross-spectrum. The energy or frequency-dependent time lags obtained from the {\em AstroSat} (LAXPC) data are consistent with a zero-lag considering 1 $\sigma$ uncertainty.

\subsubsection{{\em {\em NICER}}}
\label{section3.2.2}

\begin{figure}
    \hspace*{-1.4cm}
    \begin{tabular}{lr}    
    \includegraphics[width=0.55\textwidth]{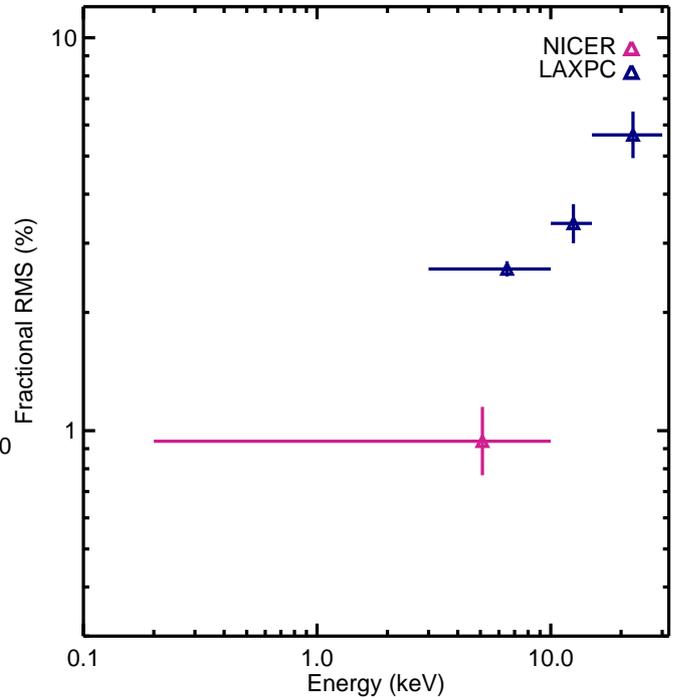}
    \end{tabular}
    \caption{Fractional rms as a function of photon energy in the integrated frequency range from the simultaneous {\em NICER} (pink triangle) and {\em AstroSat} (triangles) observation on 14 July, 2017 (N5 and A2) (see Section \ref{section3.2.1} and \ref{section3.2.2}).} 
    \label{figure7}
\end{figure}

Similar to LAXPC, with {\em NICER} data, we extract the RMS normalized power spectra for each 256 s segment of the light curves for all 8 {\em NICER} observations following the same methodology mentioned in section \ref{section3.2.1}. We fit the resultant {\em NICER} power spectra with a power law model and a Lorentzian. Following the method mentioned in Section \ref{section3.2.1}, we perform F-test to check the significance of the Lorentzian component. According to the F-test, except for observations N1 and N7, power spectral fitting for all the other NICER observations does not require a Lorentzian. Figure~\ref{figure6} shows the model-fitted power spectrum for observation N7. The best-fitting parameters for all eight {\em NICER} observations are mentioned in Table \ref{table4}. Figure~\ref{figure7} shows the energy dependence of the fractional RMS variability in the integrated frequency range from the simultaneous observation of GX 349+2 using {\em NICER} (N5) and {\em AstroSat} LAXPC (A2). The variability is observed to show an increasing trend with energy. \\

%\newpage
\begin{figure}
    \hspace*{-1.4cm}
    \begin{tabular}{lr}    
    \includegraphics[width=0.55\textwidth]{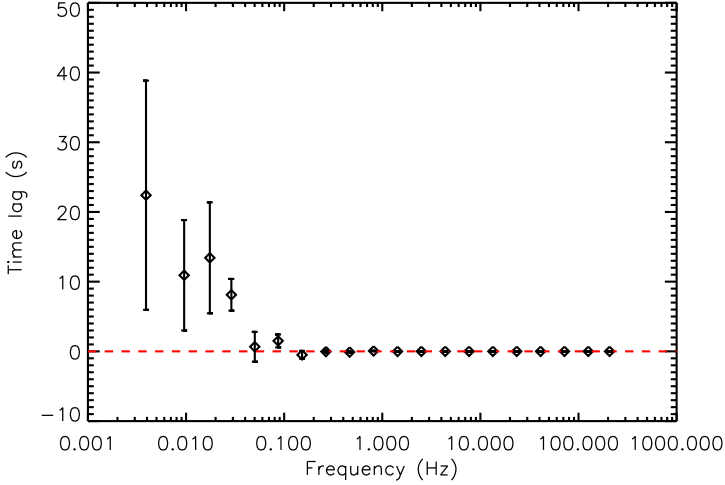}
    \end{tabular}
    \caption{Frequency-dependent hard (positive) time lags between 3–4 keV and 4–12 keV band light curves of N7 observation ({\em NICER}) (see Section \ref{section3.2.2}).} 
    \label{figure8}
\end{figure}

Following the same method mentioned in section \ref{section3.2.1}, we compute the energy and frequency-dependent time lags for all eight {\em NICER} observations. As shown in Figure \ref{figure8}, we extracted the frequency-dependent time lags in the 4-12 keV energy band, considering 3-4 keV as a reference energy band for the observation N7. Our time lag analysis shows the presence of hard time lag at the low-frequency ranges.\\

\begin{table}
\centering
\caption{RMS variability in the integrated energy and frequency ranges for state a and state b during all five {\em AstroSat} observations (see Section \ref{section3.2.2}).}
\begin{tabular}{|c|c|c|c|}
\hline
Obs No. &State & RMS Variability (\%)&LAXPC Flux ($10^{-9}$ erg/s/$cm^{2}$)\\ 
\hline
A1 &a&$6.61^{+0.44}_{-0.46}$&$12.04^{+0.13}_{-0.12}$\\
&b&$4.36^{+0.19}_{-0.20}$&$4.30^{+0.04}_{-0.04}$\\
\hline
A2 &a&$19.51^{+0.88}_{-0.92}$&$8.24^{+0.08}_{-0.08}$\\
&b&$3.35^{+0.13}_{-0.13}$&$4.08^{+0.04}_{0.04}$\\
\hline
A3&a&$3.79^{0.31}_{0.34}$&$6.22^{+0.07}_{-0.06}$\\
&b&$1.52^{0.17}_{0.19}$&$4.36^{+0.05}_{-0.04}$\\
\hline
A4&a&$3.60^{+0.60}_{-0.71}$&$9.20^{+0.01}_{-0.01}$\\
&b&$2.68^{+0.41}_{-0.48}$&--\\
\hline
A5&a&$4.24^{+0.39}_{-0.43}$&$6.79^{+0.08}_{-0.07}$\\
&b&$2.22^{+0.23}_{-0.26}$&$5.12^{+0.05}_{-0.05}$\\
\hline
\label{table5}
\end{tabular}
\begin{flushleft}
\footnotesize{$-$ denotes no  available simultaneous SXT and LAXPC observations of GX 349+2.  }\\
\end{flushleft}
\end{table}

\newpage

\subsection{Spectral Analysis}
\label{section3.3}

\subsubsection{{\em AstroSat}}
\label{section3.3.1}

We first examine the joint broadband emission spectra of GX 349+2 using LAXPC20 and SXT data. For the spectral fitting and analysis, the \texttt{XSPEC v 12.12.0c} spectral fitting package distributed as a part of the \texttt{Heasoft 6.29c} is used extensively. We consider photons within 1 keV and 7 keV for SXT and within 4 keV and 25 keV for LAXPC to avoid background effects and the loss of sensitivity. It is to be noted here that for the spectral fitting of observation 2, we consider the LAXPC spectra in the 4–20 energy range only, as the instrument systematics is significantly present outside this range. For the spectral modelling, a 3\% systematic error is added to the spectrum to account for the instrumental uncertainty in the response \citep{2022MNRAS.510.4040H, 2021MNRAS.505..713J}. We perform the joint spectral fit for each of the high (FB) and low intensity (NB/FB vertex) states (state a and b, see Section \ref{section3.1}) separately for all five observations using the combination of a blackbody radiation model (\texttt{bbodyrad} in \texttt{XSPEC}), a multicolour disc blackbody model (MCD; \texttt{diskbb} in \texttt{XSPEC}; \citet{1984PASJ...36..741M}), and a power law model (\texttt{powerlaw} in \texttt{XSPEC}). It should be noted here that the results obtained from the fits show that adding a \texttt{diskbb} model makes the fit better, and it is statistically required for the fits. Although several other physical models, such as Thermally comptonized continuum (\texttt{Nthcomp}) model \citep{1996MNRAS.283..193Z}, Comptonization (\texttt{comptt}) model \citep{1994ApJ...434..570T}, are tested, especially for the non-thermal component, the aforementioned model combination provides the best goodness of fit values and the best parameter constraints. The emitted spectrum is modified by the presence of neutral hydrogen absorption in the interstellar medium, and this is taken care of by using \texttt{Tbabs} model. The abundances and photoelectric cross-sections are adopted from \cite{2000ApJ...542..914W}. We add a \texttt{constant} component to the model to represent the cross-calibration constant between the LAXPC20 and the SXT instruments for the joint spectral fitting. We note that the addition of an absorption edge (\texttt{edge} in \texttt{XSPEC}) at 8.5 keV (possibly instrumental) improves the fit (also mentioned in \citet{2022MNRAS.510.4040H}). 

\begin{figure*}
    \hspace*{-1.4cm}
    \begin{tabular}{lr}    
     \includegraphics[width=0.45\textwidth]{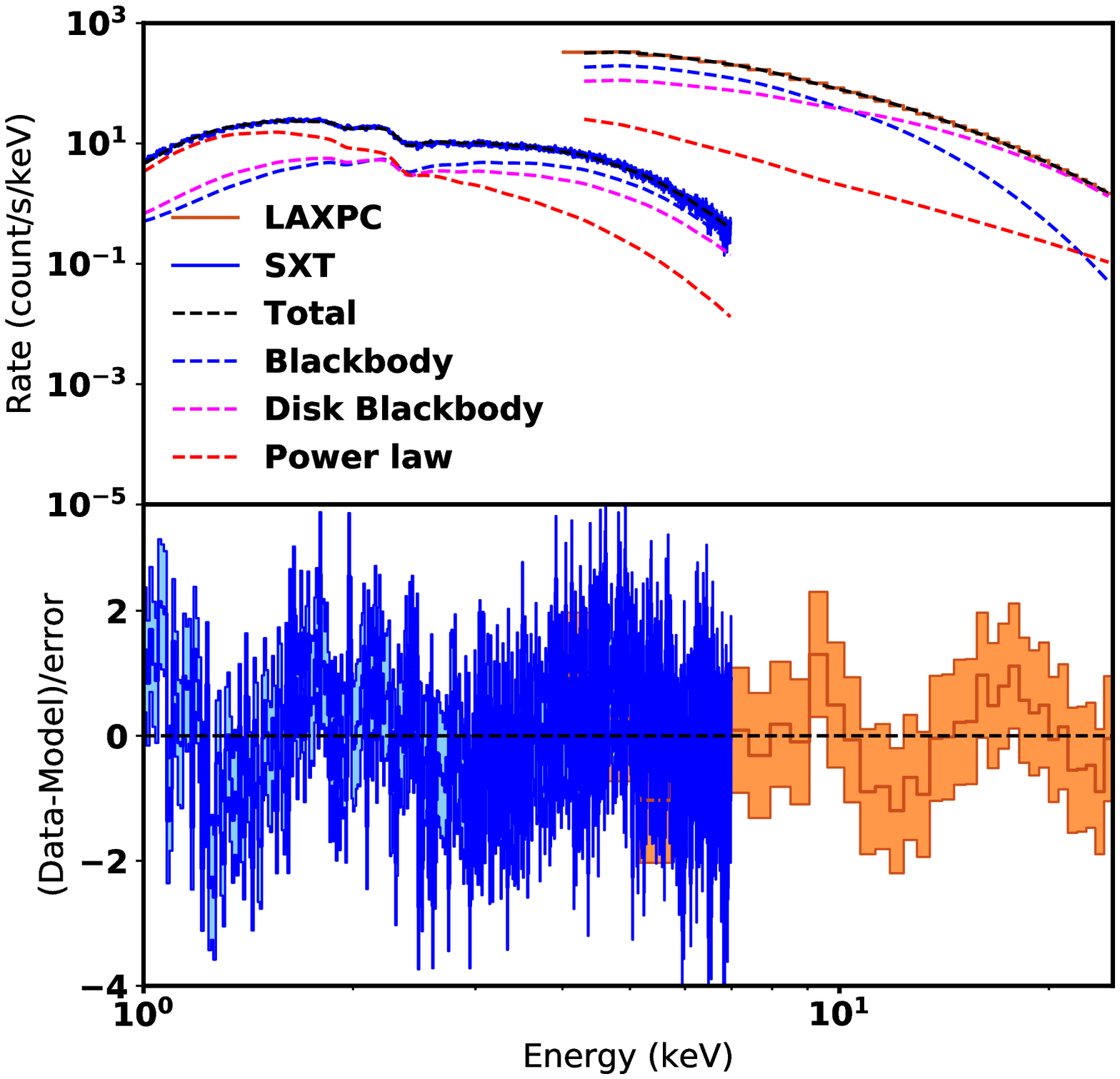} &
    \includegraphics[width=0.45\textwidth]{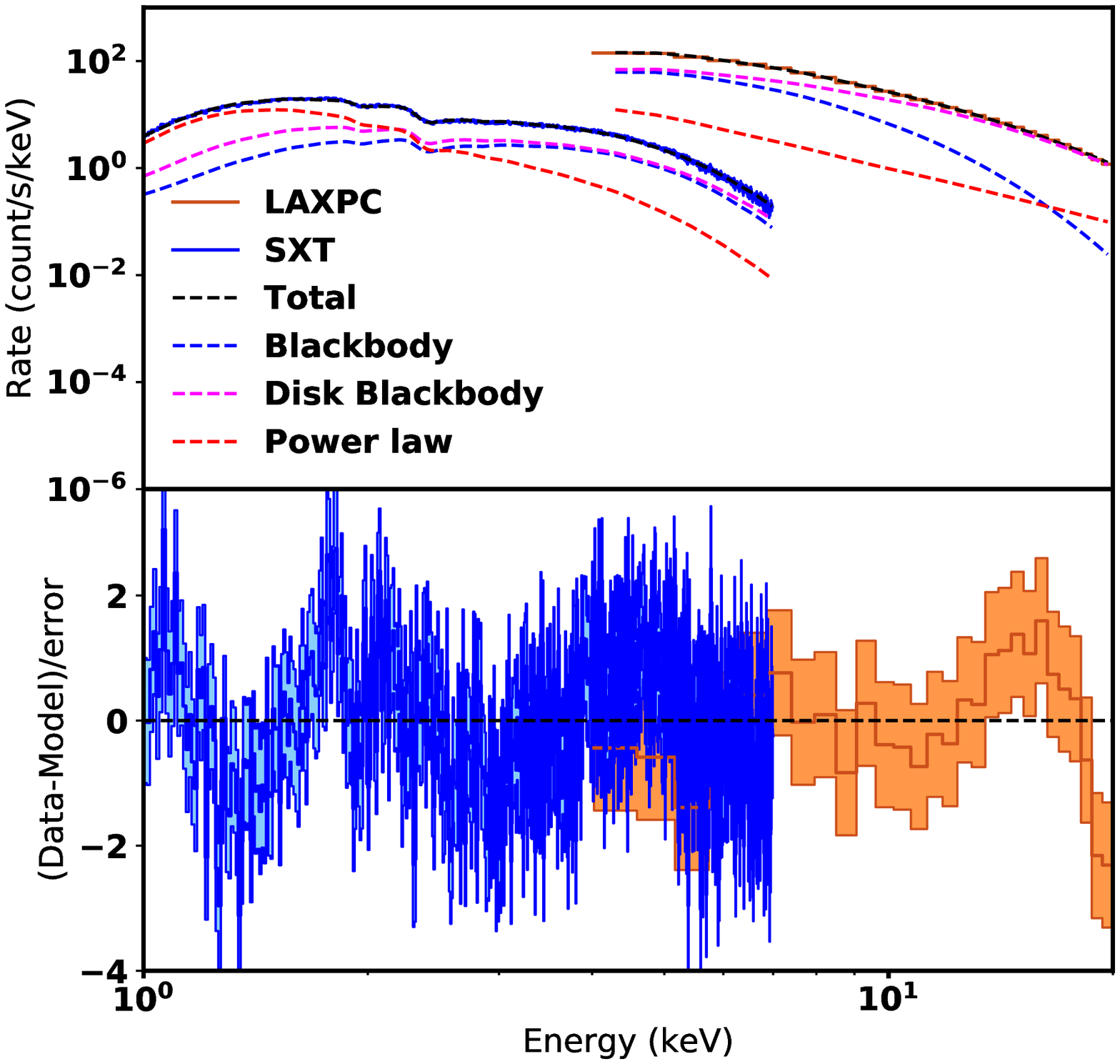} \\
    \end{tabular}
    \caption{The 1-25 keV and 1-20 keV broadband X-ray spectrum of GX 349+2 for obs 1 state a (left) and state b (right). The blue and orange data points represent the SXT and LAXPC20 data, respectively. The shaded region shows the 1 $\sigma$ error. Spectra are fitted with \texttt{const*tbabs*edge*(bbodyrad+diskbb+powerlaw)} model. The individual additive components blackbody radiation, multicolour disc blackbody and power law are also shown by the dotted blue, magenta, and red curves (see Section \ref{section3.3.1}).} 
    \label{figure9}
\end{figure*}

\begin{figure}
    \hspace*{-1.4cm}
    \begin{tabular}{lr}    
    \includegraphics[width=0.5\textwidth]{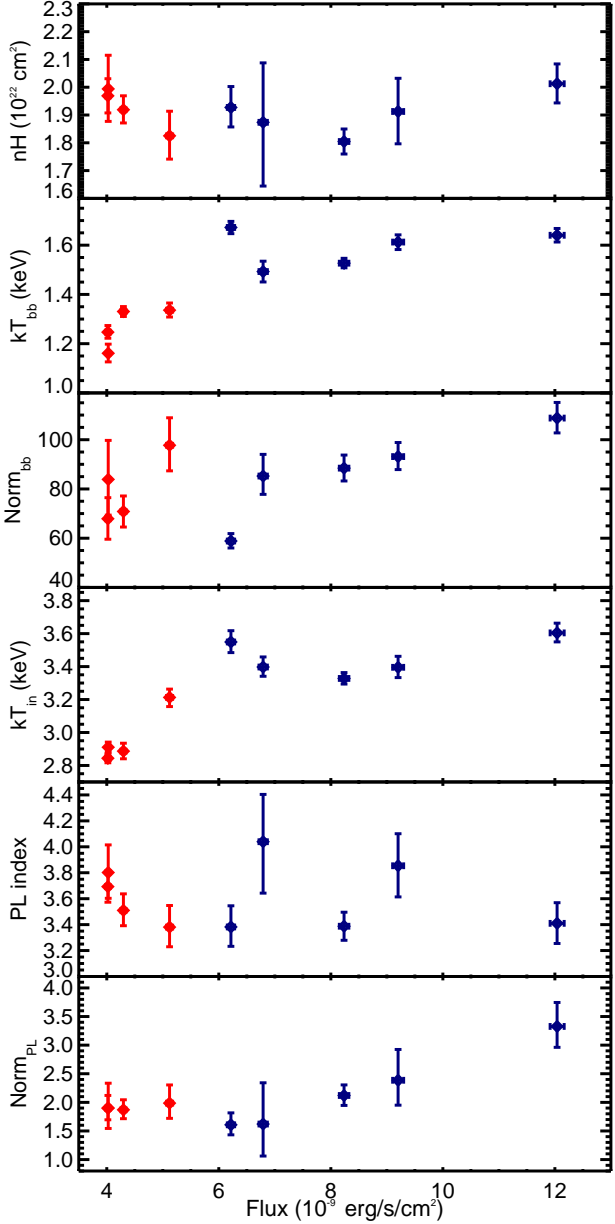} \\
    \end{tabular}
    \caption{Evolution of broadband X-ray Spectral Parameters with flux (4.0-25.0 keV) for GX 349+2 for all five {\em AstroSat} observations. First panel: Galactic neutral hydrogen column densities (nH), second panel: blackbody temperatures ($kT_{bb}$), third panel: blackbody normalizations, fourth panel: disc blackbody temperatures ($kT_{in}$), fifth panel: power law indices, sixth panel: power law normalizations, for all five {\em AstroSat} observations corresponding to the best-fitting model (see Section \ref{section3.3.1}).} 
    \label{figure10}
\end{figure}

The broadband spectra and the residuals, along with the best-fitting models for state a and state b of observation A1 are shown in Figure~\ref{figure9}. The best-fitting spectral parameters obtained from the joint fits are reported in Table~\ref{table6}. The evolution of broadband X-ray spectral parameters such as Galactic neutral hydrogen column densities (nH), blackbody temperatures ($kT_{bb}$), blackbody normalizations, temperature at the inner disc ($kT_{in}$), power law normalizations with flux in the 4-25 keV energy band for each of the states for all five observations are shown in Figure~\ref{figure10}.

\begin{table*}
    \caption{ Best-fitting spectral parameters corresponding to the best-fitting model consisting of absorbed blackbody radiation, multicolour disc blackbody and a power law component \texttt{const*tbabs*edge*(bbodyrad+diskbb+powerlaw)} for all five {\em AstroSat} observations of GX 349+2 (see Section \ref{section3.3.1}).}
    \begin{tabular}{|p{0.4cm}|c|c|c|c|c|c|c|c|c|c|}
    \hline
    Obs No. & state &$\text{nH}^{a}$&$\text{kT}^{b}$&$\text{Norm}^{c}$&$\text{kT}_{in}^{d}$&$\text {DBB}$ $\text{Norm}^{e}$&$\Gamma^{f}$&$\text{Norm}^{g}$& $\text{Flux}^{h}$ &Chi-sq \\%& RMS Variability \\ \hline\hline
    && $10^{22}$ cm$^{-2}$ &keV& &keV& &&&$10^{-9}$ erg s$^{-1}$ cm$^{-2}$&  \\
    \hline
   A1 &a&$2.01_{-0.07}^{+0.07}$&$1.64_{-0.03}^{+0.03}$&$108.78_{-6.03}^{+6.35}$&$3.60_{-0.05}^{+0.06}$&$2.14_{-0.27}^{+0.29}$&$ 3.41_{-0.16}^{+0.16}$&$3.33_{-0.36}^{+0.42}$&$12.04_{-0.12}^{+0.13}$& 1.12 (625)\\
    \\
    &b&$1.92_{-0.05}^{+0.05}$&$1.33_{-0.02}^{+0.02}$&$70.86_{-6.32}^{6.29}$&$ 2.89_{-0.05}^{+0.05}$&$2.84_{-0.35}^{+0.40}$&$3.51_{-0.12}^{+0.12}$&$1.87_{-0.15}^{+0.18}$&$4.30_{-0.04}^{+0.04}$&1.18 (618)\\
    \hline
   A2 &a&$1.80_{-0.04}^{+0.04}$&$1.53_{-0.02}^{+0.02}$&$ 88.46_{-5.22}^{5.32}$&$3.33_{-0.03}^{+0.04}$&$2.46_{-0.20}^{+0.21}$&$3.39_{-0.11}^{+0.11}$&$2.12_{-0.17}^{+0.18}$&$8.24_{-0.08}^{+0.09}$&1.31 (626)\\
    &b&$1.97_{-0.06}^{+0.06}$&$1.25_{-0.04}^{+0.04}$&$67.90_{-8.32}^{+8.56}$&$2.84_{-0.03}^{+0.03}$&$3.32_{-0.25}^{+0.26}$&$3.69_{-0.12}^{+0.12}$&$1.90_{-0.20}^{+0.22}$&$4.02_{-0.04}^{+0.04}$&1.12 (626) \\
    \hline
    A3&a&$1.93_{-0.07}^{+0.08}$&$1.67_{-0.02}^{+0.02}$&$58.86_{-2.84}^{3.02}$&$3.55_{-0.06}^{+0.07}$&$0.98_{-0.15}^{+0.17}$&$3.38_{-0.15}^{+0.16}$&$1.61_{-0.18}^{+0.21}$&$6.22_{-0.06}^{+0.07}$&1.06 (625)\\
    \\
    &b&$1.99_{-0.12}^{+0.12}$&$1.16_{-0.04}^{+0.04}$&$83.87_{-14.49}^{+15.84}$&$2.91_{-0.03}^{+0.03}$&$3.45_{-0.27}^{+0.15}$&$3.80_{-0.20}^{+0.21}$&$1.90_{-0.36}^{+0.43}$&$4.03_{-0.03}^{+0.03}$&1.13 (625) \\
\hline
A4&a&$1.91_{-0.12}^{+0.12}$&$1.61_{-0.03}^{+0.03}$&$93.15_{-5.28}^{+5.70}$&$3.40_{-0.06}^{+0.07}$&$2.16_{0.32}^{0.36}$&$3.85_{-0.24}^{+0.25}$&$2.39_{-0.44}^{+0.54}$& $9.20_{-0.10}^{+0.10}$&1.13 (625)\\
\\
&b&--&--&--&--&--&--&--&--&--\\
\hline
A5&a&$1.87_{-0.23}^{+0.21}$&$1.49_{-0.04}^{+0.04}$&$85.25_{-7.45}^{+8.79}$&$3.40_{-0.06}^{+0.06}$&$1.88_{0.25}^{0.26}$&$4.04_{-0.40}^{+0.36}$&$1.62_{-0.56}^{+0.72}$& $6.79_{-0.07}^{+0.08}$&1.02(525)\\
\\
&b&$1.82_{-0.08}^{+0.09}$&$1.34_{-0.03}^{+0.03}$&$97.73_{-10.38}^{+11.17}$& $3.21_{-0.06}^{+0.05}$&$1.84_{-0.18}^{+0.21}$&$3.38_{-0.15}^{+0.17}$&$1.99_{-0.27}^{+0.32}$&$5.12_{-0.05}^{+0.05}$&1.04 (625)\\
\hline
\end{tabular}
\begin{flushleft}
\footnotesize{$-$ denotes no  available simultaneous SXT and LAXPC observations of GX 349+2.  }\\
\footnotesize{$^a$ Neutral hydrogen column density}\\
\footnotesize{$^b$ Blackbody Temperature } \\
\footnotesize{$^c$ Blackbody Normalisation}\\
\footnotesize{$^d$ inner disc Temperature }\\
\footnotesize{$^e$ Disc Blackbody Normalisation}\\
\footnotesize{$^f$ Power-law Index}\\
\footnotesize{$^g$ Power-law Normalisation}\\
\footnotesize{$^h$ Total LAXPC (4-25 keV) Flux}\\
\end{flushleft}
\label{table6}
\end{table*}

\subsubsection{NICER}
\label{section3.3.2}

\begin{figure*}
    \hspace*{-1.4cm}
    \begin{tabular}{lr}    
    \includegraphics[width=0.45\textwidth]{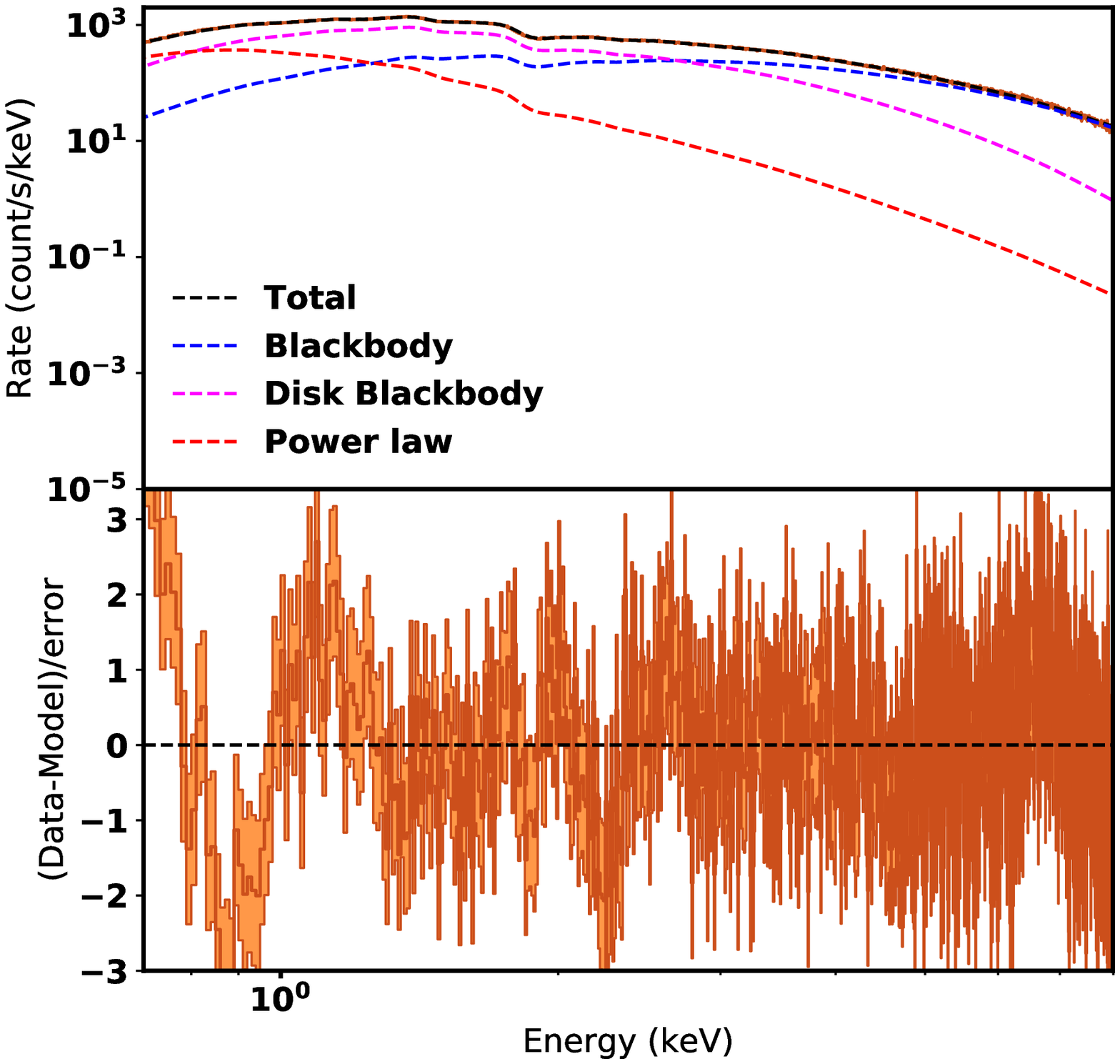} &
    \includegraphics[width=0.45\textwidth]{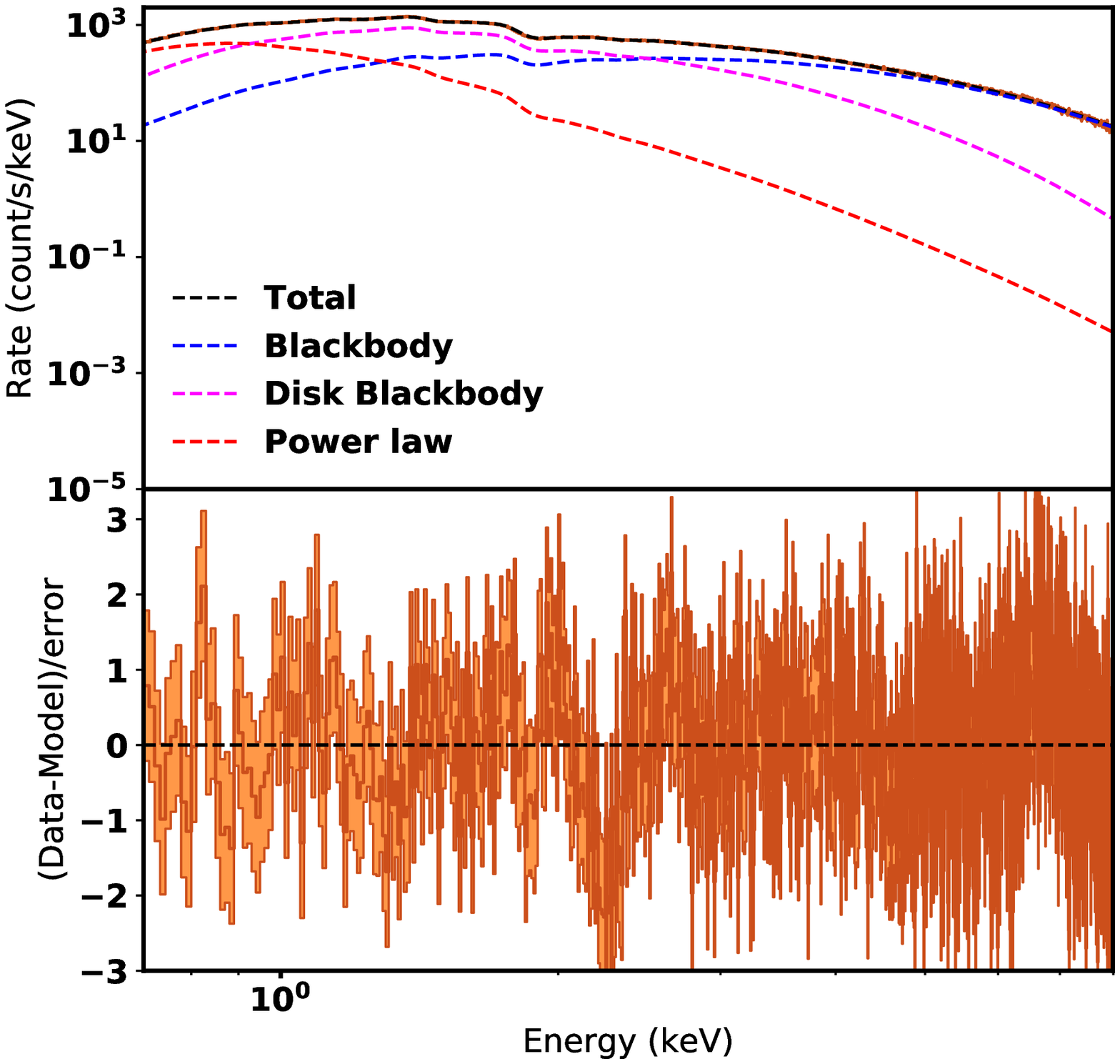} \\
    \includegraphics[width=0.45\textwidth]{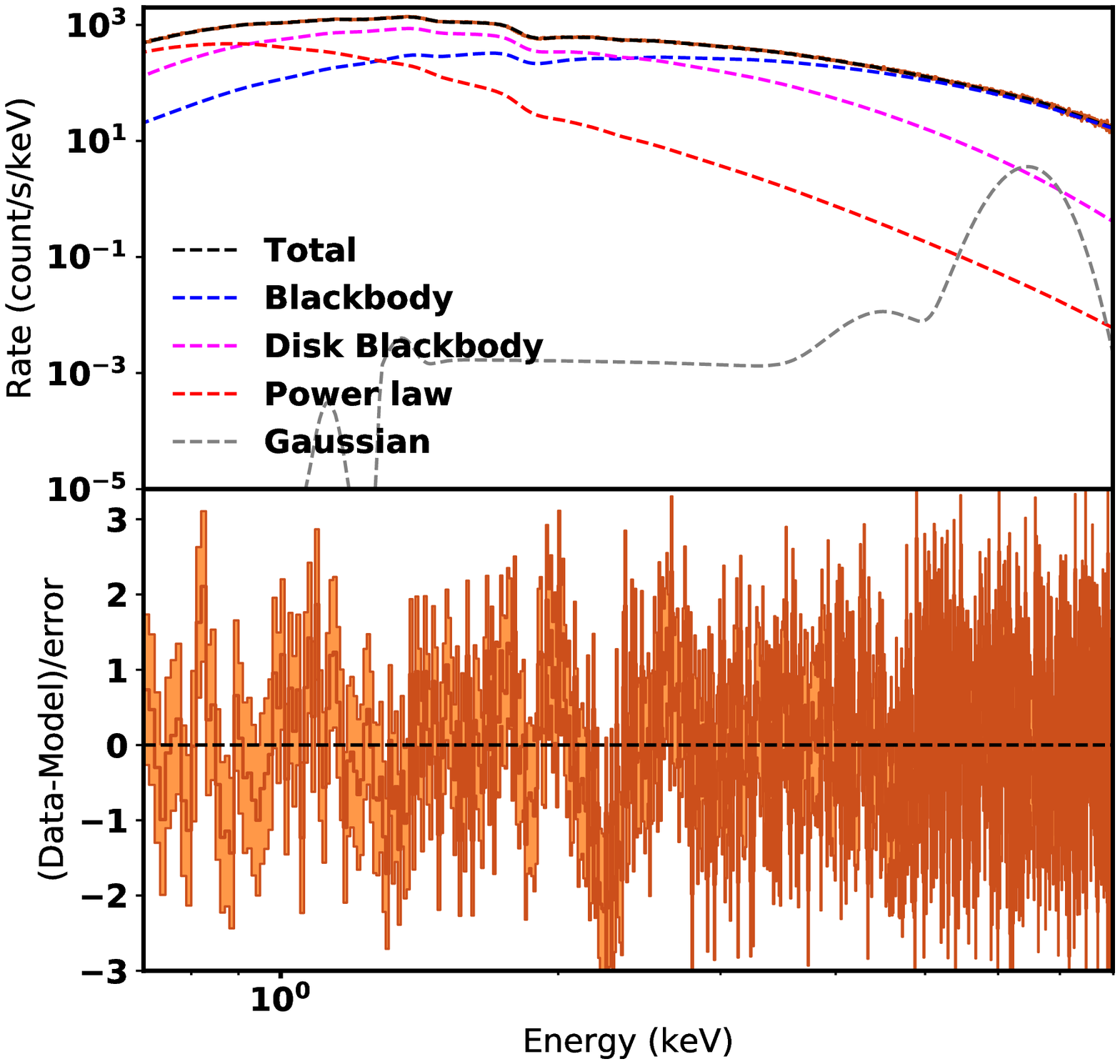} &
    \includegraphics[width=0.45\textwidth]{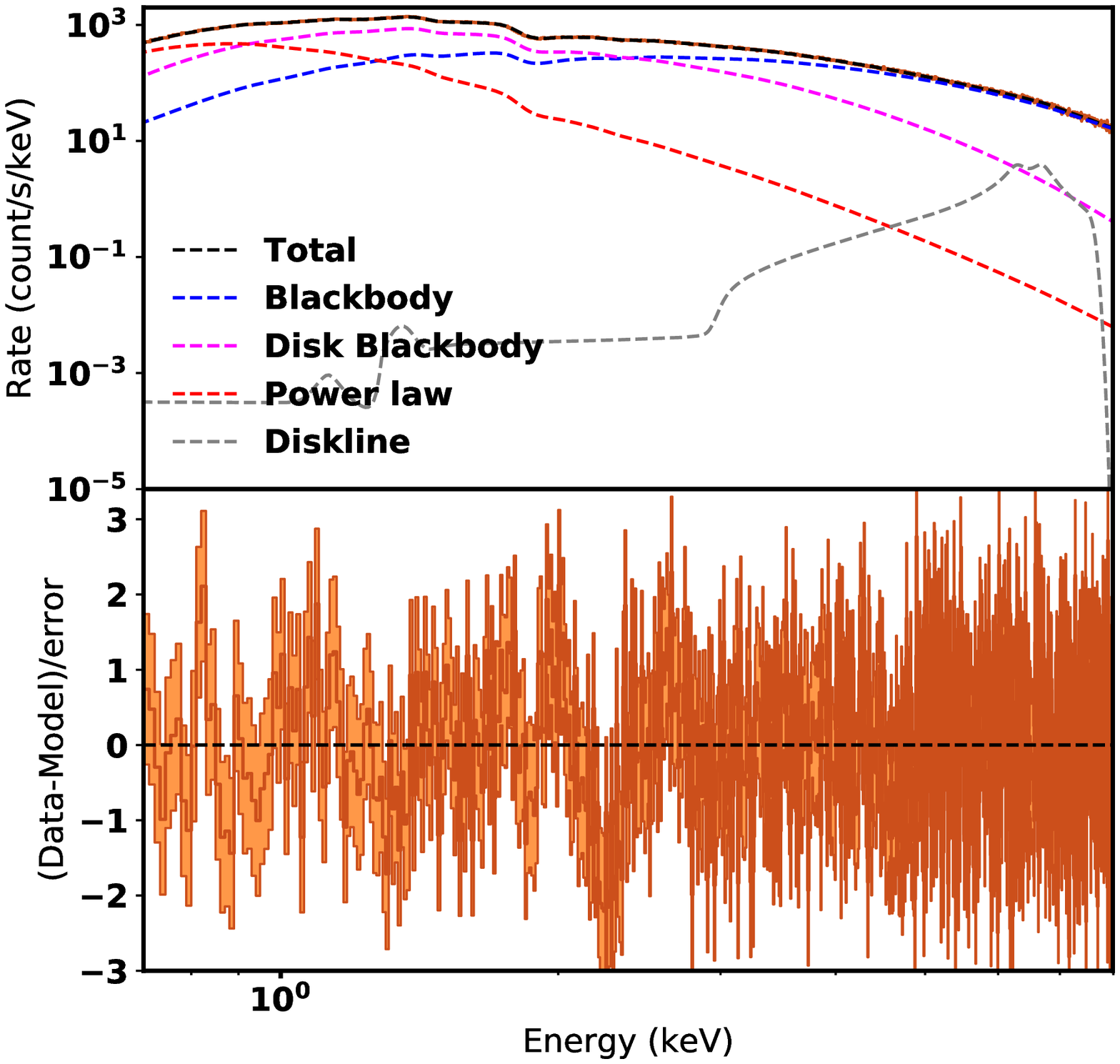} \\
    \end{tabular}
    \caption{The 0.7-8 keV X-ray spectrum of GX 349+2 for {\em NICER} observation 1 (N1). The shaded region shows the 1 $\sigma$ error. Spectra are fitted with \texttt{tbabs*(bbodyrad+diskbb+powerlaw)} model (top left), \texttt{tbabs*edge*(bbodyrad+diskbb+powerlaw)} model (top right), \texttt{tbabs*edge*(bbodyrad+diskbb+powerlaw+gauss)} model (bottom left), and \texttt{tbabs*edge*(bbodyrad+diskbb+powerlaw+diskline)} model (bottom right). The individual additive components blackbody radiation, multicolour disc blackbody, power law, Gaussian, and diskline are also shown in blue, magenta, red, and grey by the dotted curves (see Section \ref{section3.3.2}).} 
    \label{figure11}
\end{figure*}

\begin{figure}
    \hspace*{-1.4cm}
    \begin{tabular}{lr}    
    \includegraphics[width=0.46\textwidth]{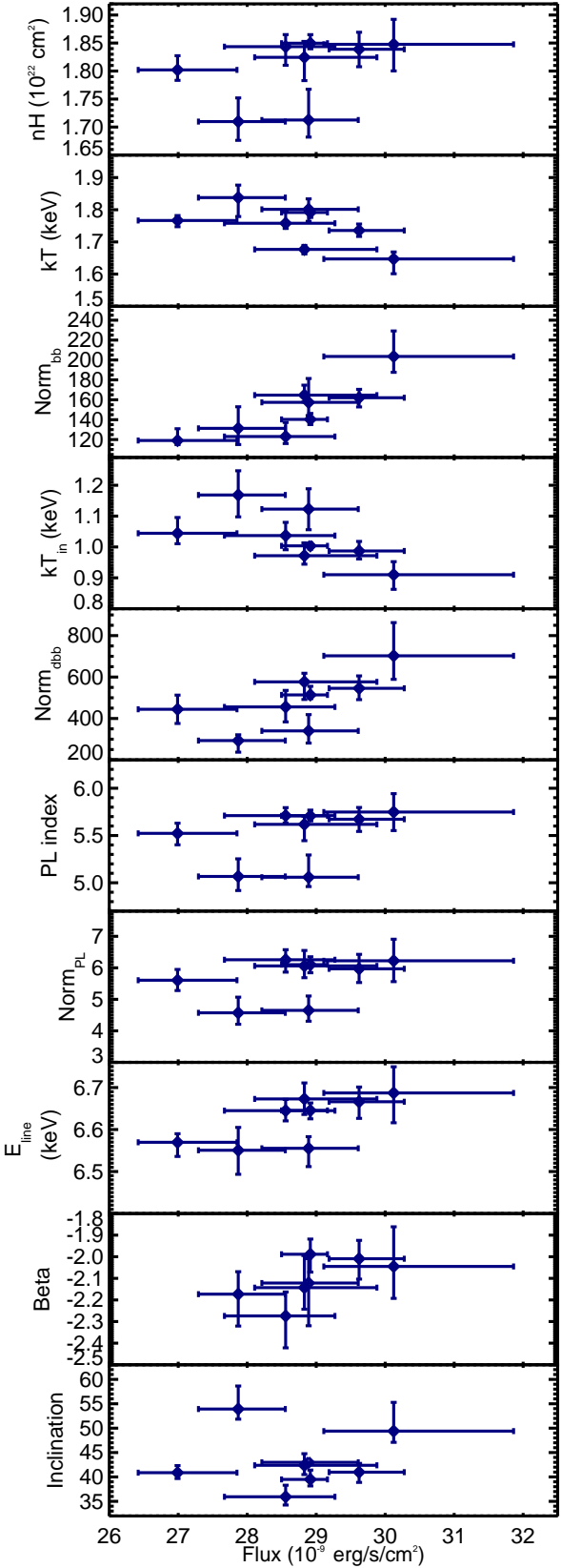} 
    \end{tabular}
    \caption{Evolution of X-ray spectral parameters with time (left) and 0.7-8 keV flux (right) for GX 349+2 for all eight {\em NICER} observations. First panel: Galactic neutral hydrogen column densities (nH), second panel: blackbody temperatures $kT_{bb}$, third panel: blackbody normalizations, fourth panel: disc blackbody temperatures ($kT_{in}$), fifth panel: disc blackbody normalizations, sixth panel: power law indices, seventh panel: power law normalizations, eighth panel: line energies, ninth panel: power law dependence of emissivities, tenth panel: inclinations for all eight {\em NICER} observations corresponding to the best-fitting model (see Section \ref{section3.3.2}).} 
    \label{figure12}
\end{figure}

\begin{figure*}
    \hspace*{-1.4cm}
    \begin{tabular}{lr}    
    \includegraphics[width=0.38\textwidth]{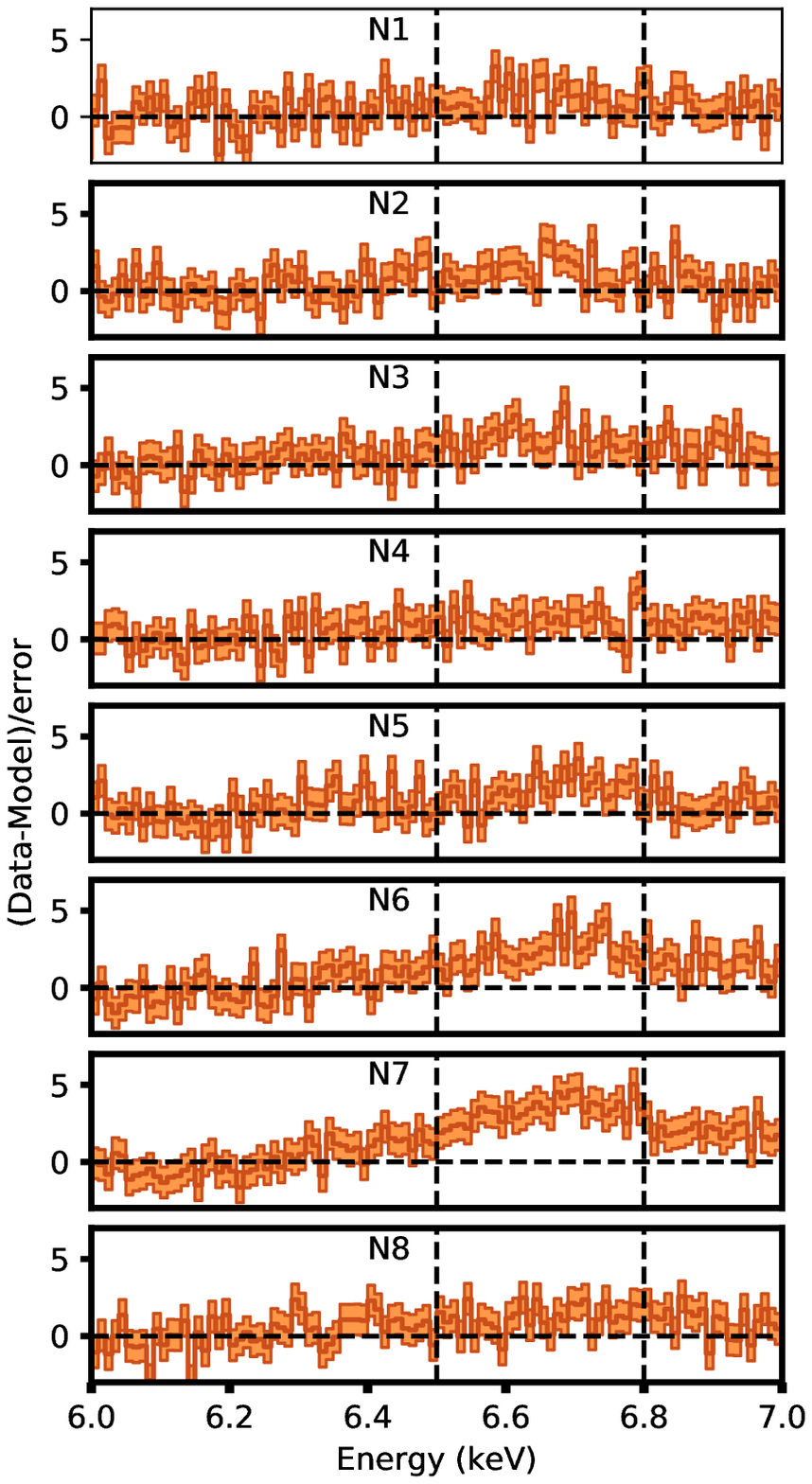}
    \includegraphics[width=0.38\textwidth]{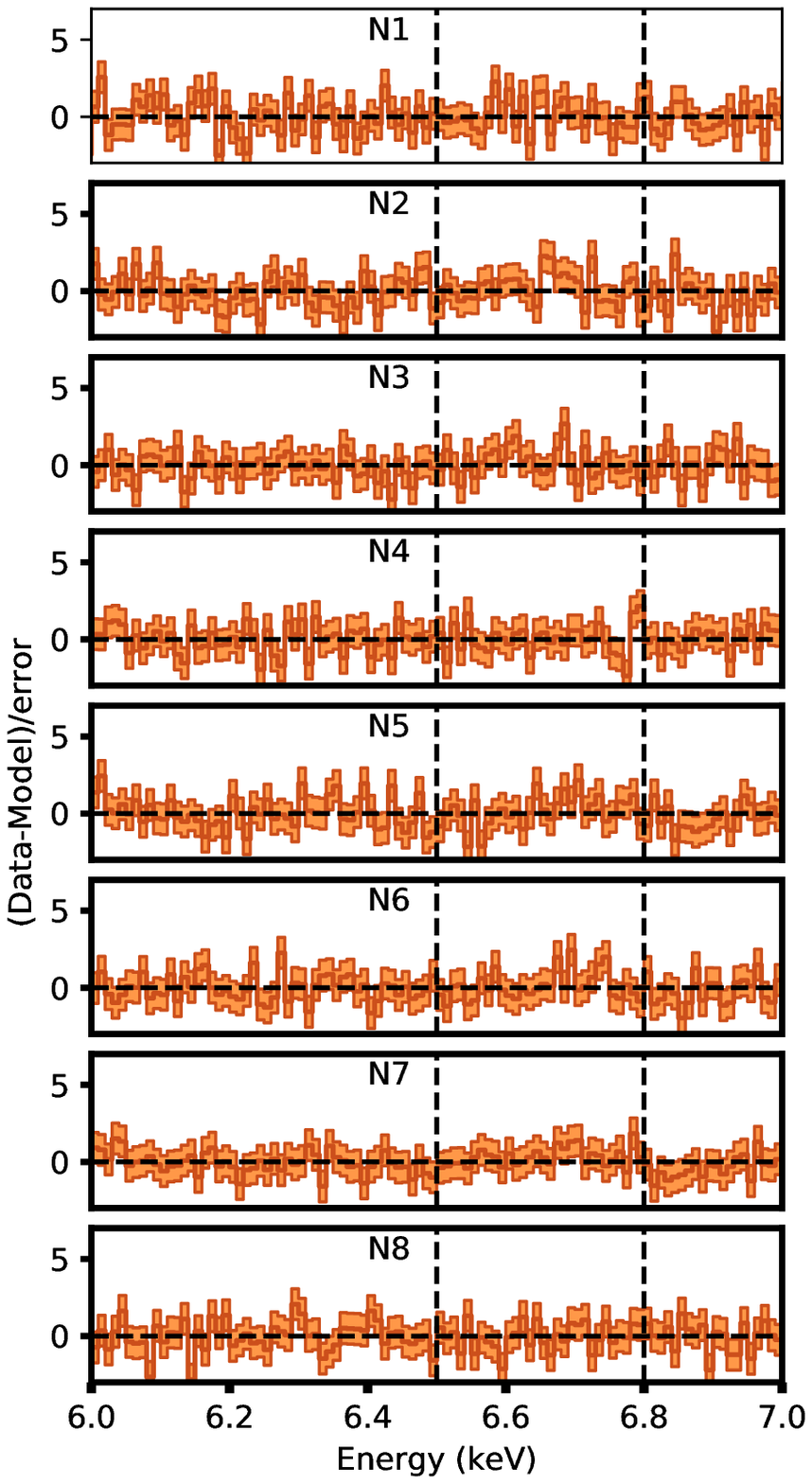}
    \includegraphics[width=0.38\textwidth]{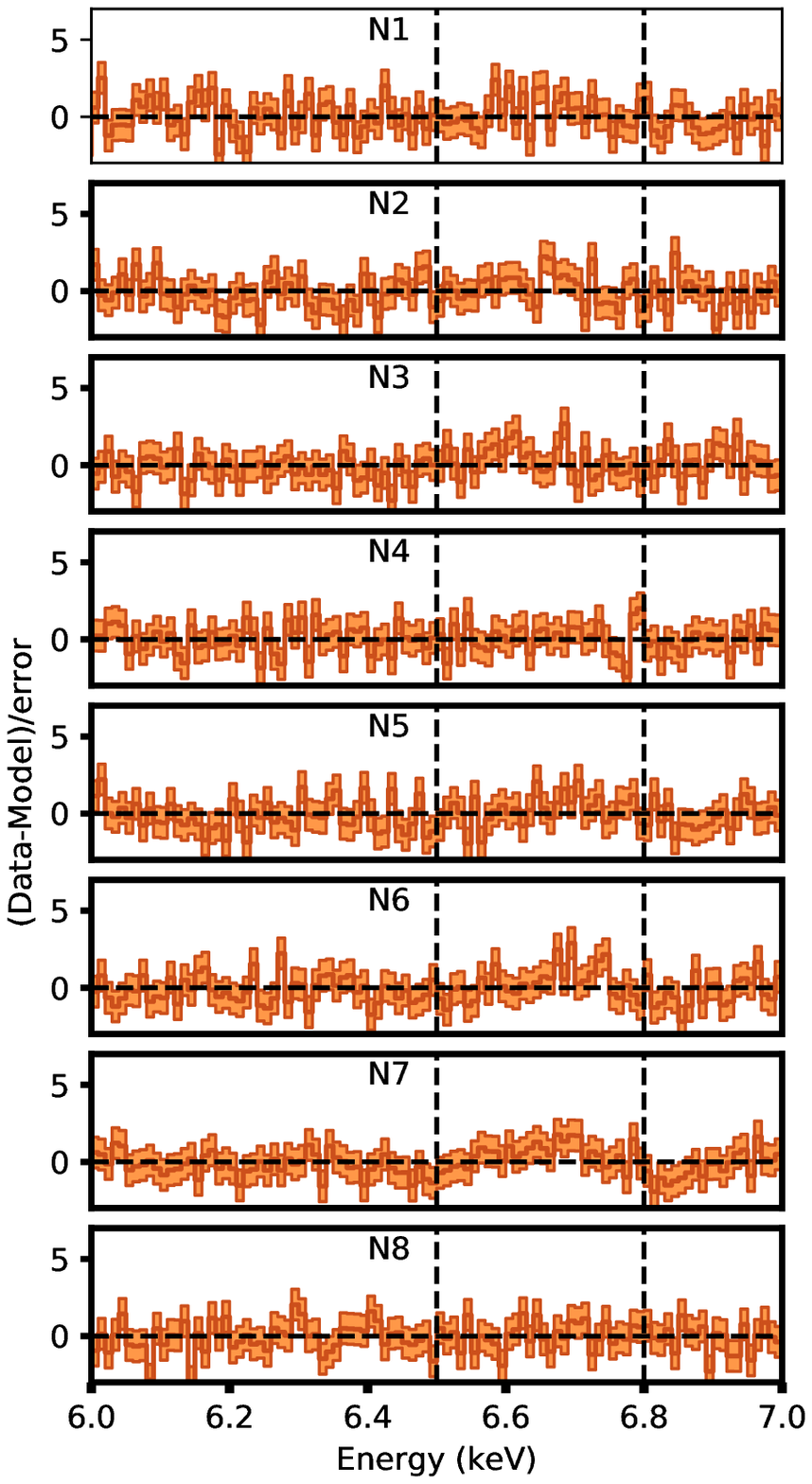}
    \end{tabular}
    \caption{The error-weighted residuals for the spectral fit in the 0.7--8.0 keV range for all eight {\em NICER} observations. The residuals against the \texttt{tbabs*edge*(bbodyrad+diskbb+powerlaw)}, \texttt{tbabs*edge*(bbodyrad+diskbb+powerlaw+gauss)}, and \texttt{tbabs*edge(bbodyrad+diskbb+powerlaw+diskline)} model are shown in the left, middle, and the right panel. The shaded region shows the 1 $\sigma$ error. The shape of the line varies during different observations, being most significant during observation N7. The dashed horizontal line shows the zero deviation level. The dashed vertical lines are used for representing a rough energy range of the detected iron (emission) line width. (see Section \ref{section3.3.2}). The residuals against the \texttt{tbabs*edge(bbodyrad+diskbb+powerlaw+diskline)} model for all eight {\em NICER} observations are shown in Figure~\ref{figurea}.} 
    \label{figure13}
\end{figure*}

\begin{figure}
    \hspace*{-1.4cm}
    \begin{tabular}{lr}    
    \includegraphics[width=0.45\textwidth]{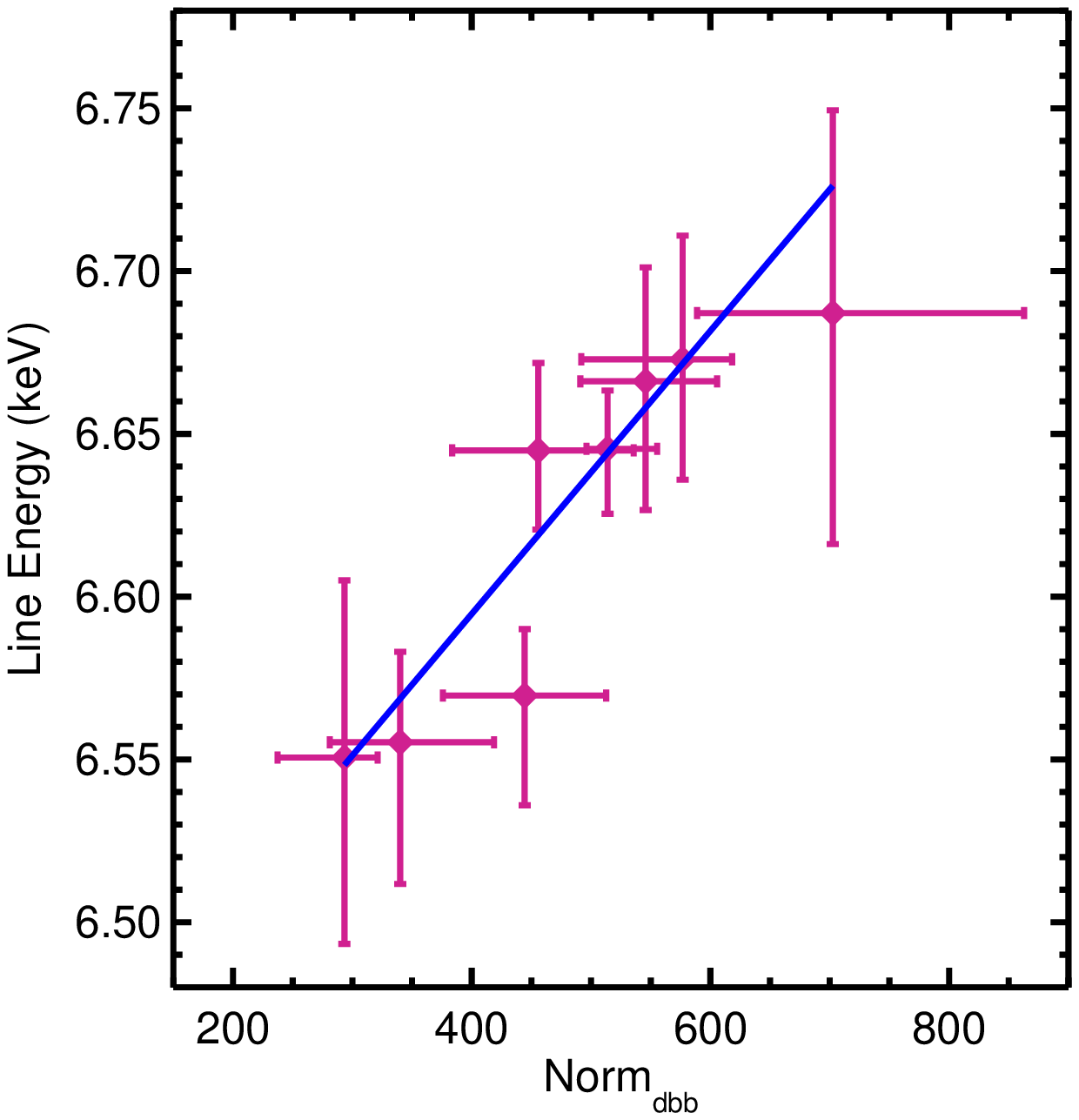}\\
    \includegraphics[width=0.45\textwidth]{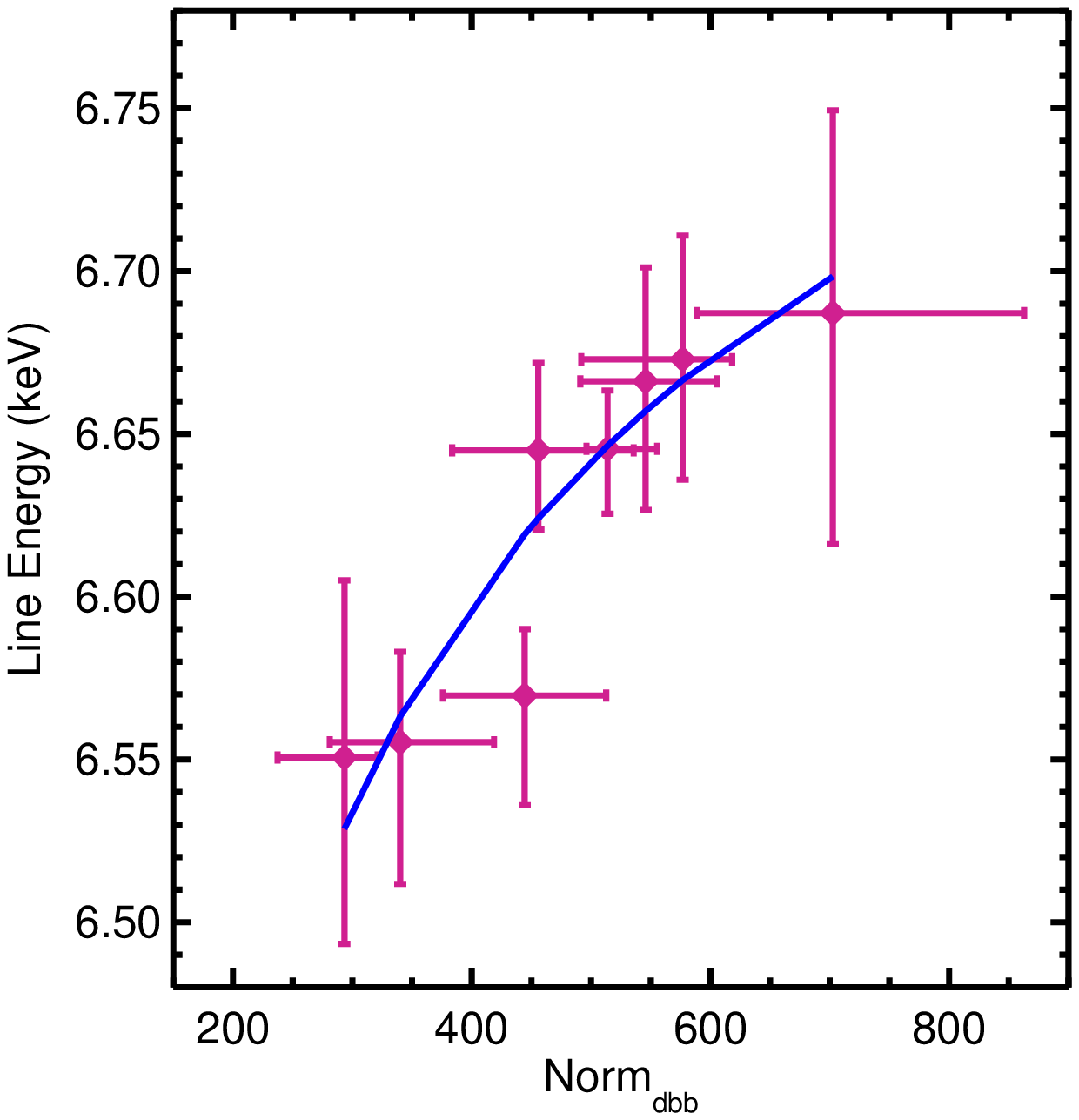}\\
    \end{tabular}
    \caption{Variation of the line energy with the disc blackbody normalizations (bottom) obtained from the spectral fitting of all eight {\em NICER} observations of GX 349+2 (see Section \ref{section3.3.2}). The best fitting model (y=a$\times$ (1-$\frac{b}{\sqrt{x}}$) and y=$a\times x+b$ is shown in blue (see Section \ref{section4}).} 
    \label{figure14}
\end{figure}

\begin{figure}
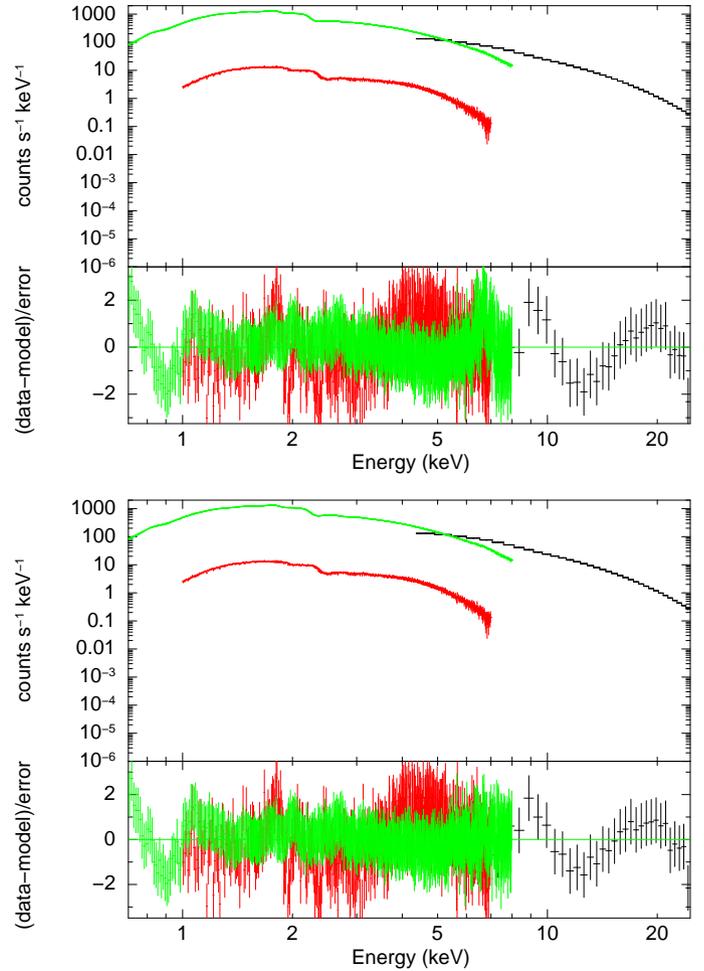

    \hspace*{-1.4cm}
    \begin{tabular}{lr}    
    \includegraphics[width=0.35\textwidth,angle =270]{Figure15a.ps}\\ 
    \includegraphics[width=0.35\textwidth,angle=270]{Figure15b.ps}
    \end{tabular}
    \caption{The 0.7-25 keV simultaneous broadband ({\em NICER}+SXT+LAXPC) X-ray spectrum of GX 349+2 for {\em AstroSat} observation 2 (A2) and {\em NICER} observation 5 (N5). Spectra are fitted with  \texttt{tbabs*edge*(bbodyrad+diskbb+powerlaw+diskline)} model (bottom). Two edge at 8.7 keV (LAXPC) and $\sim$ 0.8 keV ({\em NICER}) are added (see Section \ref{section3.3.2}).} 
    \label{figure15}
\end{figure}

Similar to {\em AstroSat}, we perform the spectral analysis of all eight {\em NICER} observations. We carry out the spectral fitting of the {\em NICER} spectrum of GX349+2, considering the energy range of 0.7-8 keV. For the spectral modelling, a 1\% systematic error is added to the spectrum to account for the instrumental uncertainty. We perform the joint spectral fit for each of the {\em NICER} observations, using the combination of a blackbody radiation model (\texttt{bbodyrad} in \texttt{XSPEC}), a multicolour disc blackbody model (MCD; \texttt{diskbb} in \texttt{XSPEC}; \citet{1984PASJ...36..741M}), and a power law model (\texttt{powerlaw} in \texttt{XSPEC}). The top left panel of Figure~\ref{figure11}  shows the model-fitted {\em NICER} spectrum during observation N1 of GX349+2, whereas the best fitting parameters for all eight observations are mentioned in Table \ref{table7}. The {\em NICER} spectra for all eight observations are observed to comprise narrow spectral features at 0.8 keV and 6.7 keV. The addition of an \texttt{edge} component at $\sim$ 0.8 keV improves the fit. The instrumental systematics can be a possible explanation of the feature observed at $\sim$ 0.8 keV (also mentioned in \cite{2021ApJ...911..123L}). The top right panel of Figure~\ref{figure11} shows the resultant model-fitted {\em NICER} spectrum, including an \texttt{edge} during observation N1, and the corresponding best-fitting parameters for all eight observations are mentioned in Table \ref{table8}. The iron (emission) feature detected at $\sim$ 6.7 keV is well represented with either an additional gaussian (\texttt{gauss}) (Figure~\ref{figure11} (bottom left) or a diskline (\texttt{diskline}) model (Figure~\ref{figure11} (bottom right)). Although we attempted fits of the data with broadband reflection models such as \texttt{RelxillNS} and \texttt{relconv*reflionx} models, given the data quality, we could not obtain well-constrained parameters. The detection of an asymmetric iron line from the Z type NS LMXB source GX 349+2 has previously been reported \citep{2018ApJ...867...64C}. We note here that between the  \texttt{diskline} and the \texttt{gauss} model, we could not statistically prefer one over the other. However, in this work, we consider the \texttt{diskline} model for the further spectral evolution study of all the NICER observations of GX 349+2, as it is more consistent with the physical scenario. The best fitting parameters for the \texttt{diskline} model are mentioned in Table \ref{table9} respectively. The left and right columns of Figure~\ref{figure12} show the evolution of broadband X-ray spectral parameters such as Galactic neutral hydrogen column densities (nH), blackbody temperatures ($kT_{bb}$), blackbody normalizations, temperature at the inner disc ($kT_{in}$), disc blackbody normalizations,  power law indices, power law normalizations, and \texttt{diskline} model parameters with flux. The \texttt{diskline} model parameters include line energies (E), power law dependence of emissivities ($\beta$), inner disc radii ($R_{in}$), outer disc radii ($R_{out}$), and inclinations (Incl). It is to be noted here that, for the fitting, we fix the outer disc radii at 1000 $R_{g}$, as otherwise, the parameters become completely unconstrained. The iron (emission) line detected $\sim$ 6.7 keV is observed to have the strongest presence during observations N6 and N7. The error-weighted fit residuals corresponding to the {\tt tbabs*edge*(diskbb+bbodyrad+powerlaw)}, the {\tt tbabs*edge*(diskbb+bbodyrad+powerlaw+gauss)}, and the {\tt tbabs*edge*(diskbb+bbodyrad+powerlaw+diskline)} models for all eight {\em NICER} observations are shown in Figure~\ref{figure13}. Figure~\ref{figure13} clearly shows that the  iron (emission) line is most prominent during observation N7. The variations of the line energy with DBB normalizations for all 8 NICER observations are shown in Figure \ref{figure14}. 

In one particular case (A2 and N5), we have simultaneous coverage during an observation by both {\em AstroSat} and {\em NICER}. We show model-fitted broadband simultaneous {\em NICER} and {\em AstroSat} spectra of GX 349+2 in Figure~\ref{figure15}. For simultaneous broadband spectroscopy, we include \texttt{edge} at 8.7 keV (LAXPC) and $\sim$ 0.8 keV ({\em NICER}).

\begin{table*}
\caption{Best-fitting spectral parameters corresponding to the best-fitting model consisting of absorbed blackbody radiation, multicolour disc blackbody, power law, and a diskline component \texttt{tbabs*edge*(bbodyrad+diskbb+powerlaw+diskline)} for all eight {\em NICER} observations of GX 349+2. The outer disc radius ( $R_{out}$) is fixed at 1000. (see Section \ref{section3.3.2}).}
\centering
\begin{tabular}{|c|c|c|c|c|c|c|c|c|}
\hline
Obs No.&N1&N2&N3&N4&N5&N6&N7&N8\\
\hline
$\text{nH}^{a}$&$1.85^{+0.04}_{-0.05}$ &$1.71^{+0.05}_{-0.03}$&$1.71^{+0.04}_{-0.03}$&$1.82^{+0.03}_{-0.04}$&$1.80_{-0.02}^{+0.02}$&$1.84^{+0.03}_{-0.03}$&$1.85^{+0.02}_{-0.01}$&$1.84^{+0.02}_{-0.03}$\\
\\
edge&$0.78^{<+0.01}_{-0.01}$&$0.78^{<+0.01}_{<-0.01}$&$0.79^{<+0.01}_{<-0.01}$&$0.78^{<+0.01}_{<-0.01}$&$0.78_{-<0.01}^{+<0.01}$&$0.78^{<+0.01}_{<-0.01}$&$0.78^{<+0.01}_{<-0.01}$&$0.77^{<+0.01}_{<-0.01}$\\ 
\\
maxtau &$0.34^{+0.04}_{-0.04}$&$0.31^{+0.03}_{-0.03}$&$0.31^{+0.02}_{-0.02}$&$0.34^{+0.03}_{-0.03}$&$0.33_{-0.03}^{+0.03}$&$0.33^{+0.02}_{-0.02}$&$0.35^{+0.01}_{-0.01}$&$0.37^{+0.03}_{-0.03}$\\
\\
$\text{kT}^{b}$&$1.65^{+0.02}_{-0.04}$&$1.80^{+0.03}_{-0.04}$&$1.84^{+0.04}_{-0.06}$&$1.68^{+0.01}_{-0.01}$&$1.77_{-0.02}^{+0.02}$&$1.74^{+0.02}_{-0.02}$&$1.79^{+0.01}_{-0.01}$&$1.76^{+0.04}_{-0.02}$\\
\\
$\text{BB}$ $\text{Norm}^{c}$&$203.53^{+25.64}_{-15.94}$&$157.41^{+23.98}_{-13.26}$&$131.28^{+21.7}_{-16.28}$&$164.69^{+10.02}_{-6.91}$&$119.08_{-4.17}^{+11.80}$&$162.04^{+8.48}_{-9.24}$&$140.28^{+6.00}_{-5.18}$&$123.05^{+14.18}_{-7.00}$\\
\\
$\text{kT}_{in}^{d}$&$0.91^{+0.04}_{-0.05}$&$1.12^{+0.06}_{-0.07}$&$1.17^{+0.08}_{-0.07}$&$0.97^{+0.04}_{-0.03}$&$1.04_{-0.03}^{+0.05}$&$0.99^{+0.03}_{-0.02}$&$1.00^{+0.01}_{-0.01}$&$1.04^{+0.04}_{-0.04}$\\
\\
$\text{DBB}$ $\text{Norm}^{e}$&$702.54^{+160.29}_{-113.66}$&$340.10^{+78.63}_{-59.00}$&$293.33^{+27.90}_{-56.03}$&$576.77^{+41.47}_{-84.96}$&$444.36_{-68.44}^{+68.31}$&$545.69^{+59.95}_{-54.67}$&$513.84^{+41.55}_{-17.69}$&$455.82^{+79.86}_{-72.26}$\\
\\
$\Gamma^{f}$&$5.75^{+0.19}_{-0.20}$&$5.06^{+0.24}_{-0.10}$&$5.07^{+0.19}_{-0.15}$&$5.62^{+0.08}_{-0.17}$&$5.52_{-0.12}^{+0.11}$&$5.67^{+0.12}_{-0.13}$&$5.71^{+0.06}_{-0.05}$&$5.71^{+0.08}_{-0.08}$\\
\\
$\text{PL}$ $\text{Norm}^{g}$&$6.22^{+0.68}_{-0.66}$&$4.65^{+0.46}_{-0.34}$&$4.57^{+0.50}_{-0.36}$&$6.06^{+0.49}_{-0.38}$&$5.61_{-0.33}^{+0.34}$&$5.97^{+0.46}_{-0.44}$&$6.10^{+0.25}_{-0.26}$&$6.26^{+0.32}_{-0.39}$\\
\\
$\text{E}^{h}$&$6.69^{+0.06}_{-0.07}$&$6.56^{+0.03}_{-0.04}$&$6.55^{+0.05}_{-0.06}$&$6.67^{+0.04}_{-0.04}$&$6.57_{-0.03}^{+0.02}$&$6.67^{+0.03}_{-0.04}$&$6.64^{+0.02}_{-0.02}$&$6.64^{+0.03}_{-0.02}$\\
\\
$\beta^{i}$&$-2.04^{+0.18}_{-0.15}$&$-2.12^{+0.11}_{-0.20}$&$-2.17^{+0.10}_{-0.15}$&$-2.14^{+0.14}_{-0.10}$&$-2.19_{-0.11}^{+0.09}$&$-2.01^{+0.08}_{-0.09}$&$-1.99^{+0.07}_{-0.08}$&$-2.27^{+0.11}_{-0.15}$\\
\\
%$R_{in}$&$6^{+7.06}_{-6}$&$6.40^{+7.10}_{-6.11}$&$6^{+2.59}_{-6}$&$9.75^{+4.86}_{-10.14}$&$6.83_{-6.60}^{+4.80}$&$6.00^{+9.39}_{-6.00}$&$6.46^{+8.83}_{-6.47}$&$10.16^{+4.62}_{-9.91}$\\
\\
$\text{R}_{in}^{j}$&$6^{+7.06}$&$6.40^{+7.10}$&$6^{+2.59}$&$9.75^{+4.86}$&$6.70^{+4.11}$&$6.00^{+9.39}$&$6.46^{+8.83}$&$10.16^{+4.62}$\\
\\
$\text{incl}^{k}$&$49.40^{+5.89}_{-2.30}$&$43^{*}$&$53.93_{-2.08}^{+4.72}$&$42.38^{+2.38}_{-1.89}$&$40.87_{-1.20}^{+1.46}$&$40.98^{+1.80}_{-2.10}$&$39.49^{+0.92}_{-1.34}$&$35.94^{+2.36}_{-1.70}$\\
\\
$\text{Norm}^{l}$ ($10^{-2}$)&$1.56^{+0.76}_{-0.33}$&$1.29^{<+0.19}_{<-0.17}$&$1.62^{+0.69}_{-0.27}$&$1.20_{+0.17}^{-0.09}$&$1.23_{-0.06}^{+0.15}$&$1.06_{-0.11}^{+0.11}$& $0.96_{-0.07}^{+0.08}$&$1.14_{-0.12}^{+0.15}$\\
\\
Chi-sq (DOF) &0.95 (714)&0.90 (715)&0.79 (714)&0.78(714)&0.77 (714)&0.51 (714)&0.48  (714)&0.90 (714)\\
\hline
\end{tabular}
\label{table9}
\begin{flushleft}
\footnotesize{ $R_{in}$ is found at the hard upper limit of the diskline model}\\
\footnotesize{$^*$ fixed parameters}\\
\footnotesize{$^a$ Neutral hydrogen column density ($10^{22} cm^{-2}$)}\\
\footnotesize{$^b$ Blackbody Temperature (keV)} \\
\footnotesize{$^c$ Blackbody Normalisation}\\
\footnotesize{$^d$ inner disc Temperature (keV) }\\
\footnotesize{$^e$ Disc Blackbody Normalisation}\\
\footnotesize{$^f$ Power-law Index}\\
\footnotesize{$^g$ Power-law Normalisation}\\
\footnotesize{$^h$ Line Energies (keV)}\\
\footnotesize{$^i$ Power Law Dependence of Emissivity}\\
\footnotesize{$^j$ inner radius ($GM/c^{2}$)}\\
\footnotesize{$^k$ Inclination (degrees)}\\
\footnotesize{$^l$ Normalisation}\\
\end{flushleft}
\end{table*}

\section{Discussion}
\label{section4}

This work presents the broadband spectro-temporal correlation study of the Sco-like Z-type NS LMXB GX 349+2 using {\em AstroSat} and {\em NICER} observations. We investigate the evolution of the source through the different spectral states along its Z-track (Figure \ref{figure3}). The source exhibits large-scale variability throughout the {\em AstroSat} observations, which is also clearly visible from the {\em AstroSat} light curves displayed in Figure \ref{figure2}. It is visible from Figure \ref{figure3} that the source was in its Flaring branch (FB) and Normal Branch/Flaring Branch (NB/FB) vertex of the Z track during the {\em AstroSat} observations. The source was observed to be in a relatively harder state during observations A1 and A2 and then moved to a relatively softer state during observations A3, A4, and A5. Although we could not distinctly identify the source spectral state during the {\em NICER} observations, the source was possibly in the NB branch of its Z-track as no such flaring activities were visible during the {\em NICER}  observations.

\subsection{Emission variability behaviour}
\label{section4.1}

The Sco-like Z source GX 349+2 is known to exhibit fast variability. Peaked noise at different frequencies and quasi-periodic oscillations (QPOs) have previously been reported from this NS LMXB source. As shown in Figure~\ref{figure4} and Table~\ref{table3}, we detect two main features from the power spectra during {\em AstroSat} observations: the very low-frequency noise (VLFN) and the broad peaked noise (low-frequency noise (LFN)/the flaring branch noise (FBN)) at $\sim$ 3-8 Hz. We detect these features almost in all regions of the Z track. The power spectra generated from the {\em NICER} observations of GX 349+2 exhibit similar features. It should be noted here that, in the case of {\em NICER}, the fit does not require an additional Lorentzian component, along with a power law for all the observations except for N1 and N7. However, the non-requirement of the additional Lorentzian does not necessarily imply the absence of peaked noise at the level of N1 and N7, as the peaked noise at the level seen in other observations, with higher signal to noise ratio (S/N), would not be detectable in these observations due to lower signal to noise ratio (S/N). The VLFN is well represented by a power law model, and the power law index is observed to be relatively steeper in the state a (FB) during {\em AstroSat} A1 and A2 observations, whereas no such trend is observed during A3, A4, and A5 (Table \ref{table3}). As reported in Table \ref{table4}, the {\em NICER} PL indices are observed to lie in similar ranges implying no significant state change of the source during all 8 {\em NICER} observations, whereas the peaked noise, when detected, is observed to comprise of a wide range of central frequencies. Typically, the origin of the VLFN is proposed to be caused by the changes in intensity, possibly associated with the changes in the mass-accretion rate or an inhomogeneous accretion flow leading to the change in the physical environment in the region near the neutron star surface or the inner part of the accretion disc \citep{2002MNRAS.336..217O}. Thus, an increase in the PL index from NB to FB and then as it moves further along the FB in case of observations A1 and A2 (PL index in case of A1 and A2 is greater than A3, A4, and A5) can be interpreted as the increase in the mass accretion rate. Moreover, the increase of VLFN RMS along the FB (during observations A1 and A2) as shown in Figure~\ref{figure5} further corroborates such interpretations. Although we do not observe any significant variation in the RMS of the peaked noise, however, the centroid frequency is observed to be less in the FB than in the NB/FB vertex. Our results are almost consistent with the observed increasing trend of centroid frequency along the normal branch with an abrupt decrease along the FB from GX 349+2 reported previously in  \cite{2002MNRAS.336..217O}. A similar decreasing trend of centroid frequency of the peaked noise from NB to FB was also reported in \cite{2003A&A...398..223A}. 
Furthermore, as mentioned in Section~\ref{section1}, the origin of these peaked noise features is highly debated. We investigate their energy dependence to find a plausible explanation for the presence of the broad peaked noise in all the power spectra of GX 349+2 during the {\em AstroSat} and {\em NICER} observations. Figure~\ref{figure5} and Figure~\ref{figure7} show that the strength (or RMS) of the LFN/FBN  and the integrated fractional RMS exhibits an increasing trend with energy. A similar, but not exact evolution was reported in the LFN fractional RMS amplitude ( 5\% at 2--5.0 keV to $>$ 15\% at 13.0--60 keV) in GX 340+0 \citep{2000ApJ...537..374J}. We examined the flux dependence of the total integrated fractional RMS as reported in Table~\ref{table5}. Except for the variation of RMS in the FB and NB/FB vertex, no such RMS-flux correlation is observed. 

The origin of the FBN can be physically explained by the adaptation of the photon bubble model, which proposes the formation of photon bubbles in the settling mound below the accretion shock at super-Eddington accretion rate. The bubbles may coalesce and finally lose their photons through diffusion into the accretion shock resulting in the quasi-periodic variability in the X-ray intensity. Thus, the accretion funnel onto the neutron star magnetic poles or the radial mass accretion flow may give rise to FBN. According to the transitional layer model \citep{2002MNRAS.336..217O}, the LFN, on the other hand, is proposed to be caused by the oscillations in the transition layer between the disc and the neutron star. Thus, the strength of these features may increase with increasing accretion rate. This is consistent with the observed trend of relatively high RMS in state a than in state b, with increasing accretion rate during observations A1 and A2 (source moves more along FB) than during A3, A4, and A5. On the contrary, the origin of noise components is known to originate in the region that produces most of the luminosity in that energy band. Hence, the increasing trend of RMS with increasing photon energy may also imply that the corona may have some contribution to the observed noise components \citep{2003A&A...398..223A}. However, the radiative processes occurring in the corona may also amplify the variability originating at the disc, resulting in the increasing strength of noise components with photon energy \citep{2021ASSL..461..263M}. 

Furthermore, our lag studies, as shown in Figure~\ref{figure8}, show the hard time lag detected in the lower (< 0.1 Hz) frequency range between 3-4 and 4-12 keV {\em NICER} light curves. The hard lags at low-frequency ranges can be physically interpreted in terms of two physical scenarios: the geometrical delay due to the light travel time between the accretion dic and the corona in the standard accretion disc corona (ADC) model scenario \citep{2006A&A...460..233C,2009A&A...494.1059J,2010A&A...512A...9B,2012A&A...546A..35C,2014MNRAS.438.2784C} and the propagation of accretion fluctuations \citep{1997MNRAS.292..679L,2001MNRAS.327..799K,2006MNRAS.367..801A}. The time lags obtained in our analysis are more consistent with the letter scenario, where the fluctuations propagate through the accretion disc before they reach the corona, causing soft photons to respond before hard photons.

\subsection{Spectral evolution}
\label{section4.2}

From the broadband spectral analysis of the NS LMXB GX 349+2 using SXT (1--7 keV) and LAXPC (4--25 keV), we identified the source to be in two states: the high-intensity flaring branch (FB) and the low-intensity normal branch/flaring branch (NB/FB) vertex. The broadband {\em AstroSat} spectra for all five observations are well described with two thermal (\texttt{bbodyrad} and \texttt{diskbb}) and one non-thermal model (\texttt{powerlaw}) component. Figure~\ref{figure10} and Table \ref{table6} show the evolution of spectral parameters for each of the states during all five {\em AstroSat} observations. The blackbody radiation component with a temperature ranging from 1.25--1.67 keV can be interpreted as the thermal emission possibly coming from the NS surface. However, considering a source distance of 9.2 kpc \citep{2018ApJ...867...64C}, the radius of the region emitting blackbody radiation is estimated to range from $7.06_{-0.18}^{+0.18}$--$9.60_{-0.28}^{+0.24}$ km. Hence, a more physical explanation of the blackbody component may possibly be a hotspot on the NS boundary layer (BL) or on the NS surface. The origin of the multi-coloured disc blackbody model can be interpreted as the emission from the accretion disc with the temperature at the inner disc radius ranging from 2.84--3.60 keV.
%, and the seed photons from the accretion disc are possibly undergoing Comptonization in the corona. The hard non-thermal emission from the corona is represented by the power-law component. 
The seed photons emitted from the accretion disc can undergo Comptonization in the corona resulting in the hard non-thermal spectral emission represented by the power law component.
Alternatively, the origin of the non-thermal emission may also be due to a bulk motion of matter near the NS surface \citep{1998ApJ...493..863T,2000astro.ph.11447P}.
Figure ~\ref{figure10} shows the source flux is relatively lower in the NB/FB (red diamonds) vertex than in the FB (blue diamonds). The blackbody temperatures and normalizations are observed to increase with flux as the source moves from the NB/FB vertex to the FB. Thus, with the increasing accretion rate, the source is possibly moving to the FB from the NB/FB vertex resulting in the further heating of the NS boundary layer, resulting in an increased temperature and radius of the brightened hotspot present at the disc-boundary layer interface or the NS surface itself. Furthermore, the increasing temperature and hence decreasing inner disc radius (typically T $\propto$ $R^{-3/4}$ for thin disc approximation) firmly confirms the approach of the inner accretion disc towards the central NS, causing further heating up of the accretion disc. The increasing PL normalization, in that case, possibly describes the increasing Comptonization contributions with increasing mass influx.

The radius and temperature of the blackbody estimated from the {\em NICER} spectral analysis are almost consistent with the {\em AstroSat} results. 
%Although we could not constrain the \texttt{diskbb} normalizations from the {\em AstroSat} spectral analysis, 
Although the \texttt{diskbb} normalizations obtained from the {\em AstroSat} spectral analysis are not well constrained, the better sensitivity of {\em NICER} in the soft X-ray band allows a more precise estimate of the inner disc radius of the accreting system. Further addition of a reflection model such as {\tt diskline} component provides relatively better fit statistics for all eight {\em NICER} observations. The inner disc radius ($R_{in}$) obtained from the best-fit \texttt{diskline} model ranges from $\sim$ 6 -- 10.16 $R_{g}$, which is comparable to the expected values of the NS surface radius and the radius of the boundary layer. Thus, in that case, the inner accretion disc possibly touches the NS BL layer or the NS surface itself \citep{1999AstL...25..269I,2010AstL...36..848I}.  A similar estimate of the inner disc radius constraints of Z-type NS LMXB has been reported previously from the disc reflection modeling for the scenarios where the disc remains very close to the NS surface, providing an upper limit to the NS radius \citep{2022ApJ...927..112L}. Furthermore, our estimates are almost consistent with the previously reported estimation of the inner disc radius of other NS LMXB sources \citep{2017ApJ...836..140L, 2018ApJ...858L...5L, 2020ApJ...895...45L}. Since phenomenon like magnetically channeled accretion requires the presence of a strong magnetic field NS, the non-detection of X-ray pulsations and cyclotron lines from GX 349+2 indicates the presence of a relatively low magnetic field NS emphasizing the aforementioned scenario.
%Figure~\ref{figure12} shows an increasing, although not very significant, trend in nH with flux considering all eight {\em NICER} observations.
%The blackbody temperatures and the inner disc temperatures are observed to range from 1.65 to 1.84 keV and from 0.91 to 1.17 keV. Moreover, the estimated radius of the blackbody emitting hotspot and the inner disc radius, if isotropic, ranges from $10.09_{-0.16}^{+0.35}$ km to $13.12_{-0.52}^{+0.81}$ km and $18.42_{+1.85}^{+0.86}$ km to $28.51_{-2.41}^{+3.09}$ km (see section \ref{section3.3.2}). 
%It appears from Figure~\ref{figure13} that as the disc moves outwards, the temperature of the inner disc decreases. Consequently, the temperature of the hotspot falls off as it expands. On the contrary, the PL index and PL normalizations do not show many variations. Thus, the evolution of the observed flux in the {\em NICER} 0.7-8 keV band can be explained as an increasing radiative contribution of the blackbody emitting hotspot as it expands.  
From NICER analysis, as shown in Figure~\ref{figure12}, it is also observed that inner disc temperature decreases as the disc moves outwards. Furthermore, as the disc recedes, the hotspot is found to expand in radius with its temperature cooling off. On the contrary, the non-thermal power law components do not show significant variations in the  {\em NICER} analysis, possibly due to its poorer sensitivity at the harder energies. Thus, the evolution of the observed flux in the {\em NICER} (0.7-8 keV) band for these 8 observations can be attributed primarily due to an increasing radiative contribution of the blackbody emitting hotspot as it expands.

It is to be noted here that we find a discrepancy between the inner disc radius estimates obtained from the \texttt{diskbb} and \texttt{diskline} model fits. This can be explained by the fact that the \texttt{diskbb} provides an apparent estimate of the inner disc, which depends on both $\kappa$, the ratio of the colour temperature to the effective temperature and the correction factor $\xi$ \citep{1998PASJ...50..667K}. Alternatively, the {\tt diskline} model takes relativistic effects into consideration, and the reflections from the disc may depend on many factors, including the disc ionisations, thermodynamics, geometry, or inclination angles \citep{1989MNRAS.238..729F}.

Figure~\ref{figure11} and Figure~\ref{figure13} show the detection of the iron (emission) line in the {\em NICER} spectrum of GX 349+2. We study the correlation between the line energy and the other spectral parameters of GX 349+2 for all eight {\em NICER} observations. %Figure~\ref{figure14} shows the evolution of the energy of the line spectrum with the DBB normalizations. 
In general, as the accretion disc is mildly relativistic and the materials move faster near the central object, the observed line profiles are broadened and skewed \citep{2000PASP..112.1145F}. The combined influences of the Doppler effect and gravitational redshifting modulate the contribution from each radius of the accretion disc to lower energy. Such relativistically broadened line profiles are clearly observable in the residuals in the spectral continuum fitting, as shown by the panels in the left column of Figure~\ref{figure13}. 
A strong dependence of the energy of the line on the DBB normalization (tracer of the inner disc radius) is observed, as shown in Figure~\ref{figure14}. A correlation test performed between the DBB normalization and the line energy shows a positive correlation with the Spearman rank correlation coefficient of 1. This can be well explained by an approaching inner disc resulting in additional contribution from the lower radii and associated enhanced relativistic modulation of the line profile. Such an effect will exhibit higher redshifts of the line profile resulting in decreasing line energy with decreasing inner disc radius.   Figure~\ref{figure14} shows that the evolution is consistent with a linear (y=$a\times x+b$)  as well as the y=a$\times$ (1-$\frac{b}{\sqrt{x}}$) relation as predicted by the gravitational redshift, where a $\sim$ $E_{\circ}$ and b $\sim$ $(2GM/f^{2}Rc^{2})$ ($\frac{\sqrt{cosi}}{D_{10}}$). The linear relation provides a reasonable fit because fewer data points typically tend to follow a linear relation. Although both models provide a statistically preferred fit, we consider the relation predicted by the gravitational redshift to estimate the physical parameters. The estimated mass and rest energy from the fitting is $2.34\pm0.71$ $M_{solar}$ and $6.99\pm0.12$ keV. We note that for this calculation, we consider the color factor (f), source distance, and inclination (i) as 1.7 \citep{2012ApJ...757...11M,2000MNRAS.313..193M,1995ApJ...445..780S},  9.2 kpc \citep{2018ApJ...867...64C} and $43^{\circ}$ (average estimate from NICER spectral analysis, see Section \ref{section3.3.2}). The centroid energy of the iron line is known to range from 6.4-6.5 keV for Fe-I XX to 6.9 for Fe- XXVI, depending on the degree of ionisation \citep{1989ESASP.296...19S}. Thus, the higher value of the estimated line energy ($E_{\circ}$) can be explained by the higher degree of ionisation. Furthermore, we jointly use the continuum (\texttt{diskbb} model) and line spectroscopy (\texttt{diskline} model) to estimate the mass of the NS. A similar method is used by \cite{2019MNRAS.487.4221S} to constrain the mass and radius of the central BH of MAXI J1535–571. However, in this work, we demonstrate the studies of the evolution and correlation between the continuum (inner disc radius) and line parameters (line energy). Thus, it provides a potentially robust method to constrain the parameters of the central object of accreting systems, including NS and BH.
%measurement of the gravitational redshift at the surface of a neutron star provides a direct constraint on the mass-to-radius ratio and hence the equation of state. Thus, such a shift observed in the line energy may provide a good constraint on the mass of the NS at a particular disk radius.
%It is visible from the strong correlation observed between the inner accretion disc radius and the line energy in our analysis. 
A detailed investigation of such reflection line profiles can help trace the evolution of the inner disc and the related relativistic effects along the `Z' tracks of the source. However, the inner accretion disc radius and inclination angle estimates obtained from the \texttt{diskline} model in our analysis do not show substantially wide variations as the source exhibits a modest spectral change during all eight {\em NICER} observations. Although {\em AstroSat} could trace the spectral change of GX 349+2, we could not detect any reflection features in the corresponding spectra due to its limited spectral sensitivity. The upper panel of Figure~\ref{figure15}  shows the non-detection of the residuals near $\sim$ 6.7 keV in the AstroSat spectrum of GX 349+2, which is significantly visible in the simultaneous NICER observation. Thus, further observations, with good spectral resolution and well sampling of the entire `Z' track, are required to get an overall idea of the inner accretion physics and source geometry during the flaring and non-flaring state of Z-type NS LMXB sources, including GX 349+2. 

 Figure~\ref{figure15} shows the joint simultaneous {\em NICER} and {\em AstroSat} model-fitted spectra obtained for observation A2 and N5. The inner disc temperature from only {\em AstroSat} (relatively high energy sensitive $\sim$ 1-25 keV ) and PL indices obtained from only {\em NICER} analysis (relatively low energy sensitive $\sim$ 0.7-8 keV ) show a systematic shift towards higher values compared to the joint broadband (0.7-25 keV) simultaneous {\em NICER} and {\em AstroSat} spectral fitting parameters. Furthermore, the joint spectral fitting is observed to put constraints on the parameters with greater precision with less uncertainty. A similar effect was reported during previous {\em NICER} and {\em AstroSat} observations \citep{2022MNRAS.509.3989K, 2022MNRAS.510.1577G}. Thus, such studies demand a joint broadband analysis and manifest its requirement to get a better well-constrained estimate of the spectral parameters. 

%Figure~\ref{figure15} shows the joint simultaneous {\em NICER} and {\em AstroSat} model-fitted spectra obtained for observation A2 and N5. We see that the inner disc temperature and PL indices obtained from only {\em AstroSat} and only {\em NICER} analysis show a systematic shift towards higher values compared to the joint simultaneous {\em NICER} and {\em AstroSat} spectral fitting parameters. Thus, such studies demand a joint broadband analysis and manifest its requirement to get a better well-constrained estimate of the spectral parameters \citep{2022MNRAS.509.3989K, 2022MNRAS.510.1577G}. 

%This argument is firmly established by the detection of lag in the 4--12 keV band ($\sim$ the iron emission line energy range).           

\section{Conclusions}
\label{section5}

We have presented the broadband spectro-temporal correlation study of Sco-like  Z type NS LMXB GX 349+2 using {\em AstroSat} and {\em NICER}. In our broadband analysis, we detected a large-scale source variability, which helped to get an insight into the origin of different variability components and how they can be used to trace the physics of accretion and geometry of Z-type NS LMXBs. Our study reports the presence of hard lag at very low frequencies represented by the accretion disc fluctuation revealing the first evidence of the detection of VLFN originating from the accretion disc in NS LMXBs. The simultaneous continuum and line evolution study of the iron line detected in our NICER data provides a new powerful way to constrain the NS mass and probe the properties of the central object in BH/NS systems. Moreover, it manifests the requirement of further broadband and sensitive study of Z-type NS LMXB sources to understand several long-standing questions, including the reason behind the flaring behaviour during FB, the source geometry, inclinations, the evolution of the companion, the role of the magnetic field,  and the physics of the accretion process, its evolution, and geometry along the tracks traced by the Z-type NS LMXB sources.

\section*{Acknowledgements}
This paper includes data collected by the {\em AstroSat} mission, which are publicly available from the ISRO Science Data Archive for the {\em AstroSat} Mission. The data are made available to the users by the Indian Space Science Data Centre (ISSDC), ISRO (https://astrobrowse.issdc.gov.in/astro\_archive/archive/Home.jsp). We thank the SXT and LAXPC Payload Operation Centres at TIFR for releasing the data via the ISSDC data archive and providing the necessary software tools. This work has used software and data provided by the High Energy Astrophysics Science Archive Research Center (HEASARC), a service of the Astrophysics Science Division at NASA/GSFC. We acknowledge support from ISRO under the AstroSat archival Data utilization program (DS 2B-13013(2)/4/2020-Sec.2).

%%%%%%%%%%%%%%%%%%%%%%%%%%%%%%%%%%%%%%%%%%%%%%%%%%
\section*{Data Availability}

This paper includes data publicly available from the Indian Space Science Data Centre (ISSDC), ISRO website [\url{https://astrobrowse.issdc.gov.in/astro\_archive/archive/Home.jsp}]. The NICER data are available at [\url{https://heasarc.gsfc.
nasa.gov/docs/archive.html}].



%\clearpage
\bibliographystyle{mnras}
\bibliography{ref.bib} % if your bibtex file is called example.bib

\begin{thebibliography}{}
\makeatletter
\relax
\def\mn@urlcharsother{\let\do\@makeother \do\$\do\&\do\#\do\^\do\_\do\%\do\~}
\def\mn@doi{\begingroup\mn@urlcharsother \@ifnextchar [ {\mn@doi@}
  {\mn@doi@[]}}
\def\mn@doi@[#1]#2{\def\@tempa{#1}\ifx\@tempa\@empty \href
  {http://dx.doi.org/#2} {doi:#2}\else \href {http://dx.doi.org/#2} {#1}\fi
  \endgroup}
\def\mn@eprint#1#2{\mn@eprint@#1:#2::\@nil}
\def\mn@eprint@arXiv#1{\href {http://arxiv.org/abs/#1} {{\tt arXiv:#1}}}
\def\mn@eprint@dblp#1{\href {http://dblp.uni-trier.de/rec/bibtex/#1.xml}
  {dblp:#1}}
\def\mn@eprint@#1:#2:#3:#4\@nil{\def\@tempa {#1}\def\@tempb {#2}\def\@tempc
  {#3}\ifx \@tempc \@empty \let \@tempc \@tempb \let \@tempb \@tempa \fi \ifx
  \@tempb \@empty \def\@tempb {arXiv}\fi \@ifundefined
  {mn@eprint@\@tempb}{\@tempb:\@tempc}{\expandafter \expandafter \csname
  mn@eprint@\@tempb\endcsname \expandafter{\@tempc}}}

\bibitem[\protect\citeauthoryear{{Agrawal} \& {Bhattacharyya}}{{Agrawal} \&
  {Bhattacharyya}}{2003}]{2003A&A...398..223A}
{Agrawal} V.~K.,  {Bhattacharyya} S.,  2003, \mn@doi [\aap]
  {10.1051/0004-6361:20021709}, \href
  {https://ui.adsabs.harvard.edu/abs/2003A&A...398..223A} {398, 223}

\bibitem[\protect\citeauthoryear{{Agrawal} \& {Sreekumar}}{{Agrawal} \&
  {Sreekumar}}{2003}]{2003MNRAS.346..933A}
{Agrawal} V.~K.,  {Sreekumar} P.,  2003, \mn@doi [\mnras]
  {10.1111/j.1365-2966.2003.07147.x}, \href
  {https://ui.adsabs.harvard.edu/abs/2003MNRAS.346..933A} {346, 933}

\bibitem[\protect\citeauthoryear{{Agrawal} et~al.,}{{Agrawal}
  et~al.}{2017}]{2017JApA...38...30A}
{Agrawal} P.~C.,  et~al., 2017, \mn@doi [Journal of Astrophysics and Astronomy]
  {10.1007/s12036-017-9451-z}, \href
  {https://ui.adsabs.harvard.edu/abs/2017JApA...38...30A} {38, 30}

\bibitem[\protect\citeauthoryear{{Antia} et~al.,}{{Antia}
  et~al.}{2017}]{2017ApJS..231...10A}
{Antia} H.~M.,  et~al., 2017, \mn@doi [\apjs] {10.3847/1538-4365/aa7a0e}, \href
  {https://ui.adsabs.harvard.edu/abs/2017ApJS..231...10A} {231, 10}

\bibitem[\protect\citeauthoryear{{Ar{\'e}valo} \& {Uttley}}{{Ar{\'e}valo} \&
  {Uttley}}{2006}]{2006MNRAS.367..801A}
{Ar{\'e}valo} P.,  {Uttley} P.,  2006, \mn@doi [\mnras]
  {10.1111/j.1365-2966.2006.09989.x}, \href
  {https://ui.adsabs.harvard.edu/abs/2006MNRAS.367..801A} {367, 801}

\bibitem[\protect\citeauthoryear{{Ba{\l}uci{\'n}ska-Church}, {Gibiec},
  {Jackson}  \& {Church}}{{Ba{\l}uci{\'n}ska-Church}
  et~al.}{2010}]{2010A&A...512A...9B}
{Ba{\l}uci{\'n}ska-Church} M.,  {Gibiec} A.,  {Jackson} N.~K.,   {Church}
  M.~J.,  2010, \mn@doi [\aap] {10.1051/0004-6361/200913199}, \href
  {https://ui.adsabs.harvard.edu/abs/2010A&A...512A...9B} {512, A9}

\bibitem[\protect\citeauthoryear{{Belloni} \& {Hasinger}}{{Belloni} \&
  {Hasinger}}{1990}]{1990A&A...230..103B}
{Belloni} T.,  {Hasinger} G.,  1990, \aap, \href
  {https://ui.adsabs.harvard.edu/abs/1990A&A...230..103B} {230, 103}

\bibitem[\protect\citeauthoryear{{Bhattacharyya} et~al.,}{{Bhattacharyya}
  et~al.}{2021}]{2021JApA...42...17B}
{Bhattacharyya} S.,  et~al., 2021, Journal of Astrophysics and Astronomy, 42

\bibitem[\protect\citeauthoryear{{Cackett} et~al.,}{{Cackett}
  et~al.}{2008}]{2008ApJ...674..415C}
{Cackett} E.~M.,  et~al., 2008, \mn@doi [\apj] {10.1086/524936}, \href
  {https://ui.adsabs.harvard.edu/abs/2008ApJ...674..415C} {674, 415}

\bibitem[\protect\citeauthoryear{{Chakraborty} \&
  {Bhattacharyya}}{{Chakraborty} \&
  {Bhattacharyya}}{2011}]{2011ApJ...730L..23C}
{Chakraborty} M.,  {Bhattacharyya} S.,  2011, \mn@doi [\apjl]
  {10.1088/2041-8205/730/2/L23}, \href
  {https://ui.adsabs.harvard.edu/abs/2011ApJ...730L..23C} {730, L23}

\bibitem[\protect\citeauthoryear{{Chakraborty}, {Bhattacharyya}  \&
  {Mukherjee}}{{Chakraborty} et~al.}{2011}]{2011MNRAS.418..490C}
{Chakraborty} M.,  {Bhattacharyya} S.,   {Mukherjee} A.,  2011, \mn@doi
  [\mnras] {10.1111/j.1365-2966.2011.19499.x}, \href
  {https://ui.adsabs.harvard.edu/abs/2011MNRAS.418..490C} {418, 490}

\bibitem[\protect\citeauthoryear{{Church} \&
  {Ba{\l}uci{\'n}ska-Church}}{{Church} \&
  {Ba{\l}uci{\'n}ska-Church}}{2012}]{2012MmSAI..83..170C}
{Church} M.~J.,  {Ba{\l}uci{\'n}ska-Church} M.,  2012, \memsai, \href
  {https://ui.adsabs.harvard.edu/abs/2012MmSAI..83..170C} {83, 170}

\bibitem[\protect\citeauthoryear{{Church}, {Halai}  \&
  {Ba{\l}uci{\'n}ska-Church}}{{Church} et~al.}{2006}]{2006A&A...460..233C}
{Church} M.~J.,  {Halai} G.~S.,   {Ba{\l}uci{\'n}ska-Church} M.,  2006, \mn@doi
  [\aap] {10.1051/0004-6361:20065035}, \href
  {https://ui.adsabs.harvard.edu/abs/2006A&A...460..233C} {460, 233}

\bibitem[\protect\citeauthoryear{{Church}, {Gibiec}, {Ba{\l}uci{\'n}ska-Church}
   \& {Jackson}}{{Church} et~al.}{2012}]{2012A&A...546A..35C}
{Church} M.~J.,  {Gibiec} A.,  {Ba{\l}uci{\'n}ska-Church} M.,   {Jackson}
  N.~K.,  2012, \mn@doi [\aap] {10.1051/0004-6361/201218987}, \href
  {https://ui.adsabs.harvard.edu/abs/2012A&A...546A..35C} {546, A35}

\bibitem[\protect\citeauthoryear{{Church}, {Gibiec}  \&
  {Ba{\l}uci{\'n}ska-Church}}{{Church} et~al.}{2014}]{2014MNRAS.438.2784C}
{Church} M.~J.,  {Gibiec} A.,   {Ba{\l}uci{\'n}ska-Church} M.,  2014, \mn@doi
  [\mnras] {10.1093/mnras/stt2364}, \href
  {https://ui.adsabs.harvard.edu/abs/2014MNRAS.438.2784C} {438, 2784}

\bibitem[\protect\citeauthoryear{{Coughenour}, {Cackett}, {Miller}  \&
  {Ludlam}}{{Coughenour} et~al.}{2018}]{2018ApJ...867...64C}
{Coughenour} B.~M.,  {Cackett} E.~M.,  {Miller} J.~M.,   {Ludlam} R.~M.,  2018,
  \mn@doi [\apj] {10.3847/1538-4357/aae098}, \href
  {https://ui.adsabs.harvard.edu/abs/2018ApJ...867...64C} {867, 64}

\bibitem[\protect\citeauthoryear{{Di Salvo}, {Robba}, {Iaria}, {Stella},
  {Burderi}  \& {Israel}}{{Di Salvo} et~al.}{2001}]{2001ApJ...554...49D}
{Di Salvo} T.,  {Robba} N.~R.,  {Iaria} R.,  {Stella} L.,  {Burderi} L.,
  {Israel} G.~L.,  2001, \mn@doi [\apj] {10.1086/321353}, \href
  {https://ui.adsabs.harvard.edu/abs/2001ApJ...554...49D} {554, 49}

\bibitem[\protect\citeauthoryear{{Ding}, {Zhang}, {Wang}, {Li}, {Qu}  \&
  {Huang}}{{Ding} et~al.}{2016}]{2016MNRAS.455.2959D}
{Ding} G.~Q.,  {Zhang} W.~Y.,  {Wang} Y.~N.,  {Li} Z.~B.,  {Qu} J.~L.,
  {Huang} C.~P.,  2016, \mn@doi [\mnras] {10.1093/mnras/stv2459}, \href
  {https://ui.adsabs.harvard.edu/abs/2016MNRAS.455.2959D} {455, 2959}

\bibitem[\protect\citeauthoryear{{Fabian}, {Rees}, {Stella}  \&
  {White}}{{Fabian} et~al.}{1989}]{1989MNRAS.238..729F}
{Fabian} A.~C.,  {Rees} M.~J.,  {Stella} L.,   {White} N.~E.,  1989, \mn@doi
  [\mnras] {10.1093/mnras/238.3.729}, \href
  {https://ui.adsabs.harvard.edu/abs/1989MNRAS.238..729F} {238, 729}

\bibitem[\protect\citeauthoryear{{Fabian}, {Iwasawa}, {Reynolds}  \&
  {Young}}{{Fabian} et~al.}{2000}]{2000PASP..112.1145F}
{Fabian} A.~C.,  {Iwasawa} K.,  {Reynolds} C.~S.,   {Young} A.~J.,  2000,
  \mn@doi [\pasp] {10.1086/316610}, \href
  {https://ui.adsabs.harvard.edu/abs/2000PASP..112.1145F} {112, 1145}

\bibitem[\protect\citeauthoryear{Gendreau, Arzoumanian  \& Okajima}{Gendreau
  et~al.}{2012}]{10.1117/12.926396}
Gendreau K.~C.,  Arzoumanian Z.,   Okajima T.,  2012, in Takahashi T.,  Murray
  S.~S.,   den Herder J.-W.~A.,  eds, ~ Vol. 8443, Space Telescopes and
  Instrumentation 2012: Ultraviolet to Gamma Ray. SPIE, pp 322 -- 329,
  \mn@doi{10.1117/12.926396}, \url {https://doi.org/10.1117/12.926396}

\bibitem[\protect\citeauthoryear{Gendreau et~al.,}{Gendreau
  et~al.}{2016}]{10.1117/12.2231304}
Gendreau K.~C.,  et~al., 2016, in den Herder J.-W.~A.,  Takahashi T.,   Bautz
  M.,  eds, ~ Vol. 9905, Space Telescopes and Instrumentation 2016: Ultraviolet
  to Gamma Ray. SPIE, pp 420 -- 435, \mn@doi{10.1117/12.2231304}, \url
  {https://doi.org/10.1117/12.2231304}

\bibitem[\protect\citeauthoryear{{G{\"u}ver} et~al.,}{{G{\"u}ver}
  et~al.}{2022}]{2022MNRAS.510.1577G}
{G{\"u}ver} T.,  et~al., 2022, \mn@doi [\mnras] {10.1093/mnras/stab3422}, \href
  {https://ui.adsabs.harvard.edu/abs/2022MNRAS.510.1577G} {510, 1577}

\bibitem[\protect\citeauthoryear{{Hasinger} \& {van der Klis}}{{Hasinger} \&
  {van der Klis}}{1989}]{1989A&A...225...79H}
{Hasinger} G.,  {van der Klis} M.,  1989, \aap, \href
  {https://ui.adsabs.harvard.edu/abs/1989A&A...225...79H} {225, 79}

\bibitem[\protect\citeauthoryear{{Homan} et~al.,}{{Homan}
  et~al.}{2007}]{2007ApJ...656..420H}
{Homan} J.,  et~al., 2007, \mn@doi [\apj] {10.1086/510447}, \href
  {https://ui.adsabs.harvard.edu/abs/2007ApJ...656..420H} {656, 420}

\bibitem[\protect\citeauthoryear{{Husain}, {Misra}  \& {Sen}}{{Husain}
  et~al.}{2022}]{2022MNRAS.510.4040H}
{Husain} N.,  {Misra} R.,   {Sen} S.,  2022, \mn@doi [\mnras]
  {10.1093/mnras/stab3780}, \href
  {https://ui.adsabs.harvard.edu/abs/2022MNRAS.510.4040H} {510, 4040}

\bibitem[\protect\citeauthoryear{{Iaria}, {D'A{\'\i}}, {di Salvo}, {Robba},
  {Riggio}, {Papitto}  \& {Burderi}}{{Iaria}
  et~al.}{2009}]{2009A&A...505.1143I}
{Iaria} R.,  {D'A{\'\i}} A.,  {di Salvo} T.,  {Robba} N.~R.,  {Riggio} A.,
  {Papitto} A.,   {Burderi} L.,  2009, \mn@doi [\aap]
  {10.1051/0004-6361/200911936}, \href
  {https://ui.adsabs.harvard.edu/abs/2009A&A...505.1143I} {505, 1143}

\bibitem[\protect\citeauthoryear{{Inogamov} \& {Sunyaev}}{{Inogamov} \&
  {Sunyaev}}{1999}]{1999AstL...25..269I}
{Inogamov} N.~A.,  {Sunyaev} R.~A.,  1999, Astronomy Letters, \href
  {https://ui.adsabs.harvard.edu/abs/1999AstL...25..269I} {25, 269}

\bibitem[\protect\citeauthoryear{{Inogamov} \& {Sunyaev}}{{Inogamov} \&
  {Sunyaev}}{2010}]{2010AstL...36..848I}
{Inogamov} N.~A.,  {Sunyaev} R.~A.,  2010, \mn@doi [Astronomy Letters]
  {10.1134/S1063773710120029}, \href
  {https://ui.adsabs.harvard.edu/abs/2010AstL...36..848I} {36, 848}

\bibitem[\protect\citeauthoryear{{Jackson}, {Church}  \&
  {Ba{\l}uci{\'n}ska-Church}}{{Jackson} et~al.}{2009}]{2009A&A...494.1059J}
{Jackson} N.~K.,  {Church} M.~J.,   {Ba{\l}uci{\'n}ska-Church} M.,  2009,
  \mn@doi [\aap] {10.1051/0004-6361:20079234}, \href
  {https://ui.adsabs.harvard.edu/abs/2009A&A...494.1059J} {494, 1059}

\bibitem[\protect\citeauthoryear{{Jia} et~al.,}{{Jia}
  et~al.}{2020}]{2020JHEAp..25....1J}
{Jia} S.~M.,  et~al., 2020, \mn@doi [Journal of High Energy Astrophysics]
  {10.1016/j.jheap.2019.11.001}, \href
  {https://ui.adsabs.harvard.edu/abs/2020JHEAp..25....1J} {25, 1}

\bibitem[\protect\citeauthoryear{{Jithesh}, {Misra}, {Maqbool}  \&
  {Mall}}{{Jithesh} et~al.}{2021}]{2021MNRAS.505..713J}
{Jithesh} V.,  {Misra} R.,  {Maqbool} B.,   {Mall} G.,  2021, \mn@doi [\mnras]
  {10.1093/mnras/stab1307}, \href
  {https://ui.adsabs.harvard.edu/abs/2021MNRAS.505..713J} {505, 713}

\bibitem[\protect\citeauthoryear{{Jonker}, {Wijnands}, {van der Klis},
  {Psaltis}, {Kuulkers}  \& {Lamb}}{{Jonker}
  et~al.}{1998}]{1998ApJ...499L.191J}
{Jonker} P.~G.,  {Wijnands} R.,  {van der Klis} M.,  {Psaltis} D.,  {Kuulkers}
  E.,   {Lamb} F.~K.,  1998, \mn@doi [\apjl] {10.1086/311372}, \href
  {https://ui.adsabs.harvard.edu/abs/1998ApJ...499L.191J} {499, L191}

\bibitem[\protect\citeauthoryear{{Jonker} et~al.,}{{Jonker}
  et~al.}{2000}]{2000ApJ...537..374J}
{Jonker} P.~G.,  et~al., 2000, \mn@doi [\apj] {10.1086/309029}, \href
  {https://ui.adsabs.harvard.edu/abs/2000ApJ...537..374J} {537, 374}

\bibitem[\protect\citeauthoryear{{Kashyap}, {Ram}, {G{\"u}ver}  \&
  {Chakraborty}}{{Kashyap} et~al.}{2022}]{2022MNRAS.509.3989K}
{Kashyap} U.,  {Ram} B.,  {G{\"u}ver} T.,   {Chakraborty} M.,  2022, \mn@doi
  [\mnras] {10.1093/mnras/stab2838}, \href
  {https://ui.adsabs.harvard.edu/abs/2022MNRAS.509.3989K} {509, 3989}

\bibitem[\protect\citeauthoryear{{Kotov}, {Churazov}  \& {Gilfanov}}{{Kotov}
  et~al.}{2001}]{2001MNRAS.327..799K}
{Kotov} O.,  {Churazov} E.,   {Gilfanov} M.,  2001, \mn@doi [\mnras]
  {10.1046/j.1365-8711.2001.04769.x}, \href
  {https://ui.adsabs.harvard.edu/abs/2001MNRAS.327..799K} {327, 799}

\bibitem[\protect\citeauthoryear{{Kubota}, {Tanaka}, {Makishima}, {Ueda},
  {Dotani}, {Inoue}  \& {Yamaoka}}{{Kubota} et~al.}{1998}]{1998PASJ...50..667K}
{Kubota} A.,  {Tanaka} Y.,  {Makishima} K.,  {Ueda} Y.,  {Dotani} T.,  {Inoue}
  H.,   {Yamaoka} K.,  1998, \mn@doi [\pasj] {10.1093/pasj/50.6.667}, \href
  {https://ui.adsabs.harvard.edu/abs/1998PASJ...50..667K} {50, 667}

\bibitem[\protect\citeauthoryear{{Kuulkers} \& {Vander Klis}}{{Kuulkers} \&
  {Vander Klis}}{1995}]{1995NYASA.759..344K}
{Kuulkers} E.,  {Vander Klis} M.,  1995, in {B{\"o}hringer} H.,  {Morfill}
  G.~E.,   {Tr{\"u}mper} J.~E.,  eds, ~ Vol. 759, Seventeeth Texas Symposium on
  Relativistic Astrophysics and Cosmology. p.~344,
  \mn@doi{10.1111/j.1749-6632.1995.tb17560.x}

\bibitem[\protect\citeauthoryear{{Kuulkers} \& {van der Klis}}{{Kuulkers} \&
  {van der Klis}}{1998}]{1998A&A...332..845K}
{Kuulkers} E.,  {van der Klis} M.,  1998, \aap, \href
  {https://ui.adsabs.harvard.edu/abs/1998A&A...332..845K} {332, 845}

\bibitem[\protect\citeauthoryear{{Kuulkers}, {van der Klis}, {Oosterbroek},
  {Asai}, {Dotani}, {van Paradijs}  \& {Lewin}}{{Kuulkers}
  et~al.}{1994}]{1994A&A...289..795K}
{Kuulkers} E.,  {van der Klis} M.,  {Oosterbroek} T.,  {Asai} K.,  {Dotani} T.,
   {van Paradijs} J.,   {Lewin} W.~H.~G.,  1994, \aap, \href
  {https://ui.adsabs.harvard.edu/abs/1994A&A...289..795K} {289, 795}

\bibitem[\protect\citeauthoryear{{Kuulkers}, {van der Klis}, {Oosterbroek},
  {van Paradijs}  \& {Lewin}}{{Kuulkers} et~al.}{1997}]{1997MNRAS.287..495K}
{Kuulkers} E.,  {van der Klis} M.,  {Oosterbroek} T.,  {van Paradijs} J.,
  {Lewin} W.~H.~G.,  1997, \mn@doi [\mnras] {10.1093/mnras/287.3.495}, \href
  {https://ui.adsabs.harvard.edu/abs/1997MNRAS.287..495K} {287, 495}

\bibitem[\protect\citeauthoryear{{Leahy}, {Darbro}, {Elsner}, {Weisskopf},
  {Sutherland}, {Kahn}  \& {Grindlay}}{{Leahy}
  et~al.}{1983}]{1983ApJ...266..160L}
{Leahy} D.~A.,  {Darbro} W.,  {Elsner} R.~F.,  {Weisskopf} M.~C.,  {Sutherland}
  P.~G.,  {Kahn} S.,   {Grindlay} J.~E.,  1983, \mn@doi [\apj]
  {10.1086/160766}, \href
  {https://ui.adsabs.harvard.edu/abs/1983ApJ...266..160L} {266, 160}

\bibitem[\protect\citeauthoryear{{Ludlam} et~al.,}{{Ludlam}
  et~al.}{2017}]{2017ApJ...836..140L}
{Ludlam} R.~M.,  et~al., 2017, \mn@doi [\apj] {10.3847/1538-4357/836/1/140},
  \href {https://ui.adsabs.harvard.edu/abs/2017ApJ...836..140L} {836, 140}

\bibitem[\protect\citeauthoryear{{Ludlam} et~al.,}{{Ludlam}
  et~al.}{2018}]{2018ApJ...858L...5L}
{Ludlam} R.~M.,  et~al., 2018, \mn@doi [\apjl] {10.3847/2041-8213/aabee6},
  \href {https://ui.adsabs.harvard.edu/abs/2018ApJ...858L...5L} {858, L5}

\bibitem[\protect\citeauthoryear{{Ludlam} et~al.,}{{Ludlam}
  et~al.}{2020}]{2020ApJ...895...45L}
{Ludlam} R.~M.,  et~al., 2020, \mn@doi [\apj] {10.3847/1538-4357/ab89a6}, \href
  {https://ui.adsabs.harvard.edu/abs/2020ApJ...895...45L} {895, 45}

\bibitem[\protect\citeauthoryear{{Ludlam} et~al.,}{{Ludlam}
  et~al.}{2021}]{2021ApJ...911..123L}
{Ludlam} R.~M.,  et~al., 2021, \mn@doi [\apj] {10.3847/1538-4357/abedb0}, \href
  {https://ui.adsabs.harvard.edu/abs/2021ApJ...911..123L} {911, 123}

\bibitem[\protect\citeauthoryear{{Ludlam} et~al.,}{{Ludlam}
  et~al.}{2022}]{2022ApJ...927..112L}
{Ludlam} R.~M.,  et~al., 2022, \mn@doi [\apj] {10.3847/1538-4357/ac5028}, \href
  {https://ui.adsabs.harvard.edu/abs/2022ApJ...927..112L} {927, 112}

\bibitem[\protect\citeauthoryear{{Lyubarskii}}{{Lyubarskii}}{1997}]{1997MNRAS.292..679L}
{Lyubarskii} Y.~E.,  1997, \mn@doi [\mnras] {10.1093/mnras/292.3.679}, \href
  {https://ui.adsabs.harvard.edu/abs/1997MNRAS.292..679L} {292, 679}

\bibitem[\protect\citeauthoryear{{M{\'e}ndez} \& {Belloni}}{{M{\'e}ndez} \&
  {Belloni}}{2021}]{2021ASSL..461..263M}
{M{\'e}ndez} M.,  {Belloni} T.~M.,  2021, in {Belloni} T.~M.,  {M{\'e}ndez} M.,
    {Zhang} C.,  eds,  Astrophysics and Space Science Library Vol. 461,
  Astrophysics and Space Science Library. pp 263--331 (\mn@eprint {arXiv}
  {2010.08291}), \mn@doi{10.1007/978-3-662-62110-3_6}

\bibitem[\protect\citeauthoryear{{Merloni}, {Fabian}  \& {Ross}}{{Merloni}
  et~al.}{2000}]{2000MNRAS.313..193M}
{Merloni} A.,  {Fabian} A.~C.,   {Ross} R.~R.,  2000, \mn@doi [\mnras]
  {10.1046/j.1365-8711.2000.03226.x}, \href
  {https://ui.adsabs.harvard.edu/abs/2000MNRAS.313..193M} {313, 193}

\bibitem[\protect\citeauthoryear{{Miller}, {Pooley}, {Fabian}, {Nowak}, {Reis},
  {Cackett}, {Pottschmidt}  \& {Wilms}}{{Miller}
  et~al.}{2012}]{2012ApJ...757...11M}
{Miller} J.~M.,  {Pooley} G.~G.,  {Fabian} A.~C.,  {Nowak} M.~A.,  {Reis}
  R.~C.,  {Cackett} E.~M.,  {Pottschmidt} K.,   {Wilms} J.,  2012, \mn@doi
  [\apj] {10.1088/0004-637X/757/1/11}, \href
  {https://ui.adsabs.harvard.edu/abs/2012ApJ...757...11M} {757, 11}

\bibitem[\protect\citeauthoryear{{Mitsuda} et~al.,}{{Mitsuda}
  et~al.}{1984}]{1984PASJ...36..741M}
{Mitsuda} K.,  et~al., 1984, \pasj, \href
  {https://ui.adsabs.harvard.edu/abs/1984PASJ...36..741M} {36, 741}

\bibitem[\protect\citeauthoryear{{Muno}, {Remillard}  \& {Chakrabarty}}{{Muno}
  et~al.}{2002}]{2002ApJ...568L..35M}
{Muno} M.~P.,  {Remillard} R.~A.,   {Chakrabarty} D.,  2002, \mn@doi [\apjl]
  {10.1086/340269}, \href
  {https://ui.adsabs.harvard.edu/abs/2002ApJ...568L..35M} {568, L35}

\bibitem[\protect\citeauthoryear{Nowak, Vaughan, Wilms, Dove  \&
  Begelman}{Nowak et~al.}{1999}]{1998astro.ph..7278N}
Nowak M.~A.,  Vaughan B.~A.,  Wilms J.,  Dove J.~B.,   Begelman M.~C.,  1999,
  \mn@doi [\apj] {10.1086/306610}, 510, 874

\bibitem[\protect\citeauthoryear{{O'Neill}, {Kuulkers}, {Sood}  \& {van der
  Klis}}{{O'Neill} et~al.}{2002}]{2002MNRAS.336..217O}
{O'Neill} P.~M.,  {Kuulkers} E.,  {Sood} R.~K.,   {van der Klis} M.,  2002,
  \mn@doi [\mnras] {10.1046/j.1365-8711.2002.05729.x}, \href
  {https://ui.adsabs.harvard.edu/abs/2002MNRAS.336..217O} {336, 217}

\bibitem[\protect\citeauthoryear{{Papathanassiou} \&
  {Psaltis}}{{Papathanassiou} \& {Psaltis}}{2000}]{2000astro.ph.11447P}
{Papathanassiou} H.,  {Psaltis} D.,  2000, arXiv e-prints, \href
  {https://ui.adsabs.harvard.edu/abs/2000astro.ph.11447P} {pp
  astro--ph/0011447}

\bibitem[\protect\citeauthoryear{{Ponman}, {Cooke}  \& {Stella}}{{Ponman}
  et~al.}{1988}]{1988MNRAS.231..999P}
{Ponman} T.~J.,  {Cooke} B.~A.,   {Stella} L.,  1988, \mn@doi [\mnras]
  {10.1093/mnras/231.4.999}, \href
  {https://ui.adsabs.harvard.edu/abs/1988MNRAS.231..999P} {231, 999}

\bibitem[\protect\citeauthoryear{{Shimura} \& {Takahara}}{{Shimura} \&
  {Takahara}}{1995}]{1995ApJ...445..780S}
{Shimura} T.,  {Takahara} F.,  1995, \mn@doi [\apj] {10.1086/175740}, \href
  {https://ui.adsabs.harvard.edu/abs/1995ApJ...445..780S} {445, 780}

\bibitem[\protect\citeauthoryear{{Singh} et~al.,}{{Singh}
  et~al.}{2016}]{2016SPIE.9905E..1ES}
{Singh} K.~P.,  et~al., 2016, in {den Herder} J.-W.~A.,  {Takahashi} T.,
  {Bautz} M.,  eds,  Society of Photo-Optical Instrumentation Engineers (SPIE)
  Conference Series Vol. 9905, Space Telescopes and Instrumentation 2016:
  Ultraviolet to Gamma Ray. p. 99051E, \mn@doi{10.1117/12.2235309}

\bibitem[\protect\citeauthoryear{{Singh} et~al.,}{{Singh}
  et~al.}{2017}]{2017JApA...38...29S}
{Singh} K.~P.,  et~al., 2017, \mn@doi [Journal of Astrophysics and Astronomy]
  {10.1007/s12036-017-9448-7}, \href
  {https://ui.adsabs.harvard.edu/abs/2017JApA...38...29S} {38, 29}

\bibitem[\protect\citeauthoryear{{Sridhar}, {Bhattacharyya}, {Chandra}  \&
  {Antia}}{{Sridhar} et~al.}{2019}]{2019MNRAS.487.4221S}
{Sridhar} N.,  {Bhattacharyya} S.,  {Chandra} S.,   {Antia} H.~M.,  2019,
  \mn@doi [\mnras] {10.1093/mnras/stz1476}, \href
  {https://ui.adsabs.harvard.edu/abs/2019MNRAS.487.4221S} {487, 4221}

\bibitem[\protect\citeauthoryear{{Stella}}{{Stella}}{1989}]{1989ESASP.296...19S}
{Stella} L.,  1989, in {Hunt} J.,  {Battrick} B.,  eds,  ESA Special
  Publication Vol. 1, Two Topics in X-Ray Astronomy, Volume 1: X Ray Binaries.
  Volume 2: AGN and the X Ray Background. p.~19

\bibitem[\protect\citeauthoryear{{Titarchuk}}{{Titarchuk}}{1994}]{1994ApJ...434..570T}
{Titarchuk} L.,  1994, \mn@doi [\apj] {10.1086/174760}, \href
  {https://ui.adsabs.harvard.edu/abs/1994ApJ...434..570T} {434, 570}

\bibitem[\protect\citeauthoryear{{Titarchuk} \& {Zannias}}{{Titarchuk} \&
  {Zannias}}{1998}]{1998ApJ...493..863T}
{Titarchuk} L.,  {Zannias} T.,  1998, \mn@doi [\apj] {10.1086/305157}, \href
  {https://ui.adsabs.harvard.edu/abs/1998ApJ...493..863T} {493, 863}

\bibitem[\protect\citeauthoryear{{Wijnands} \& {van der Klis}}{{Wijnands} \&
  {van der Klis}}{1998}]{1998AIPC..431..381W}
{Wijnands} R.,  {van der Klis} M.,  1998, in {Holt} S.~S.,  {Kallman} T.~R.,
  eds,  American Institute of Physics Conference Series Vol. 431, Accretion
  processes in Astrophysical Systems: Some like it hot! - eigth AstroPhysics
  Conference. pp 381--384 (\mn@eprint {arXiv} {astro-ph/9712186}),
  \mn@doi{10.1063/1.55921}

\bibitem[\protect\citeauthoryear{{Wijnands} \& {van der Klis}}{{Wijnands} \&
  {van der Klis}}{1999}]{1999ApJ...522..965W}
{Wijnands} R.,  {van der Klis} M.,  1999, \mn@doi [\apj] {10.1086/307698},
  \href {https://ui.adsabs.harvard.edu/abs/1999ApJ...522..965W} {522, 965}

\bibitem[\protect\citeauthoryear{{Wijnands} et~al.,}{{Wijnands}
  et~al.}{1998a}]{1998ApJ...493L..87W}
{Wijnands} R.,  et~al., 1998a, \mn@doi [\apjl] {10.1086/311138}, \href
  {https://ui.adsabs.harvard.edu/abs/1998ApJ...493L..87W} {493, L87}

\bibitem[\protect\citeauthoryear{{Wijnands}, {M{\'e}ndez}, {van der Klis},
  {Psaltis}, {Kuulkers}  \& {Lamb}}{{Wijnands}
  et~al.}{1998b}]{1998ApJ...504L..35W}
{Wijnands} R.,  {M{\'e}ndez} M.,  {van der Klis} M.,  {Psaltis} D.,  {Kuulkers}
  E.,   {Lamb} F.~K.,  1998b, \mn@doi [\apjl] {10.1086/311564}, \href
  {https://ui.adsabs.harvard.edu/abs/1998ApJ...504L..35W} {504, L35}

\bibitem[\protect\citeauthoryear{{Wilms}, {Allen}  \& {McCray}}{{Wilms}
  et~al.}{2000}]{2000ApJ...542..914W}
{Wilms} J.,  {Allen} A.,   {McCray} R.,  2000, \mn@doi [\apj] {10.1086/317016},
  \href {https://ui.adsabs.harvard.edu/abs/2000ApJ...542..914W} {542, 914}

\bibitem[\protect\citeauthoryear{{Yadav} et~al.,}{{Yadav}
  et~al.}{2016}]{2016ApJ...833...27Y}
{Yadav} J.~S.,  et~al., 2016, \mn@doi [\apj] {10.3847/0004-637X/833/1/27},
  \href {https://ui.adsabs.harvard.edu/abs/2016ApJ...833...27Y} {833, 27}

\bibitem[\protect\citeauthoryear{{Yadav}, {Agrawal}, {Antia}, {Manchanda},
  {Paul}  \& {Misra}}{{Yadav} et~al.}{2017}]{2017CSci..113..591Y}
{Yadav} J.~S.,  {Agrawal} P.~C.,  {Antia} H.~M.,  {Manchanda} R.~K.,  {Paul}
  B.,   {Misra} R.,  2017, \mn@doi [Current Science]
  {10.18520/cs/v113/i04/591-594}, \href
  {https://ui.adsabs.harvard.edu/abs/2017CSci..113..591Y} {113, 591}

\bibitem[\protect\citeauthoryear{{Zdziarski}, {Johnson}  \&
  {Magdziarz}}{{Zdziarski} et~al.}{1996}]{1996MNRAS.283..193Z}
{Zdziarski} A.~A.,  {Johnson} W.~N.,   {Magdziarz} P.,  1996, \mn@doi [\mnras]
  {10.1093/mnras/283.1.193}, \href
  {https://ui.adsabs.harvard.edu/abs/1996MNRAS.283..193Z} {283, 193}

\bibitem[\protect\citeauthoryear{{Zhang}, {Strohmayer}  \& {Swank}}{{Zhang}
  et~al.}{1998}]{1998ApJ...500L.167Z}
{Zhang} W.,  {Strohmayer} T.~E.,   {Swank} J.~H.,  1998, \mn@doi [\apjl]
  {10.1086/311423}, \href
  {https://ui.adsabs.harvard.edu/abs/1998ApJ...500L.167Z} {500, L167}

\bibitem[\protect\citeauthoryear{{van der Klis}}{{van der
  Klis}}{2000}]{2000ARA&A..38..717V}
{van der Klis} M.,  2000, \mn@doi [\araa] {10.1146/annurev.astro.38.1.717},
  \href {https://ui.adsabs.harvard.edu/abs/2000ARA&A..38..717V} {38, 717}

\bibitem[\protect\citeauthoryear{{van der Klis}}{{van der
  Klis}}{2004}]{2004astro.ph.10551V}
{van der Klis} M.,  2004, arXiv e-prints, \href
  {https://ui.adsabs.harvard.edu/abs/2004astro.ph.10551V} {pp
  astro--ph/0410551}

\makeatother
\end{thebibliography}

%\begin{thebibliography}{99}
%\end{thebibliography}

%%%%%%%%%%%%%%%%%%%%%%%%%%%%%%%%%%%%%%%%%%%%%%%%%%

%%%%%%%%%%%%%%%%% APPENDICES %%%%%%%%%%%%%%%%%%%%%

\appendix

\section{Some extra material}

%If you want to present additional material which would interrupt the flow of the main paper, it can be placed in an Appendix which appears after the list of references.

%%%%%%%%%%%%%%%%%%%%%%%%%%%%%%%%%%%%%%%%%%%%%%%%%%
\begin{table*}
\centering
\caption{ Best-fitting spectral parameters corresponding to the best-fitting model consisting of absorbed blackbody radiation, multicolour disc blackbody and a power law component \texttt{tbabs*(bbodyrad+diskbb+powerlaw)} for all eight {\em NICER} observations of GX 349+2 (see Section \ref{section3.3.2}).}
\begin{tabular}{|p{0.4cm}|c|c|c|c|c|c|c|c|}
\hline
Obs No.& $\text{nH}^{a}$ &$\text{kT}^{b}$&$\text{BB}$ $\text{Norm}^{c}$&$\text{kT}_{in}^{d}$&$\text{DBB}$ $\text{Norm}^{e}$&$\Gamma^{f}$&$\text{Norm}^{g}$& Chi-sq (DOF) \\
&$10^{22}$ cm$^{-2}$ &keV& &keV &&&  \\
\hline
N1&$1.63_{-0.05}^{+0.05}$&$1.78_{-0.03}^{+0.03}$&$146.72_{-12.92}^{+12.38}$&$1.06_{-0.04}^{+0.05}$&$377.01_{-55.91}^{+62.35}$&$4.78_{-0.19}^{+0.19}$&$2.93_{-0.36}^{+0.38}$&1.25 (721)\\
\\
N2&$1.52_{-0.04}^{+0.04}$&$1.83_{-0.03}^{+0.04}$&$149.46_{-7.20}^{+13.23}$&$1.16_{-0.05}^{+0.06}$&$282.14_{-43.87}^{+48.25}$&$4.29_{-0.19}^{+0.18}$&$2.18_{-0.28}^{+0.30}$&1.25 (721)\\
\\
N3&$1.43_{-0.04}^{+0.04}$&$1.88_{-0.03}^{+0.03}$&$124.07_{-10.02}^{+9.81}$&$1.20_{-0.04}^{+0.05}$&$245.02_{-28.82}^{+31.75}$&$4.14_{-0.15}^{+0.15}$&$1.60_{-0.20}^{+0.21}$&1.42 (721)\\
\\
N4&$1.59_{-0.04}^{+0.04}$&$1.79_{-0.03}^{+0.03}$&$121.70_{-10.78}^{+11.06}$&$1.11_{-0.04}^{+0.04}$&$340.44_{-42.10}^{+48.68}$&$4.64_{-0.17}^{+0.17}$&$2.73_{-0.31}^{+0.34}$&1.21 (721)\\
\\
N5&$1.59_{-0.04}^{+0.04}$&$1.85_{-0.03}^{+0.04}$&$98.26_{-8.99}^{+9.61}$&$1.13_{-0.04}^{+0.04}$&$320.75_{-35.62}^{+42.34}$&$4.70_{-0.15}^{+0.16}$&$2.61_{-0.28}^{+0.32}$&1.24 (721)\\
\\
N6&$1.52_{-0.04}^{+0.03}$&$1.78_{-0.01}^{+0.01}$&$145.09_{-5.89}^{+5.69}$&$1.06_{-0.02}^{+0.02}$&$385.94_{-28.97}^{+30.68}$&$4.64_{-0.13}^{+0.12}$&$2.01_{-0.22}^{+0.22}$&1.58 (721)\\
\\
N7&1.11&1.81&135.34&1.07&338.40&9.5&0.01&2.33  (721)\\
\\
N8&$1.67_{-0.04}^{+0.04}$&$1.87_{-0.04}^{+0.05}$&$94.90_{-10.90}^{+10.96}$&$1.15_{-0.04}^{+0.05}$&$304.17_{-40.60}^{+45.70}$&$4.91_{-0.16}^{+0.16}$&$3.24_{-0.33}^{+0.36}$&1.32 (721)\\

\hline
\end{tabular}
\begin{flushleft}
\footnotesize{$^a$ Neutral hydrogen column density}\\
\footnotesize{$^b$ Blackbody Temperature } \\
\footnotesize{$^c$ Blackbody Normalisation}\\
\footnotesize{$^d$ inner disc Temperature }\\
\footnotesize{$^e$ Disc Blackbody Normalisation}\\
\footnotesize{$^f$ Power-law Index}\\
\footnotesize{$^g$ Power-law Normalisation}\\
\end{flushleft}
\label{table7}
\end{table*}

\begin{table*}
\caption{Best-fitting spectral parameters corresponding to the best-fitting model consisting of absorbed blackbody radiation, multicolour disc blackbody and a power law component \texttt{tbabs*edge*(bbodyrad+diskbb+powerlaw)} for all eight {\em NICER} observations of GX 349+2 (see Section \ref{section3.3.2}).}
\centering
\begin{tabular}{|p{0.6cm}|p{0.8cm}|c|c|c|c|c|c|c|c|c|c|}
\hline
Obs No. & $\text{nH}^{a}$ & edge & maxtau & $\text{kT}^{b}$ & $\text{BB}$ $\text{Norm}^{c}$ & $\text{kT}_{in}^{d}$ & $\text{DBB}$ $\text{Norm}^{e}$ & $\Gamma^{f}$ & $\text{Norm}^{g}$ & Chi-sq \\
&$10^{22}$ cm$^{-2}$ &&&keV& &keV &&&  \\
\hline
N1&$1.88^{+0.04}_{-0.04}$&$0.78^{+0.01}_{<-0.01}$&$0.36^{+0.04}_{-0.04}$&$1.71^{+0.02}_{-0.02}$&$179.53^{+10.38}_{-10.49}$&$0.92^{+0.03}_{-0.03}$&$702.31^{+95.49}_{-85.34}$&$5.89^{+0.16}_{-0.16}$&$6.75^{+0.64}_{-0.62}$& 1.07(719)\\
\\
N2&$1.81^{+0.04}_{-0.04}$&$0.78^{<+0.01}_{<-0.01}$&$0.34^{+0.03}_{-0.03}$&$1.76^{+0.02}_{-0.02}$&$183.27^{+10.52}_{-11.33}$&$0.99^{+0.04}_{-0.03}$&$542.81^{+74.96}_{-70.20}$&$5.51^{+0.15}_{-0.16}$&$5.81^{+0.55}_{-0.52}$& 1.03 (719)\\
\\
N3&$1.82^{+0.03}_{-0.03}$&$0.78^{<+0.01}_{<-0.01}$&$0.35^{+0.02}_{-0.02}$&$1.80^{+0.02}_{-0.02}$&$156.90^{+7.77}_{-8.15}$&$1.01^{+0.03}_{-0.03}$&$497.36^{+53.68}_{-50.50}$&$5.59^{+0.12}_{-0.12}$&$5.92^{+0.45}_{-0.43}$&1.00 (719)\\
\\
N4&$1.88^{+0.03}_{-0.03}$&$0.78^{<+0.01}_{<-0.01}$&$0.36^{+0.03}_{-0.03}$&$1.70^{+0.02}_{-0.02}$&$160.44^{+8.72}_{-9.01}$&$0.94^{+0.02}_{-0.02}$&$676.49^{+73.31}_{-68.45}$&$5.87^{+0.13}_{-0.13}$&$6.92^{+0.56}_{-0.53}$&0.95 (719)\\
\\
N5&$1.89_{-0.03}^{+0.03}$&$0.78_{<-0.01}^{<+0.01}$&$0.37_{-0.03}^{+0.03}$&$1.74_{-0.02}^{+0.02}$&$136.32_{-8.16}^{+8.02}$&$0.95_{-0.02}^{+0.02}$&$638.06_{-61.48}^{+66.05}$&$5.92_{-0.13}^{+0.12}$&$6.86_{-0.53}^{+0.53}$&0.97 (719)\\
\\
N6&$1.94^{+0.02}_{-0.02}$&$0.78^{<+0.01}_{<-0.01}$&$0.37^{+0.02}_{-0.02}$&$1.72^{+0.01}_{-0.01}$&$175.37^{+4.88}_{-4.98}$&$0.91^{+0.02}_{-0.01}$&$774.14^{+54.16}_{-51.52}$&$6.12^{+0.09}_{-0.10}$&$7.55^{+0.46}_{-0.44}$&1.04 (719)\\
\\
N7&$1.99^{+0.02}_{-0.02}$&$0.78^{<+0.01}_{<+0.01}$&$0.40^{+0.02}_{-0.02}$&$1.74^{<+0.01}_{<-0.01}$&$167.06^{+3.32}_{-3.40}$&$0.88^{<+0.01}_{-<0.01}$&$860.54^{+46.40}_{-44.76}$&$6.32^{+0.08}_{-0.08}$&$8.36^{+0.42}_{-0.41}$& 1.41(719)\\
\\
N8&$1.92^{+0.03}_{-0.03}$&$0.78^{<+0.01}_{<-0.01}$&$0.40^{+0.04}_{-0.03}$&$1.75^{+0.03}_{-0.02}$&$134.94^{+8.83}_{-9.13}$&$0.96^{+0.03}_{-0.02}$&$615.21^{+68.76}_{-64.38}$&$6.04^{+0.12}_{-0.13}$&$7.43^{+0.58}_{-0.55}$&1.07(719)\\
\hline
\end{tabular}
\begin{flushleft}
\footnotesize{$^a$ Neutral hydrogen column density}\\
\footnotesize{$^b$ Blackbody Temperature } \\
\footnotesize{$^c$ Blackbody Normalisation}\\
\footnotesize{$^d$ inner disc Temperature }\\
\footnotesize{$^e$ Disc Blackbody Normalisation}\\
\footnotesize{$^f$ Power-law Index}\\
\footnotesize{$^g$ Power-law Normalisation}\\
\end{flushleft}
\label{table8}
\end{table*}

\begin{figure*}
    \hspace*{-1.4cm}
    \begin{tabular}{lr}    
    \includegraphics[width=0.38\textwidth]{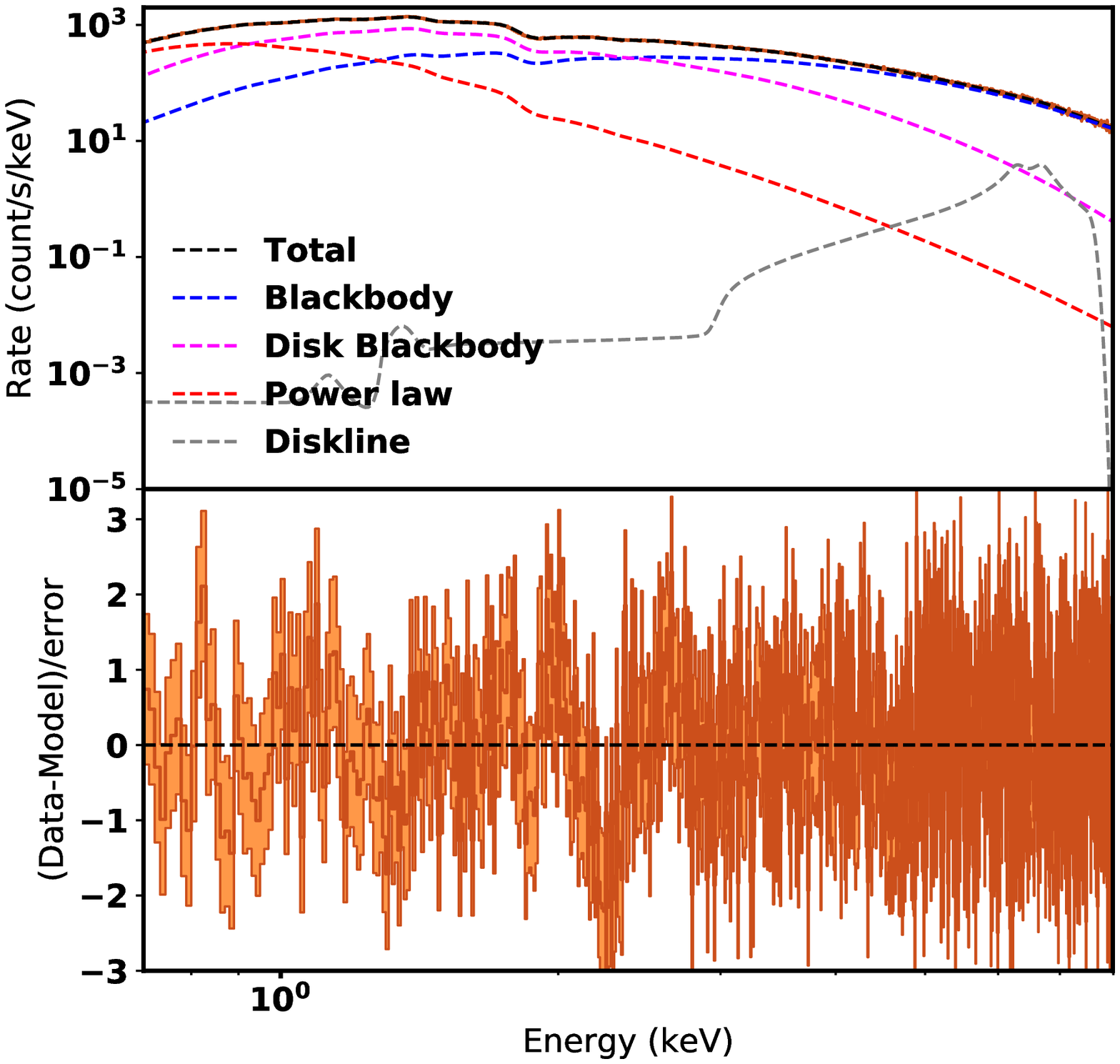}
    \includegraphics[width=0.38\textwidth]{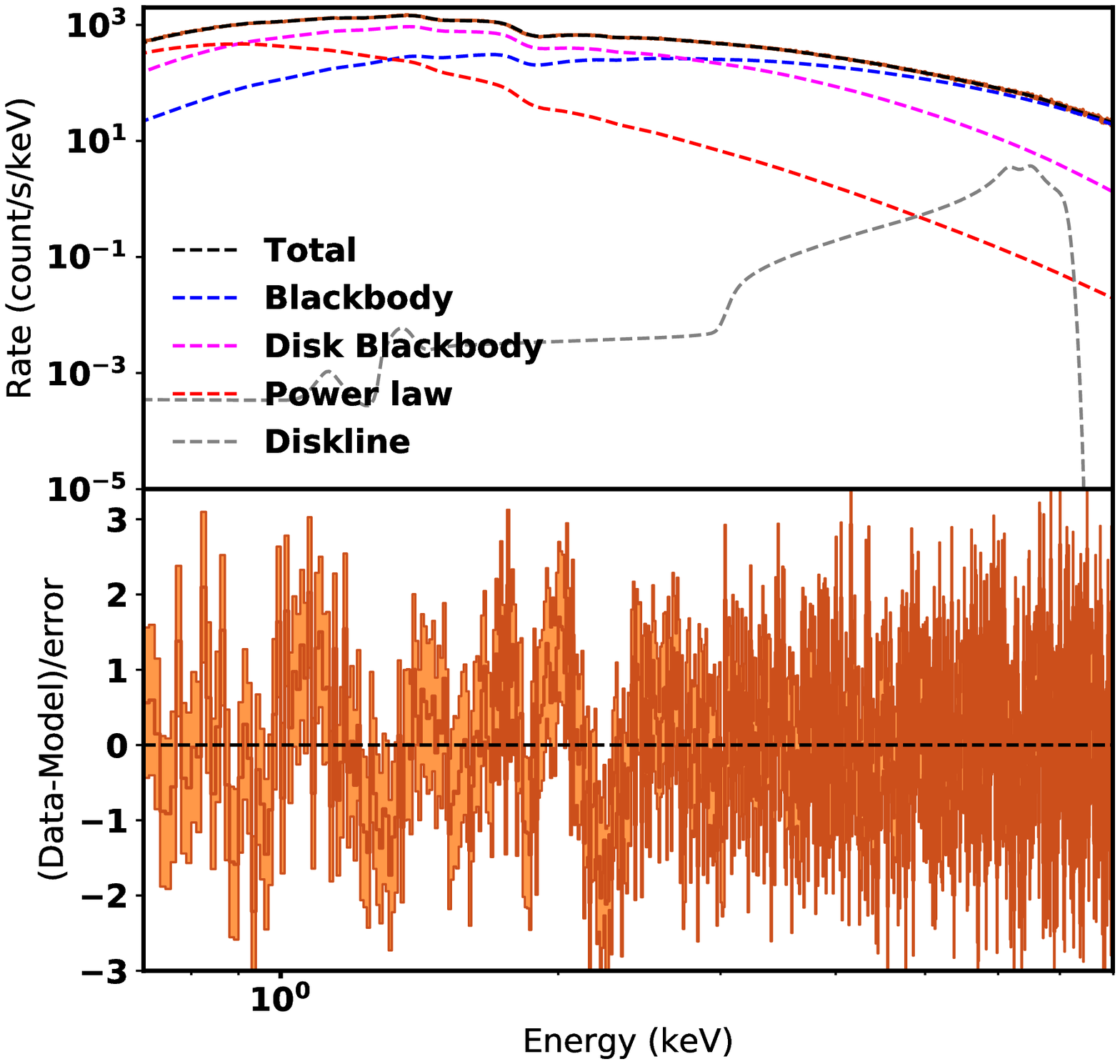}
    \includegraphics[width=0.38\textwidth]{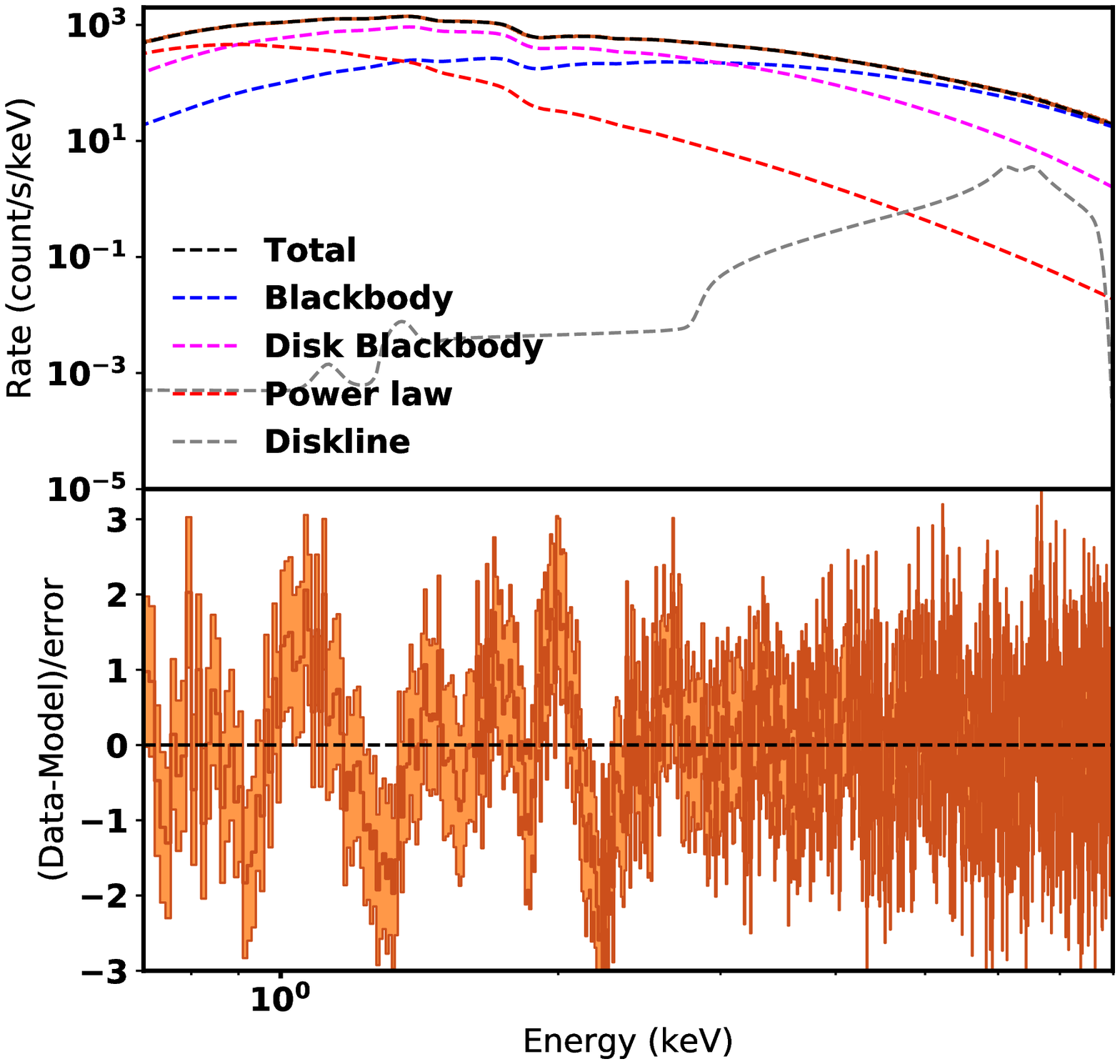}\\
     \includegraphics[width=0.38\textwidth]{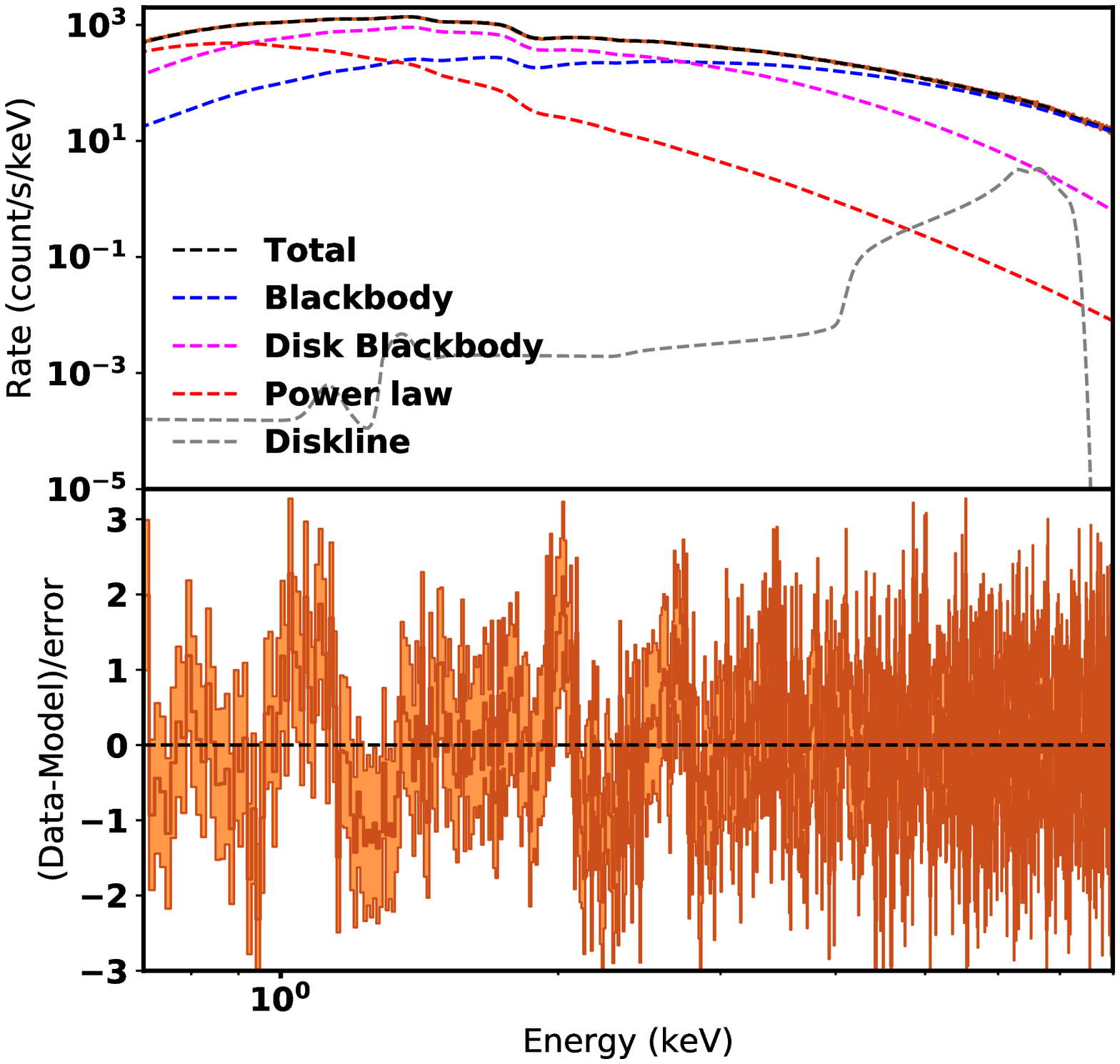}
      \includegraphics[width=0.38\textwidth]{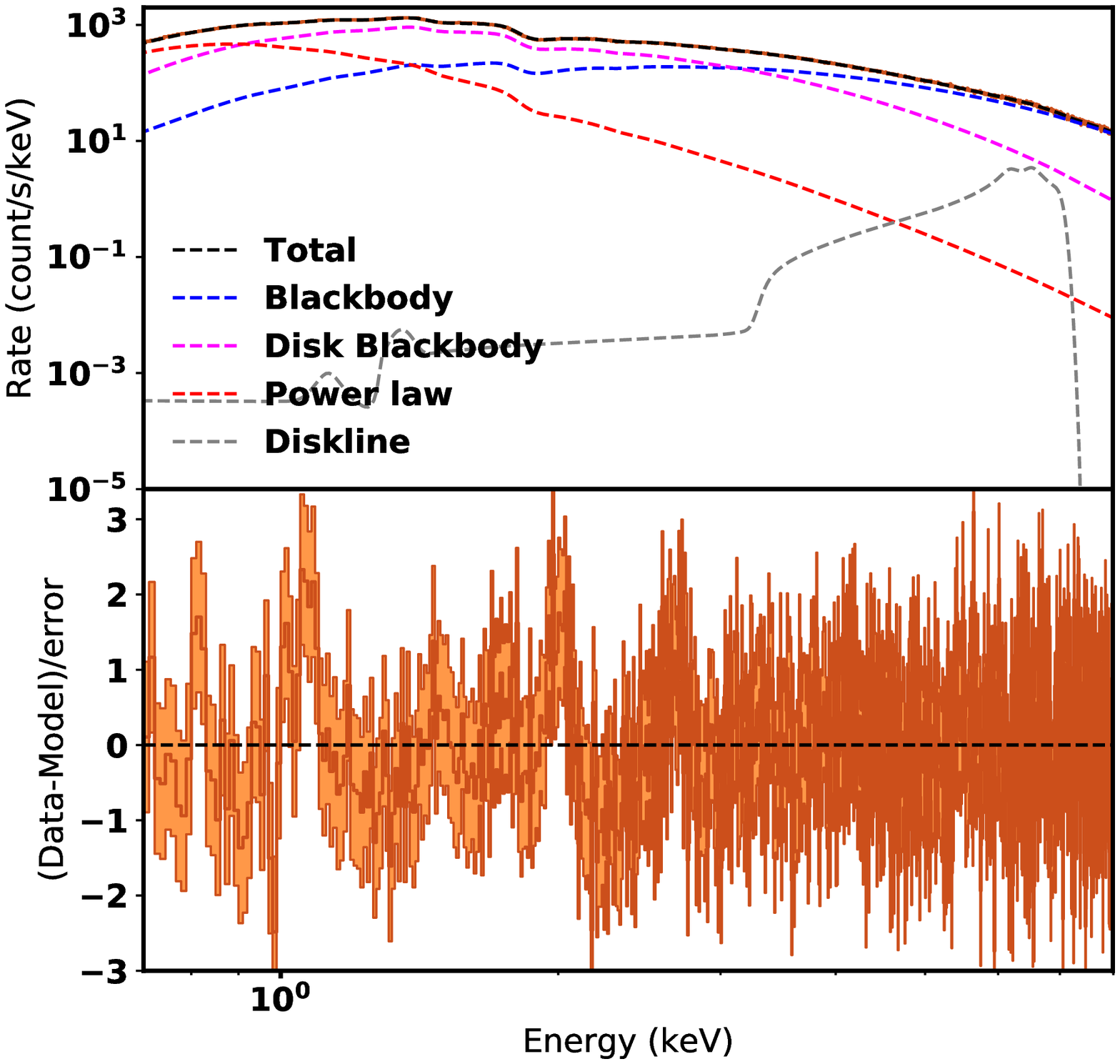}
     \includegraphics[width=0.38\textwidth]{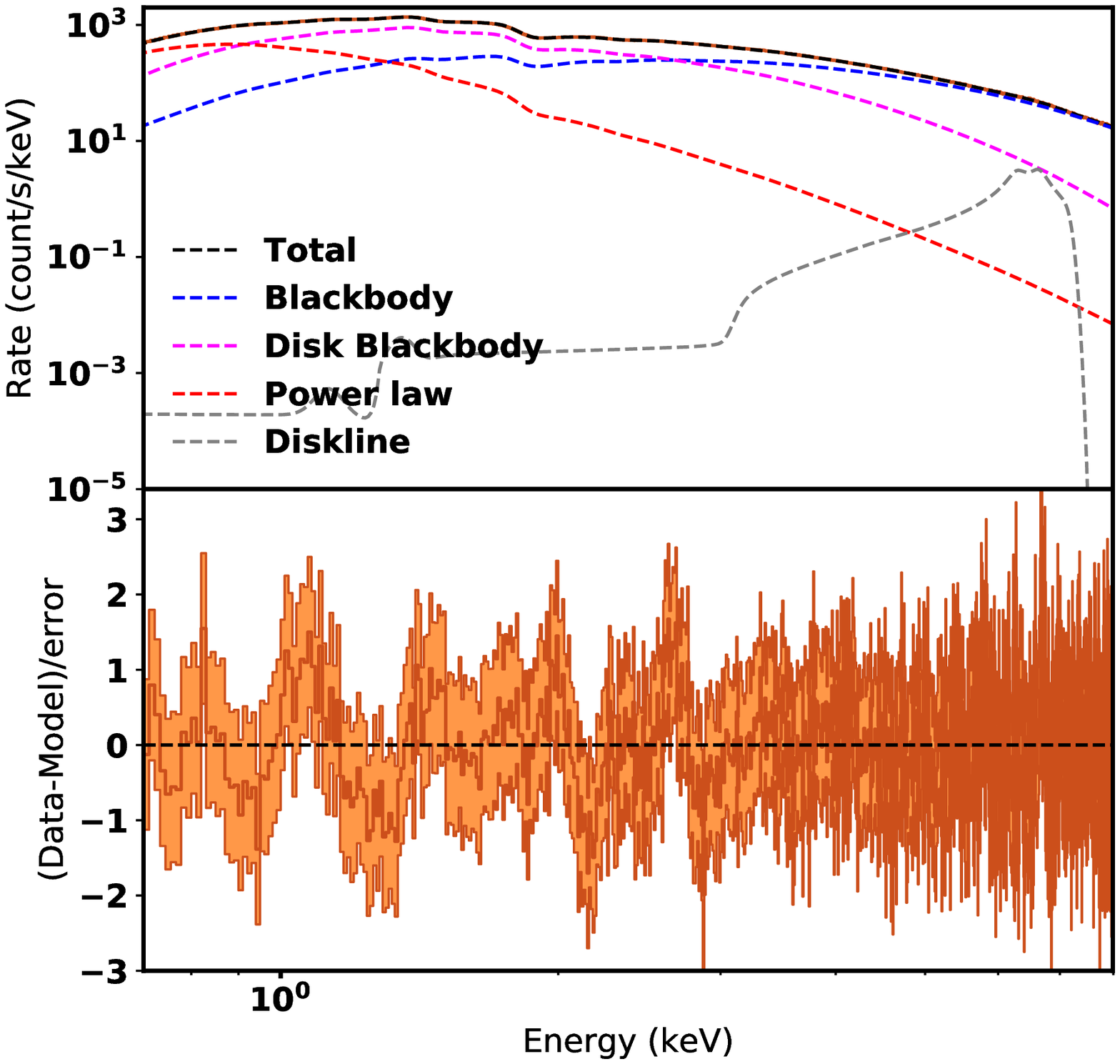}\\
     \includegraphics[width=0.38\textwidth]{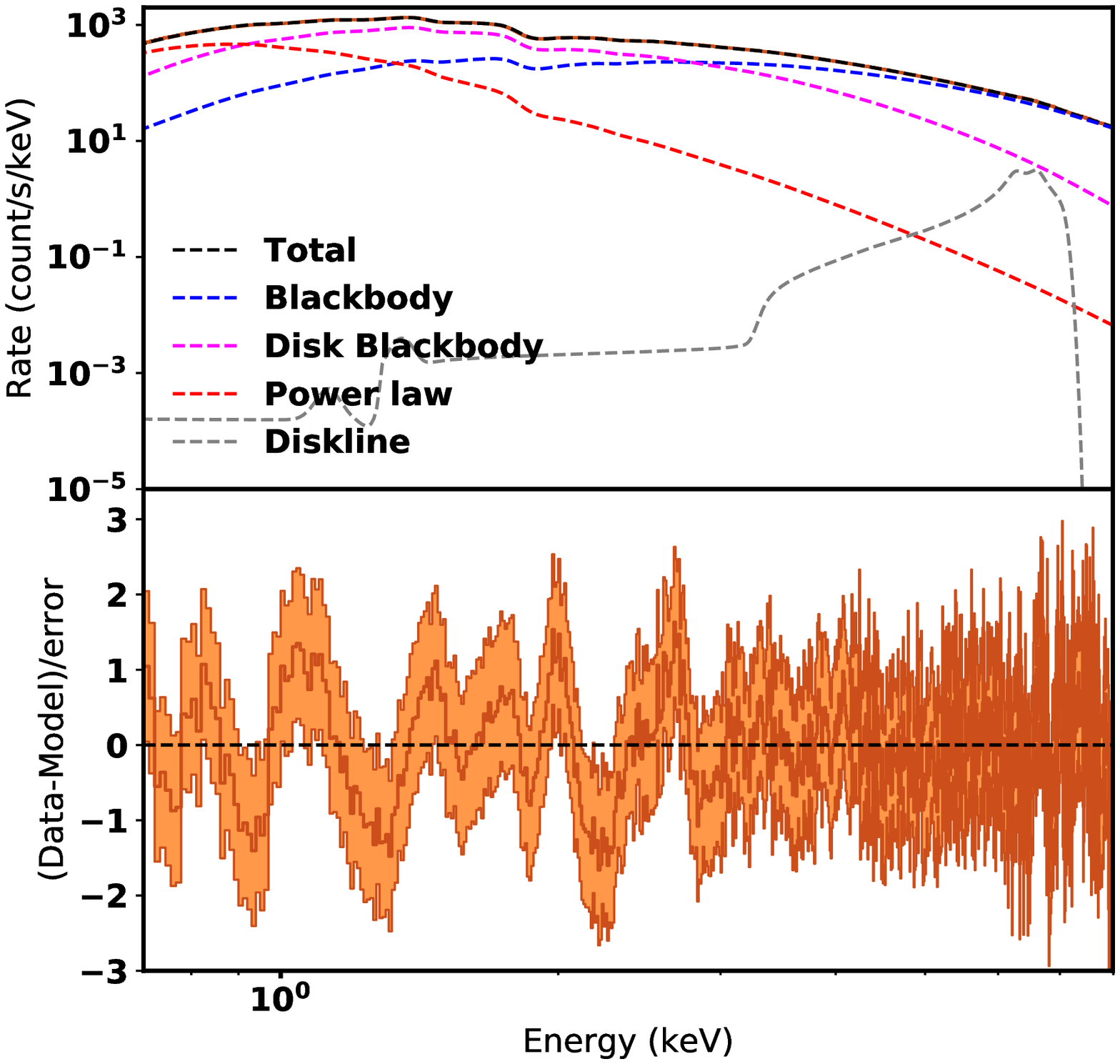}
     \includegraphics[width=0.38\textwidth]{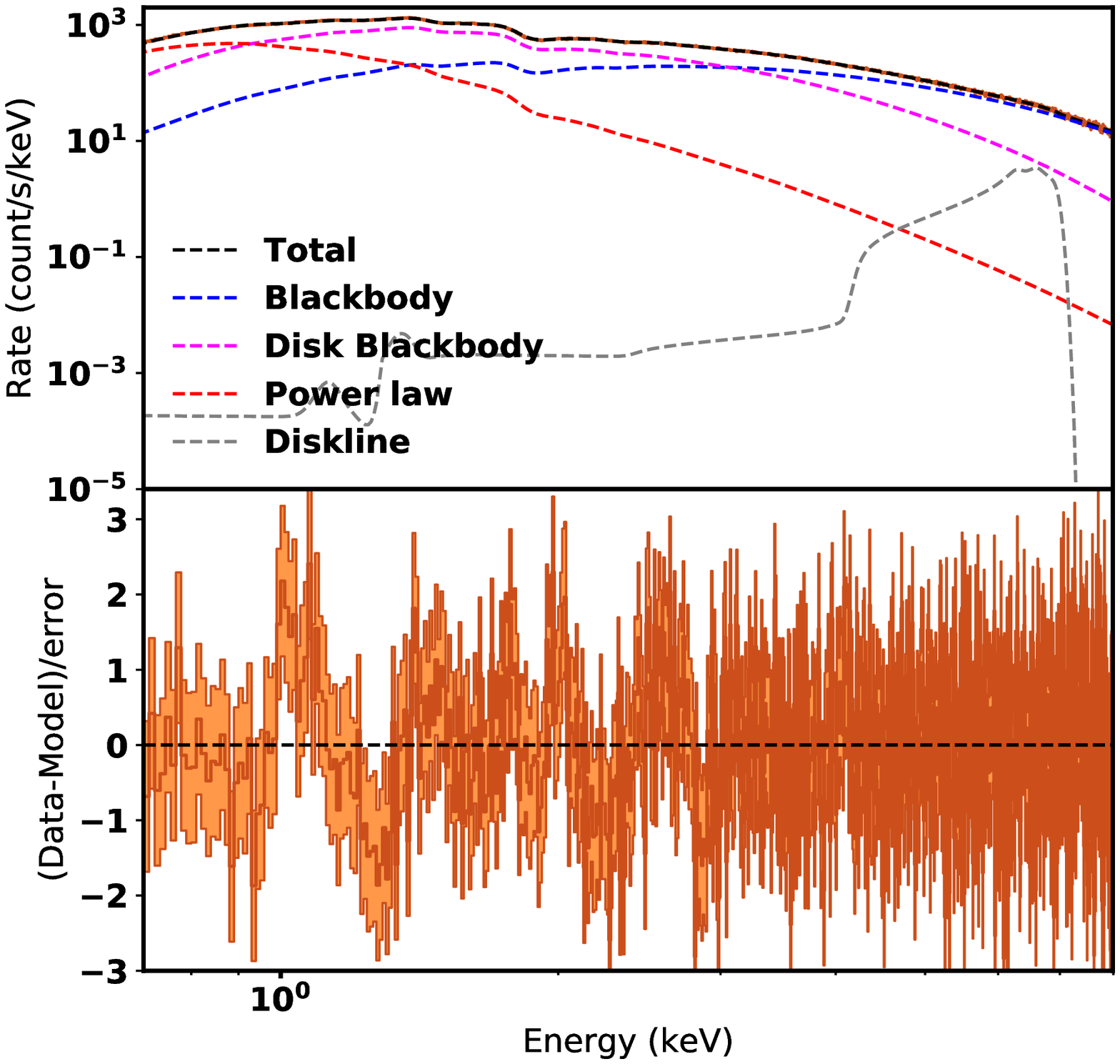}\\
    \end{tabular}
    \caption{The 0.7-8 keV X-ray spectrum of GX 349+2 for all 8 {\em NICER} observations  (left to right, top to bottom). The shaded region shows the 1 $\sigma$ error. Spectra are fitted with \texttt{tbabs*edge*(bbodyrad+diskbb+powerlaw+diskline)} model (upper panel) and the error weighted residuals (lower panel). The individual additive components blackbody radiation, multicolour disc blackbody, power law, Gaussian, and diskline are also shown in blue, magenta, red, and grey by the dotted curves (see Section \ref{section3.3.2}).} 
    \label{figurea}
\end{figure*}

% Don't change these lines
\bsp	% typesetting comment
\label{lastpage}
\end{document}